\documentclass[smallcondensed, nospthms]{svjour3}

\usepackage{amsmath}
\usepackage{amssymb}
\newcommand{\bfm}[1]{\mbox{\boldmath{$#1$}}}
\newcommand{\ds}{\displaystyle}	

\allowdisplaybreaks

\usepackage{mathtools}

\DeclarePairedDelimiter\floor{\lfloor}{\rfloor}

\usepackage{amsthm}

\usepackage{graphicx}
\usepackage{epstopdf}
\usepackage{subfigure}
\usepackage{multirow}

\usepackage{natbib}
\bibpunct{(}{)}{;}{a}{}{,}

\usepackage{appendix}
\usepackage[table, x11names]{xcolor}
\usepackage{longtable}
\usepackage{tabu}
\usepackage{color}
\usepackage[justification=centering]{caption}

\def\H{{\mathcal H}}
\def\R{{\mathcal R}}
\def\O{{\mathcal O}}

\def\s{\text{s}}
\def\S{\text{S}}
\def\M{\text{M}}
\def\J{\text{J}}

\begin{document}

\title{The dynamical structure of the MEO region:}

\subtitle{long-term stability, chaos, and transport}

\titlerunning{The dynamical structure of the MEO region}

\author{J\'{e}r\^{o}me Daquin \and 
			Aaron J. Rosengren \and \\ 
			Florent Deleflie \and 
			Elisa Maria Alessi \and \\
			Giovanni B. Valsecchi \and 
			Alessandro Rossi}
\authorrunning{J. Daquin et al.}

\institute{J. Daquin 
\at Thales Servies, 3 Impasse de l'Europe, 31400 Toulouse, France \\ 
\email{jerome.daquin@imcce.fr}
\and A. J. Rosengren \and E. M. Alessi \and  G. B. Valsecchi \and A. Rossi
\at IFAC-CNR, Via Madonna del Piano 10, 50019 Sesto Fiorentino (FI), Italy
\and J. Daquin \and F. Deleflie
\at IMCCE/Observatoire de Paris, Universit\'{e} Lille1, 1 Impasse de l'Observatoire, 59000 Lille, France
\and G. B. Valsecchi
\at IAPS-INAF, Via Fosso del Cavaliere 100, 00133 Roma, Italy}

\date{Received: date / Accepted: date}

\maketitle

\begin{abstract}
It has long been suspected that the Global Navigation Satellite Systems exist in a background of complex resonances and chaotic motion; yet, the precise dynamical character of these phenomena remains elusive. Recent studies have shown that the occurrence and nature of the resonances driving these dynamics depend chiefly on the frequencies of nodal and apsidal precession and the rate of regression of the Moon's nodes. Woven throughout the inclination and eccentricity phase space is an exceedingly complicated web-like structure of lunisolar secular resonances, which become particularly dense near the inclinations of the navigation satellite orbits. A clear picture of the physical significance of these resonances is of considerable practical interest for the design of disposal strategies for the four constellations. Here we present analytical and semi-analytical models that accurately reflect the true nature of the resonant interactions, and trace the topological organization of the manifolds on which the chaotic motions take place. We present an atlas of FLI stability maps, showing the extent of the chaotic regions of the phase space, computed through a hierarchy of more realistic, and more complicated, models, and compare the chaotic zones in these charts with the analytical estimation of the width of the chaotic layers from the heuristic Chirikov resonance-overlap criterion. As the semi-major axis of the satellite is receding, we observe a transition from stable Nekhoroshev-like structures at three Earth radii, where regular orbits dominate, to a Chirikov regime where resonances overlap at five Earth radii. From a numerical estimation of the Lyapunov times, we find that many of the inclined, nearly circular orbits of the navigation satellites are strongly chaotic and that their dynamics are unpredictable on decadal timescales.


\end{abstract}

\section{Introduction}

Among the earliest problems in astrodynamics were, quite naturally, those concerned with the motion of an artificial satellite in the gravitational field of the oblate Earth; and indeed their solutions in the hands of Brouwer, Garfinkel, Kozai, and others were intimately bound with the rigorous development of artificial satellite theory \citep{aJ88}. These considerations placed emphasis on the construction of increasingly more accurate analytical theories, in the style of our forebears, valid in idealized situations or on short mission timescales. Such an approach was of course justified by the need of astrodynamical practice, but in recent years astrodynamics has had to face new problems concerning the long-term motion of space debris which forced it to abandon the somewhat utilitarian investigations mentioned above. This renaissance has been spurred on not only by the sobering implications of space debris, but also by the ability to perform computer simulations of ever-greater sophistication and by the realization that chaos has played a fundamental role in the dynamical evolution of the Solar System \citep{aM02}.

The proliferation of space debris orbiting Earth has stimulated a deeper understanding of the dynamical environments occupied by artificial satellites \citep{sB01}, and the smaller disturbing forces arising from lunisolar gravity have became particularly interesting now \citep{sB99}. For near-Earth satellites, the perturbing effects of the Sun and Moon, often assumed negligible in comparison to that of the Earth's oblateness, may have small consequences early but profound consequences late. The subtle effects that accumulate over long periods of time are known as resonances and while their basic theory was developed during the early Sputnik era \citep{pM59}, a detailed and systematic investigation of their dynamical effects is hitherto missing. The problem is particularly timely since, in the past decade, a number of numerical studies have shown that the medium-Earth orbits (MEOs) and graveyard regions of the navigation satellites are unstable \citep[q.v.][and references therein]{fD11}. This instability manifests itself as an apparent chaotic growth of the eccentricity on decadal to centennial timescales. 

To identify the underlying dynamical mechanism responsible for the noted instability in the MEO region, \citet{aR15} investigated the main resonances that organize and control the long-term dynamics of navigation satellites. They confirmed the complex resonant structure identified nearly twenty years ago by \citet{tEkH97}, which has been unjustifiably overlooked by the space debris community, and showed that these lunisolar resonant harmonics together with those studied by \citet{fDaM93} interact to produce a dense stochastic web in inclination and eccentricity phase space. The close spacing of the resonant harmonics suggested the potential for significant chaos in the orbital eccentricities and inclinations. This was partially demonstrated numerically through stroboscopic Poincar\'{e} maps, showing these resonances to be the preferential routes for transport in phase space, by which orbits can explore large regions jumping from one resonance domain to another \citep{aR15}. The nature of the chaotic layer, however, was not explored. 

The present investigation is a first attempt towards understanding the chaotic structure of the phase space near lunisolar secular resonances. This is studied analytically by examining the width of the chaotic layer from the overlap of neighboring resonances \citep{bC79_universal,rM08} and numerically using the fast Lyapunov indicator (FLI) \citep{cF97}. To cast light on the speculations made in \citet{aR15} and to gain insight into the structure, extent, and evolution of the chaotic regions, we compute FLI stability maps using a hierarchy of more realistic dynamical models. We make a detailed comparison between the analytical and numerical results, and test the predictions of the heuristic Chirikov criterion in systems of more than two degrees of freedom.

\section{The Chirikov resonance-overlap criterion} \label{sec:chirikov_criterion}

The long-term evolution of resonant orbits can be quite complex, and what has become clear over the past few decades is that the phenomenon of chaos is tied fundamentally to the dynamics of resonances \citep{bC79_universal,aM02}. As an instance we can mention the problem of motion of asteroids in the main belt between Mars and Jupiter on which so much of our astronomical knowledge is based; and here the light thrown by purely dynamical considerations upon, for example, the problem of Kirkwood's gaps in the distribution of the asteroidal semi-major axes reveals the sorts of resonant interactions that have helped to shape the Solar System. The physical basis of chaos and instability in nearly integrable, multidimensional Hamiltonian systems, of which the Sun-Jupiter-Saturn-Asteroid four-body problem is one example, is the overlapping of nonlinear resonances \citep{bC79_universal}. 

The overlapping of resonances as the mechanism of onset and evolution of chaos has been known since the late 1960s. This important result from nonlinear dynamics can be made intelligible without entering into the lengthy and subtle mathematical discussion demanded by an exhaustive treatment of the subject \citep{rM08}. Astronomers long ago noted the connection between the dynamics of a single (isolated) resonance and the libration of a mathematical pendulum. An expansion of the Hamiltonian around the resonant location and a canonical transformation reduces the Hamiltonian to a pendulum-like system. The resonance angle librates when the frequencies are nearly commensurate; the domain or width of the resonance being the distance from exact resonance to the separatrix (a homoclinic curve separating the domain of circulation and libration). Whereas a single resonance exhibits a well-behaved, pendulum-like motion, when the system is moving within a region of the phase space where two or more resonances are present, the single resonance theory breaks down and any attempt to describe the interaction rigorously is bound to fail. The nature of this breakdown manifests itself in numerical solutions as zones of chaotic motion near the boundary of each interacting resonance. The Chirikov resonance-overlap criterion simply states that chaos will ensue if two dynamically relevant harmonic angles, when neglecting the interaction between them, are both analytically calculated to be librating in the same region of phase space \citep{bC79_universal}. 
 
\subsection{Location of lunisolar resonance centers} \label{sec:skeleton}

Our aim here, following \citet{tEkH97} and \citet{aR15}, is to identify the significant resonances so that we can apply the resonance-overlap criterion and analytically determine stability boundaries. We focus on the MEO region between roughly three and five Earth radii and, as a result, are permitted to make certain approximations and assumptions. There are two principal classes of resonances which affect the motion of satellites in these orbital altitudes: tesseral resonances, where the orbital periods (or mean motions) of the satellites are commensurable to the Earth's rotation rate \citep{mL88,aCcG14}, and lunisolar secular resonances, which involve commensurabilities amongst the slow frequencies of orbital precession of a satellite and the perturbing body \citep{tEkH97}. The dynamical system with tesseral, zonal, and lunisolar perturbations is, however, characterized by two different timescales and two independent modes. Consequently, we focus here only on the resonant effects of secular origin, which dominate the eccentricity and inclination evolution on long timescales \citep{kB15}. 

The analytical methodology for computing the lunisolar resonant effects is based on a theoretical series development of the lunar and solar disturbing functions (the negative potential functions of the perturbing accelerations). These harmonic series expansions are expressed most simply, in their dependence on the orbital elements, when the latter are defined with respect to the ecliptic plane for the Moon and with respect to the equatorial plane for the satellite and the Sun \citep{sH80}. For our purposes, we can truncate these series to second order in the ratio of semi-major axes, so that the lunar and solar potentials are approximated by quadrupole fields. In the secular approximation, short-periodic terms which depend on the fast orbital phases (i.e., the mean anomalies of both the satellite and the perturbing bodies) can be readily averaged over and dropped from the disturbing functions \citep{vA06}. 

The secular and quadrupole order of approximation for the lunar disturbing function expansion follows from \citet{mL89} and has the form
\begin{align}
	\label{eq:Lane}
	\R_\M = \sum_{m = 0}^2 \sum_{s = 0}^2 \sum_{p = 0}^2 h_{2-2p, m, \pm s}^\M  
		\cos \Phi_{2-2p, m, \pm s}^\M,
\end{align}
with {\itshape harmonic angle}
\begin{equation}
	\Phi_{2-2p, m, \pm s}^\M = (2 - 2 p) \omega + m \Omega \pm s (\Omega_\M - \pi/2) - y_s \pi, 
\end{equation} 
and associated {\itshape harmonic coefficient}
\begin{equation}
\label{eq:harm_coeff_lunar}
\begin{array}{l}
	\ds h_{2-2p, m, \pm s}^\M
	= \frac{\mu_\M a^2}{a_\M^3 (1 - e_\M^2)^{3/2}}
		(-1)^{\floor*{m/2}} \epsilon_m \frac{(2 - s)!}{(2 + m)!}
		F_{2,m,p} (i) F_{2,s,1} (i_\M) H_{2,p,2p-2} (e) (-1)^{m + s} U_2^{m, \mp s} (\epsilon),
\end{array}
\end{equation}
where $m$, $s$, and $p$ are integers, $\floor{\cdot}$ is the floor operator, and the quantities $\epsilon_m$ and $y_s$ are such that 
\begin{align*}
	\epsilon_m & = \left\{ \begin{array}{cl} 1 & \text{ if } m = 0, \\ 2 & \text{ if } m \neq 0, \end{array} \right. \\
	y_s & = \left\{ \begin{array}{cl} 0 & \text{ if $s$ is even}, \\ 1/2 & \text{ if $s$ is odd}. \end{array} \right.
\end{align*}
The semi-major axis $a$, eccentricity $e$, inclination $i$, argument of perigee $\omega$, and longitude of ascending node $\Omega$ are the satellite's orbital elements relative to the Earth's equator. The subscript $\M$ refers to the Moon's orbital parameters ($i_\M$ and $\Omega_\M$ are measured relative to the ecliptic plane), $\mu_\M$ is the Newtonian gravitational constant times the Moon's mass, and $\epsilon$ is the obliquity of the ecliptic. The quantities $F_{2,m,p} (i)$ and $F_{2,s,1} (i_\M)$ are the Kaula inclination functions \citep{wK66}, $H_{2,p,2p-2} (e)$ is the zero-order Hansen coefficient $X_0^{2,2-2p} (e)$ \citep{sH80}, and $ U_2^{m,\pm s}$ is the Giacaglia function, required for the mixed-reference-frame formalism \citep{gG74}:
\begin{align}
	\label{eq:Giacaglia}
	U_2^{m, \pm s} 
	= \frac{(-1)^{m \mp s}}{(2 \pm s)!} (\cos \epsilon/2)^{m \pm s} 
		(\sin \epsilon/2)^{\pm s - m} \frac{\mathrm{d}^{2 \pm s}}{\mathrm{d}\, Z^{2 \pm s}} 
		\left\{ Z^{2 - m} (Z - 1)^{2 + m} \right\},
\end{align}
where $Z = \cos^2 \epsilon/2$. Note that the expansion \eqref{eq:Lane} is valid for arbitrary inclination and for any eccentricity $e < 1$, and, moreover, this formulation allows us to consider the lunar inclination as a constant and the longitude of the ascending node of the Moon's orbit as a linear function of time \cite[qq.v.][]{mL89,sH80}. 

The form of the solar disturbing function is considerably simpler, with the secular and quadrupole order of approximation following from Kaula's classical spherical harmonic expansion:
\begin{align}
	\label{eq:Kaula}
	\R_\S = \sum_{m = 0}^2 \sum_{p = 0}^2 h_{2-2p, m}^\S \cos \Phi_{2-2p, m}^\S,
\end{align}
with harmonic angle
\begin{equation}
	\Phi_{2-2p, m}^\S = (2 - 2 p) \omega + m (\Omega - \Omega_\S), 
\end{equation} 
and associated harmonic coefficient
\begin{equation}
\label{eq:harm_coeff_solar}
\begin{array}{l}
	\ds h_{2-2p, m}^\S
	= \frac{\mu_\S a^2}{a_\S^3 (1 - e_\S^2)^{3/2}}
		\epsilon_m \frac{(2 - m)!}{(2 + m)!}
		F_{2,m,p} (i) F_{2,m,1} (i_\S) H_{2,p,2p-2} (e).
\end{array}
\end{equation}
Here equatorial elements are used for both the satellite and the Sun. The apparent orbit of the Sun can be considered, for all practical purposes, as Keplerian. 

A lunar secular resonance occurs when a specific linear combination of the averaged precession frequencies of the angles appearing Eq.~\ref{eq:Lane} is zero. That is,
\begin{align}
	\label{eq:lunar_res_con}
	\dot\psi_{2 - 2 p, m, \pm s} = (2 - 2 p) \dot\omega + m \dot\Omega \pm s \dot\Omega_\M \approx 0.
\end{align}
For the Moon, the rate of change of $\Omega_\M$, due to the perturbations of the Sun, is approximately $-0.053$ deg/d. Since $\dot\Omega_\S \equiv 0$, it follows that secular resonances of the solar origin are characterized by the simpler relation
\begin{align}
	\label{eq:solar_res_con}
	(2 - 2 p) \dot\omega + m \dot\Omega \approx 0. 
\end{align} 
Under these approximations, the occurrence and nature of the secular resonances driving the dynamics depend only on the frequencies of nodal and apsidal precession and the rate of regression of the Moon's nodes. The Moon's argument of perigee does not explicitly appear in the expansion: this important result does not require the additional assumption made by previous researchers \citep{fDaM93,tEkH97} that the Moon's orbit be circular. 

\begin{figure}[b!]
	\captionsetup{justification=justified}
	\centering
	\includegraphics[width=10cm,height=13cm]{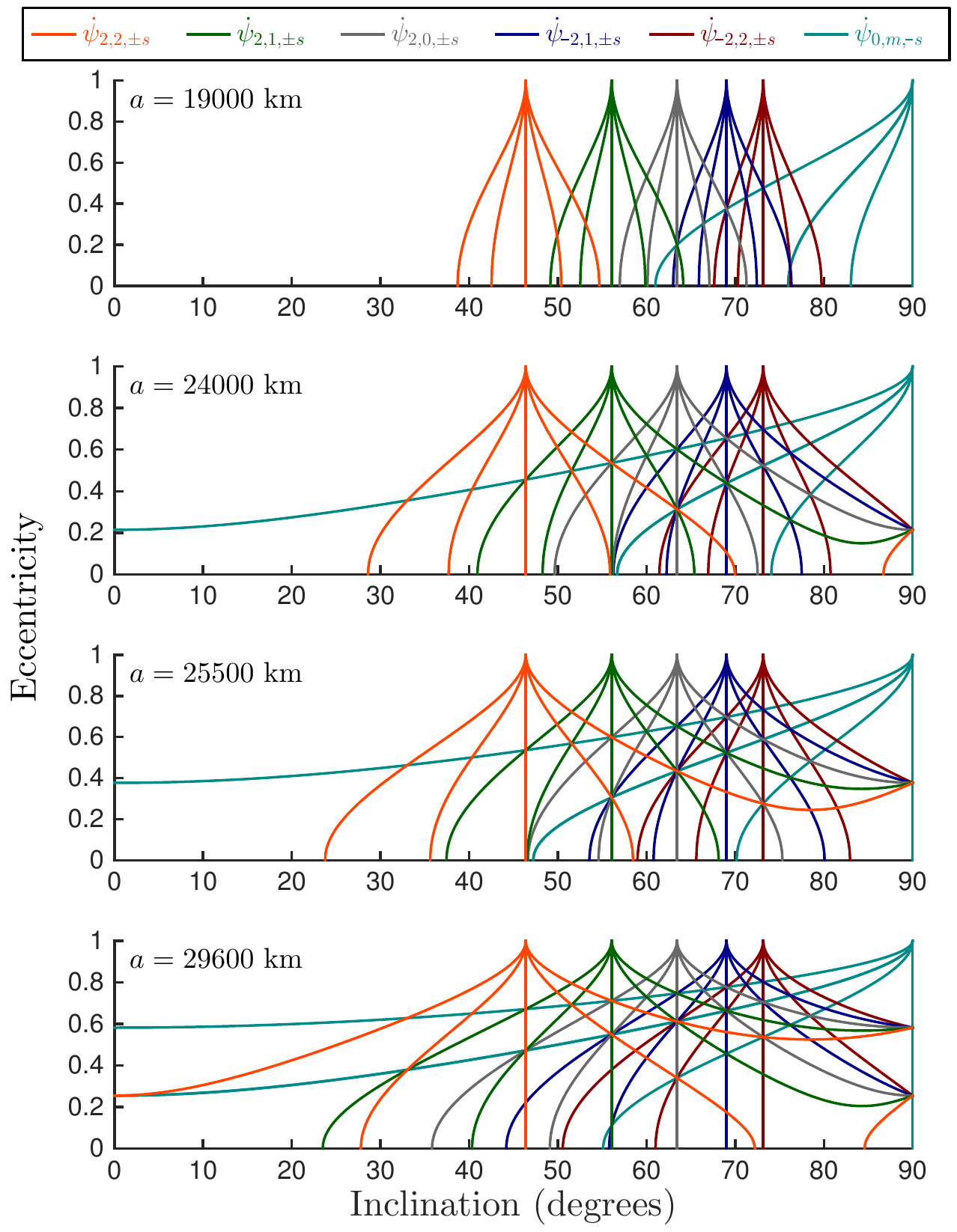} 
	\caption{The location of resonance centers of the form $\dot\psi_{2 - 2 p, m, \pm s} = (2 - 2 p) \dot\omega + m \dot\Omega \pm s \dot\Omega_M = 0$, where only the effects of the $J_2$ perturbation on $\omega$ and $\Omega$ have been considered \citep[adapted from][]{aR15}. These resonances form the dynamical backbone of the phase space, organizing and controlling the long-term orbital motion of MEO satellites.}
	\label{fig:skeleton}
\end{figure}

For the medium-Earth orbits considered here (i.e., semi-major axes between three and five Earth radii), the oblateness effect is at least an order of magnitude larger than that of lunisolar gravity \citep{tEkH97}. We are therefore permitted in the first approximation to neglect the changes in $\omega$ and $\Omega$ from lunisolar perturbations in order to make a more efficient mathematical treatment of the problem possible without at the same time deviating too drastically from the actual conditions prevailing in the system. The secular rates of change of a satellite's argument of perigee and longitude of ascending node cased by the $J_2$ harmonic are such that \citep{gC62}
\begin{equation}
\label{eq:oblateness}
\begin{array}{l}
	\ds \dot\omega = \frac{3}{4} J_2 n \left( \frac{R}{a} \right)^2 \frac{5 \cos^2 i - 1}{(1 - e^2)^2}, \\[1.25em]
	\ds \dot\Omega = -\frac{3}{2} J_2 n \left( \frac{R}{a} \right)^2 \frac{\cos i}{(1 - e^2)^2},
\end{array}
\end{equation}
where $R$ is the mean equatorial radius of the Earth and $n$ is the satellite's mean motion. As the semi-major axis is constant after averaging, Eqs.~\ref{eq:lunar_res_con}-\ref{eq:oblateness} define analytical curves of lunisolar secular resonances in the inclination and eccentricity phase space (Fig. ~\ref{fig:skeleton}). It is particularly noteworthy that the solar resonances and the lunar resonance with $s = 0$, both occurring at $46.4^\circ$, $56.1^\circ$, $63.4^\circ$, $69.0^\circ$, $73.2^\circ$ and $90^\circ$, are independent of the orbit eccentricity \citep{gC62,sH80}; these inclination-dependent-only resonances are intimately related to the critical inclination problem \citep[q.v.][]{aJ88}. Each of the critical inclinations split into a multiplet-like structure of lunar resonance curves corresponding to $s \in \left\{ 1, 2 \right\}$, emanating from unity eccentricity. As the semi-major axis of the satellite is receding from three to five Earth radii, the resonance curves began to intersect indicating locations of multiple resonances, where two or more critical arguments have vanishing frequencies. The complex web of intersecting lines suggests the potential for chaos and large excursions in the eccentricity, on account of the resonance-overlap criterion.   

\subsection{A measure of the resonant islands} \label{sec:widths}

Each of the resonances in Fig.~\ref{fig:skeleton} determines its own domain in the phase space---the resonance width \citep{bG66,mL88,sB03_fundamental}. Based on the isolated resonance hypothesis, we derive here a measure of the libration islands, expressed in terms of the Delaunay and Keplerian variables. The former set of canonical variables are the most frequently employed in astronomical perturbation methods, and allow us to appeal to the Hamiltonian formalism. The conjugate coordinates are the Keplerian angles $M$, $\omega$, and $\Omega$ and the conjugate momenta are functions of the orbital elements $a$, $e$, and $i$, as follows:
\begin{align}
	\label{eq:Delaunay}
	\setlength\arraycolsep{9pt}
	\begin{array}{ll}
		L = \sqrt{\mu a}, 			& l = M, \\
		G = L \sqrt{1 - e^2}, 		& g = \omega, \\
		H = G \cos i, 					& h = \Omega,
	\end{array}
\end{align} 
in which $\mu$ denotes the Earth's gravitational parameter. 

We take the averaged Hamiltonian $\H$ to include the second zonal harmonic of the Earth's gravitational potential and the lowest-order term, i.e., the second harmonic, in the expansion of each of the solar and lunar disturbing functions:
\begin{align}
	\label{eq:Hamiltonian}
	\H (G, H, g, h, t; L) = \H_\text{kep} + \H_{\J_2} + \H_\M + \H_\S, 
\end{align}
where $\H_\text{kep} = -\mu^2/2 L^2$ is the Kepler Hamiltonian, which depends only on the mean semi-major axis, treated as a parameter ($M$ is a cyclic variable and consequently $a$ is constant), $\H_{J_2}$ is the classical Hamiltonian for the averaged oblateness perturbation \citep{sB01}
\begin{align}
	\H_{\J_2} (G, H; L) = \frac{J_2 R^2 \mu^4}{4 L^3} \frac{G^2 - 3 H^2}{G^5},
\end{align}
$\H_\M (G, H, g, h, t; L) = -\R_\M$, and $\H_\S (G, H, g, h, t; L) = -\R_\S$, given by Eqs.~\ref{eq:Lane} and \ref{eq:Kaula}, respectively. Equation~\ref{eq:Hamiltonian} is a non-autonomous Hamiltonian of two degrees of freedom, depending periodically on time through the Moon's nodal motion. In order to render the Hamiltonian autonomous, we extend the phase space to three degrees of freedom by introducing the canonical variables $(\Gamma, \tau)$, such that $\dot\tau \equiv \partial \H/\partial \Gamma = \dot\Omega_\M$, where the new Hamiltonian (without the Keplerian part) takes the form
\begin{align}
	\label{eq:Hamiltonian_3DOF}
	\H (G, H, \Gamma, g, h, \tau; L) 
	= \H^\text{sec} (G, H; L) + \H^\text{lp} (G, H, g, h, \tau; L) + \dot\Omega_\M \Gamma, 
\end{align}
in which
\begin{align}
	\label{eq:H_sec}
	\H^\text{sec} (G, H; L) & = \H_{\J_2} + \H_\M^\text{sec} + \H_\S^\text{sec}, \\
	\H^\text{lp} (G, H, g, h, \tau; L) & = \H_\M^\text{lp} + \H_\S^\text{lp}.
\end{align}
Secular perturbations due to the Moon and the Sun are related to terms with $2 - 2 p = 0$, $m = 0$, and $s = 0$; that is, $\dot\psi_{2 - 2 p, m, \pm s} = 0$. Long-periodic lunisolar perturbations are connected with $\dot\psi_{2 - 2 p, m, \pm s} \neq 0$.

Introducing the frequency vector $\dot{\bfm \Theta} = \big( \dot{g}, \dot{h}, \dot\tau \big)$, the lunisolar resonances, Eqs.~\ref{eq:lunar_res_con} and \ref{eq:solar_res_con}, may be summarized by the existence of a vector ${\bfm n} \in \mathbb{Z}_{\star}^{3}$ such that
\begin{align}
	\label{eq:res_con}
	{\bfm n} \cdot \dot{\bfm \Theta} = n_1 \dot{g} + n_2 \dot{h} + n_3 \dot\tau \approx 0.
\end{align}
In order to study each resonance, we treat them in isolation and introduce new action-angle variables, appropriate to the resonance involved, by means of the canonical unimodular transformation (linear transformation belonging to $\textrm{SL}(3,\mathbb{Z})$). The non-null vector ${\bfm n}$ has at most two zero components, and depending on the case where $n_1 = 0$ or $n_2 = 0$, we can introduce the resonance angle $\sigma \equiv n_1 g + n_2 h + n_3 \tau$ through the following unimodular transformations:
\begin{align}
	{\bfm{\frak T}}_1 
	= \left( \begin{array}{ccc}
		n_1 & n_2 & n_3 \\[0.2em]
		0 & n_1^{-1} & 0 \\[0.2em]
		0 & 0 & 1 
	\end{array} \right), \quad 
	{\bfm{\frak T}}_2
	= \left( \begin{array}{ccc}
		1 & 0 & 0 \\[0.2em]
		n_1 & n_2 & n_3 \\[0.2em]
		0 & 0 & n_2^{-1}
	\end{array} \right).
\end{align}
The action of ${\bfm{\frak T}}_i$  on ${\bfm \Theta}$ gives $\big( {\bfm{\frak T}}_i \cdot {\bfm \Theta}^\intercal \big)_i = \sigma$, $i \in \{1,2\}$, where $\big( {\bfm{\frak T}}_i \cdot {\bfm \Theta}^\intercal \big)_i$ denotes the $i$-th components of the vector ${\bfm{\frak T}}_i \cdot {\bfm \Theta}^\intercal \in \mathbb{R}^{3}$. To keep the system canonical, new actions ${\bfm \Lambda} \in \mathbb{R}^{3}$ are introduced as 
\begin{align}
	{\bfm \Lambda} = {\bfm{\frak T}}_i^{-\intercal} \cdot \big( G, H, \Gamma \big)^\intercal, 
\end{align}
where ${\bfm{\frak T}}_i^{-\intercal}$ denotes the inverse transpose of ${\bfm{\frak T}}_i$. These two transformations lead to the new set $({\bfm \Lambda}, {\bfm \sigma})$ of action-angle variables. Using ${\bfm{\frak T}}_1$, we have
\begin{align}
	\label{eq:action_angle_T1}
	\setlength\arraycolsep{9pt}
	\begin{array}{ll}
		\Lambda_1 = n_1^{-1} G, 								& \sigma_1 \equiv \sigma = n_1 g + n_2 h + n_3 \tau, \\
		\Lambda_2 = -n_2 G + n_1 H, 						& \sigma_2 = n_1^{-1} h, \\
		\Lambda_3 = -n_1^{-1} n_3 G + \Gamma, 	& \sigma_3 = \tau,
	\end{array}
\end{align} 
and using ${\bfm{\frak T}}_2$ gives
\begin{align}
	\label{eq:action_angle_T2}
	\setlength\arraycolsep{9pt}
	\begin{array}{ll}
		\Lambda_1 = G - n_1 n_2^{-1} H, 					& \sigma_1= g, \\
		\Lambda_2 = n_2^{-1} H, 								& \sigma_2 \equiv \sigma = n_1 g + n_2 h + n_3 \tau, \\
		\Lambda_3 = -n_3 H + n_2 \Gamma, 			& \sigma_3 = n_2^{-1} \tau.
	\end{array}
\end{align}
In either case, under the isolated resonance hypothesis, the Hamiltonian, Eq.~\ref{eq:Hamiltonian_3DOF}, expressed in terms of the new variables (\ref{eq:action_angle_T1} or \ref{eq:action_angle_T2}) and neglecting the constant terms, reduces to a single degree of freedom with the resonant angle $\sigma$ and its conjugated action. Denoting these by $(X, x \equiv \sigma)$, the new Hamiltonian has the simple form
\begin{align}
	\label{eq:H_breiter}
	\H = f_0 \big( X \big) + f_1 \big( X \big) \cos \big( x + \rho \big), 
\end{align}
with 
\begin{align}
	\left\{ 
	\begin{aligned}
	f_0 \big( X \big) & = \H^\text{sec} \big( X \big) \\
	f_1 \big( X \big) & = -h_{\bfm n} \big( X \big), 
	\end{aligned}
	\right.
\end{align}
where $\rho$ is a constant phase, $\H^\text{sec}$ is given by Eq.~\ref{eq:H_sec}, and $h_{\bfm n}$ is the harmonic coefficient in the lunar and solar disturbing function expansions, associated with the harmonic angle which is in resonance. For the secular resonances with $n_3 \neq 0$, $h_{\bfm n} \equiv h_{2-2p, m, \pm s}^\M$, as given by Eq.~\ref{eq:harm_coeff_lunar}. For the lunisolar resonances where $n_3 \equiv 0$,\footnote{Note that this refinement was not taken into account in \citet{tEkH97} and is more suitable for the calculation of the widths for the inclination-dependent-only resonances, where solar perturbations play a non-negligible role (see Appendix~\ref{sec:lunisolar_res}).}
\begin{align}\label{eq:magnitude-harmonic}
	\big( h_{\bfm n} \big)^2 =  \big( h_{2-2p, m, 0}^\M \big)^2 + \big(  h_{2-2p, m}^\S \big)^2 
		+ 2  \big( h_{2-2p, m, 0}^\M \big) \big( h_{2-2p, m}^\S \big) \cos \big( m \Omega_\S \big).
\end{align}

Equation~\ref{eq:H_breiter} is the familiar form for the Hamiltonian when only one critical term is present \citep{bG66,mL88,sB03_fundamental}. An expansion of the Hamiltonian around the resonant location $X_\star$ further reduces it to a pendulum-like system, which provides the calculation of the amplitude of the libration region around the resonance.\footnote{Recall that we need the resonance width in order to determine when neighboring resonances overlap and hence when a system is unstable, \`{a} la Chirikov.} Expanding $\H$ in a Taylor series about $X_\star$, neglecting terms of $\O \big( (X-X_{\star})^{3} \big)$ and dividing through by $-\partial^2 \H^\text{sec} / \partial X^2 \big\vert_{X = X_\star}$, gives  
\begin{align}
	\label{eq:H_pendulum}
	\H (I, x) = -\frac{1}{2} I^{2} + \nu \cos \big(x + \tilde{\rho} \big),
\end{align}
where $I = X - X_\star$ and 
\begin{align}
	\label{eq:coeff}
	\nu & = \left| \frac{h_{\bfm n} \big( X_\star \big)}
		{\partial^2 \H^\text{sec} / \partial X^2 \big\vert_{X = X_\star}} \right|.
\end{align}
It should be noted that this last division changes the physical timescale. In this form, the maximum excursion in the action, $\Delta I$, measured from the line $I = 0$ to the extreme point of the separatrix (the resonance half-width), is given by
\begin{align}
	\label{eq:pend_aperture}
	\Delta I = 2 \sqrt{\nu}.
\end{align}
The width can be expressed in terms of the Delaunay variables using the appropriate transformation, ${\bfm{\frak T}}_1$ or ${\bfm{\frak T}}_2$. It turns out that both transformations, while formally needed, actually lead to the same formulas, and thus, in the following, we detail only one of them. Note that the Hamiltonian $\H$ is $\sigma_2$ and $\sigma_3$ cyclic, so that $\Lambda_2$ and $\Lambda_3$ are constants of motion. From Eq.~\ref{eq:pend_aperture} and Eq~\ref{eq:action_angle_T1}, we find 
\begin{align}
	\Delta G = \left| 2 n_1 \sqrt{\nu} \right|.
\end{align}
Since $\Lambda_2 = \Lambda_{2,\star}$, we have $H = n_1^{-1} n_2 G + n_1^{-1} \Lambda_{2,\star}$. Taking the first variation gives
\begin{align}
	\Delta H = \left| n_1^{-1} n_2 \Delta G \right| = \left| 2 n_2 \sqrt{\nu} \right|.
\end{align}

The Keplerian elements are related to the Delaunay variables through Eq.~\ref{eq:Delaunay}, where we note in particular that 
\begin{align*}
	e = \sqrt{1 - \left( \frac{G}{L} \right)^2}, \quad
	\cos i = \frac{H}{G}.
\end{align*}
From the first variations, we have
\begin{align}
	\Delta e 
	\label{eq:ecc_aperture}
	& = \left| \frac{2 n_1 \sqrt{1 - e_\star^2}  \sqrt{\nu}}{L_\star e_\star} \right|, \\
	\Delta i 
	\label{eq:inc_aperture}
	& = \left| \frac{2 \left( n_2 - n_1 \cos i_\star \right) 
		\sqrt{\nu}}{L_\star \sin i_\star \sqrt{1 - e_\star^2}} \right| 
	= \left| \frac{e_\star \Delta e}{1 - e_\star^2} 
		\left( \frac{n_2}{n_1 \sin i_\star} - \cot i_\star \right) \right|,
\end{align}
in agreement with the generating-function approach used by \citet{tEkH97}, for which we recall here that $n_1 = 2 - 2p$ and $n_2 = m$.

A few words are in order regarding the calculation of $\nu$ appearing in the above expressions. To be consistent with our computation of the resonant centers in Section \ref{sec:skeleton}, we neglect the secular contributions from the lunisolar perturbations, so that 
\begin{align}
	\H^\text{sec} (\Lambda_1, \Lambda_2) \nonumber
	& = \H_{\J_2} (\Lambda_1, \Lambda_2) \\
	& = \frac{J_2 R^2 \mu^4}{4 L^3 n_1^3} 
		\left[ (1 - 3 n_1^{-2} n_2^2) \Lambda_1^{-3} - 6 n_1^{-3} n_2 \Lambda_2 \Lambda_1^{-4}
			- 3 n_1^{-4} \Lambda_2^2 \Lambda_1^{-5} \right].
\end{align}
Since $X \equiv \Lambda_1$, we find\footnote{The form of Eq.~\ref{eq:partial_Ely} contained in \citet{tEkH97} is missing a factor of 2 in the second term in the brackets.}
\begin{align}
	\label{eq:second_partial_h}
	\frac{\partial^2 \H^\text{sec}}{\partial X^2} 
	& = \frac{3 J_2 R^2 \mu^4}{2 L^3 G^5} \left[ n_1^2 \left( 2 - 15 \frac{H^2}{G^2} \right) 
		+ 10 n_1 n_2 \frac{H}{G} - n_2^2 \right] \\
	\label{eq:partial_Ely}
	& = \frac{3 J_2 R^2}{2 a^4 (1 - e^2)^{5/2}} \left[ n_1^2 \left( 2 - 15 \cos^2 i \right) 
		+ 10 n_1 n_2 \cos i - n_2^2 \right].
\end{align}

From the adjacent locations of several resonance centers in the $i$--$e$ plane (Figure~\ref{fig:skeleton}), it is natural to suspect that Chirikov's resonance overlapping---a universal route to chaos \citep{bC79_universal}---occurs in the mesh of the web. Stricto sensu, this was never rigorously demonstrated for the MEO problem, though \citet{tEkH97} had derived all of the formulas necessary to do so. Appendix~\ref{sec:harmonic_coefficients} gives the explicit formula for the lunisolar harmonic coefficients of the 29 resonant curves appearing in Figure~\ref{fig:skeleton}. Recall that each center of the resonances (cf. Fig.~\ref{fig:skeleton}) are defined, in the $i$--$e$ phase space, by the curves  
\begin{align*}
	\mathcal{C}_{{\bfm n}} = \Big\{ (i,e) \in [0,\frac{\pi}{2}] \times [0,1] \ : \
	{\bfm n} \cdot \dot{\bfm 	\Theta} \equiv n_1 \dot{g} + n_2 \dot{h} + n_3 \dot\tau = 0 \Big\}.	
\end{align*}
The curves delimiting the maximal widths of each resonance are defined by
\begin{align}
	\label{eq:widths_curves}
	\mathcal{W}^{\pm}_{{\bfm n}} = \left\{ (i,e) \in [0,\frac{\pi}{2}] \times [0,1] \ : \ 
	{\bfm n} \cdot \dot{\bfm \Theta} \equiv n_1 \dot{g} + n_2 \dot{h} + n_3 \dot\tau 
	= \pm 2 \sqrt{\nu} \left( \frac{\partial^2 H^\text{sec}}{\partial X^2} \bigg\vert_{X=X_\star} \right) \right\}.	
\end{align}
Note that the multiplication by $\partial^2 \H^\text{sec} / \partial X^2 \big\vert_{X = X_\star}$ in Eq.~\ref{eq:widths_curves} is due to the timescale operated in Eq.~\ref{eq:H_pendulum}.

\begin{figure}[b!]
	\captionsetup{justification=justified}
	\centering
	\includegraphics[width=10cm,height=13cm]{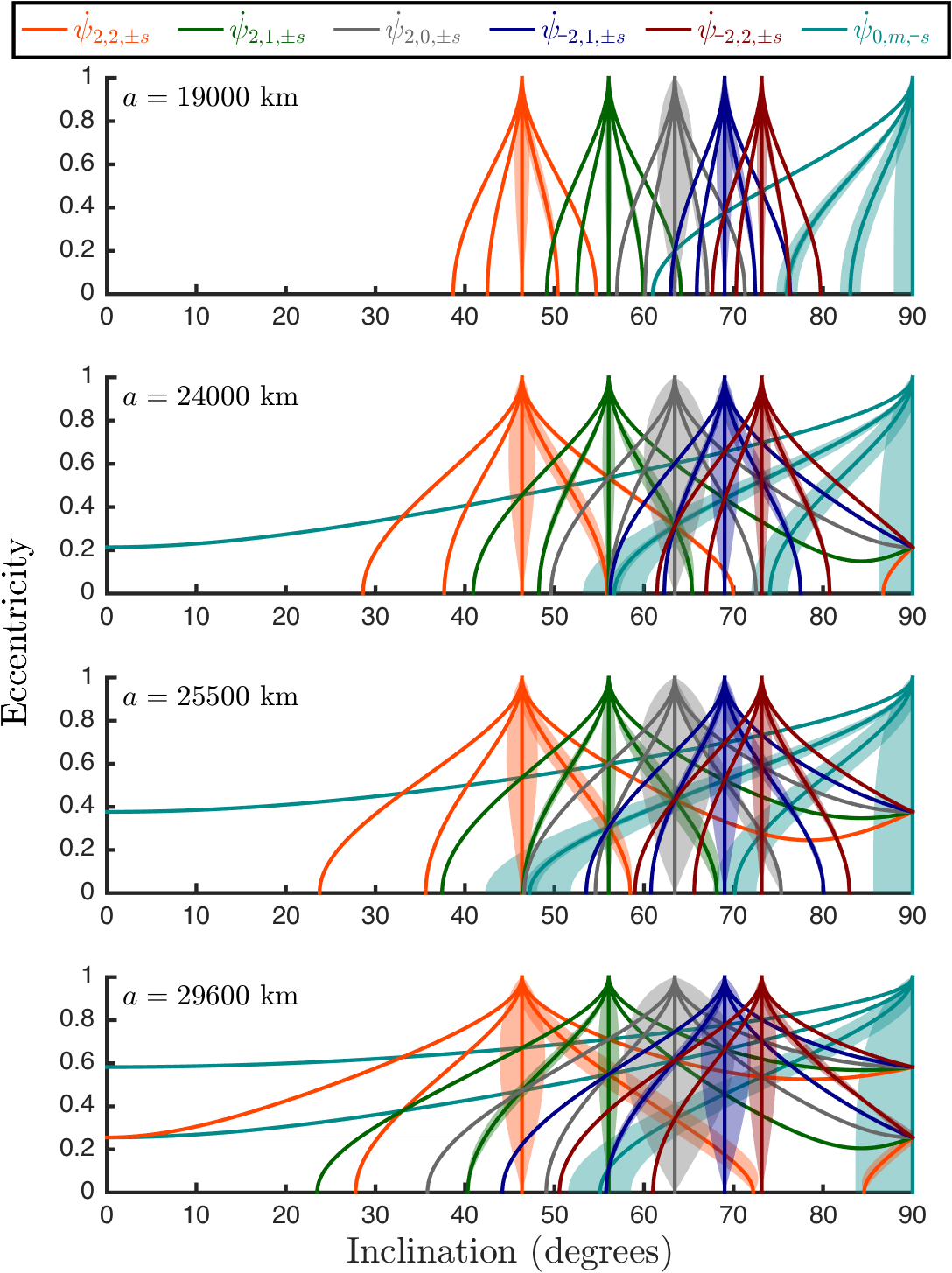} 
	\caption{Lunisolar resonance centers (solid lines) and widths (transparent shapes) for increasing values of the satellite's semi-major axis. This plot shows the regions of overlap between distinct resonant harmonics.}   
	\label{fig:chirikov_criterion}
\end{figure}

For each resonance, we compute the half-width excursion in the inclination and eccentricity phase space, according to Eq.~\ref{eq:widths_curves}. Figure~\ref{fig:chirikov_criterion} shows the centers and widths of the resonances in the region of semi-major axes between roughly 3 and 5 Earth radii, the range of validity of the present theory. At $a = 19,000$ km, the resonances are thin and the regions of overlap are quite narrow, indicating that regular orbits should dominate the phase space. As the ratio of the semi-major axis of the orbits of the satellite and the perturbing bodies (the perturbation parameter, $\varepsilon = \varepsilon (a/a_M)$) is increased, the resonance centers for $s \neq 0$ spread out and the libration regions become wider. We can observe a transition from the stability regime at $a = 19,000$ km to a Chirikov one where resonances overlap significantly at $a = 29,600$ km. It is well known that mean-motion resonances become wider at larger eccentricity \citep[q.v.][]{aM02}: the width of the lunisolar secular resonances exhibit a much more complicated dependence on the eccentricity, as well as the inclination. 

\begin{figure}[htp!]
  \centering
    \subfigure[$a_{0}=19,000$ km.]
    {\includegraphics[width=7.8cm,height=4.8cm]{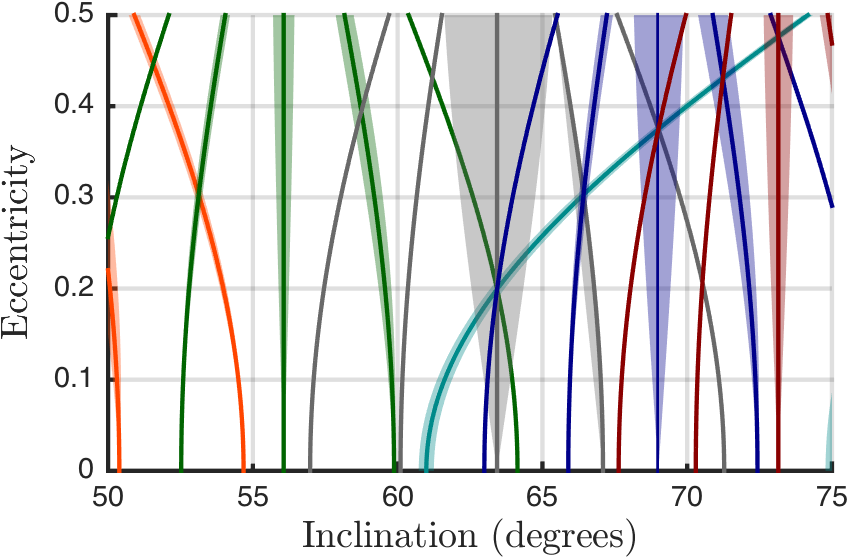}}
    \subfigure[$a_{0}=24,000$ km.]
    {\includegraphics[width=7.8cm,height=4.8cm]{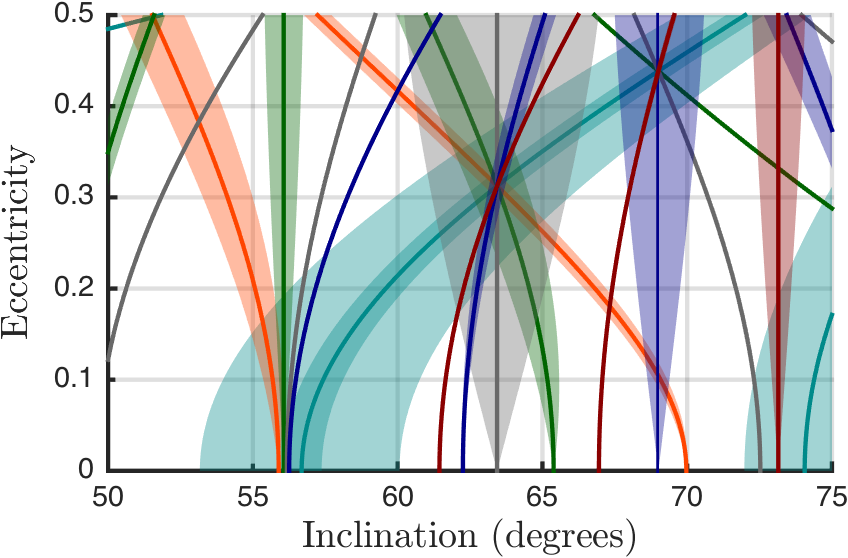}} 
    \subfigure[$a_{0}=25,500$ km.]
    {\includegraphics[width=7.8cm,height=4.8cm]{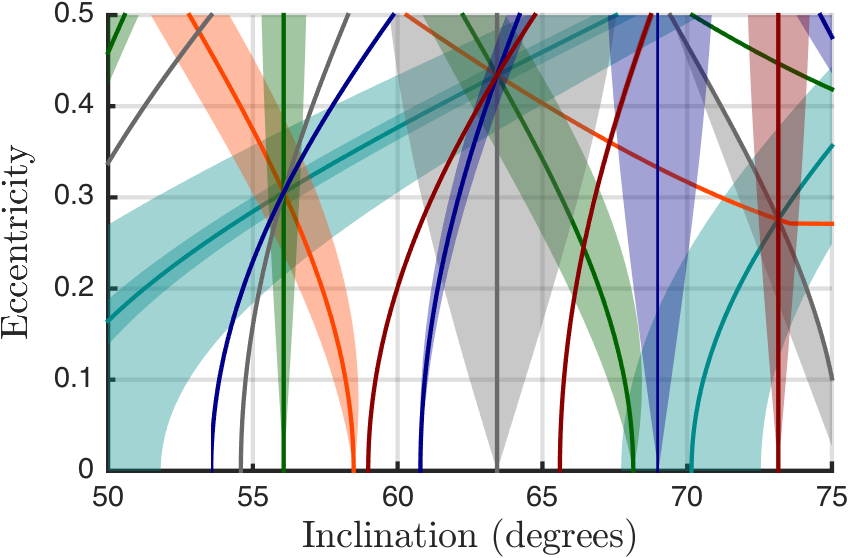}}        
    \subfigure[$a_{0}=29,600$ km.]
    {\includegraphics[width=7.8cm,height=4.8cm]{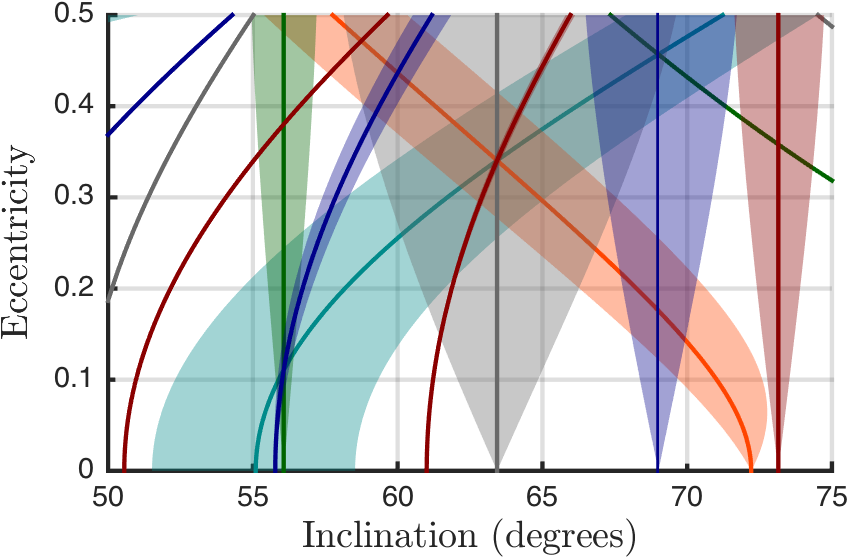}} 
  \caption{Zoomed-in portion of Figure~\ref{fig:chirikov_criterion}, showing where we concentrate our numerical calculations.} 
  \label{fig:apertures_zoom}
\end{figure}

The Chirikov resonance-overlap criterion forms the extent of our theoretical analysis. This empirical criterion gives significant qualitative and quantitative predictions about the regions in action space for which chaotic orbits can be found; yet, it contains several limitations. It lacks a rigorous theoretical grounding and, as noted by \citet{aMmG96}, it is not well tested on dynamical systems with many degrees of freedom. Chirikov's geometrical argument neglects the coupling of the resonances---to say nothing about the deformation of their separatrices---and the role played by secondary resonances, and, as a result, the criterion often underestimates the threshold of transition from order to chaos. For a more complete and detailed analysis of the phase-space stability, we must turn to numerical explorations. As such, we use the fast Lyapunov indicator in the following section. As a systematic study of the entire parameter space represents a formidable task with significant computational requirements, we chose to focus on a reduced portion of the phase space that is of considerable practical interest for the navigation satellite constellations (see Figure~\ref{fig:apertures_zoom}). 

\section{Numerical exploration of the phase space stability}
This section contains a numerical survey of the dynamical structures appearing in the MEO region by presenting an atlas of stability maps. Following a parametric approach, our main goals are (i) to give the geometry and extent of the stable, resonant, and chaotic domains, and (ii) to obtain a global picture of the resonant interactions on the emergence of chaos. In addition to quantifying the degree of hyperbolicity of generic orbits, we estimate the barriers of predictability by computing Lyapunov times. Furthermore, we demonstrate numerically that resonances and chaos are associated with transport in the $i$--$e$ phase space.  

\subsection{The numerical detection of the resonance overlapping regime}

In the past few decades, numerical investigations have played a wonderfully key role in studies on the long-term stability of dynamical systems. ``The symbiosis between mathematical results and numerical computations,'' writes \citet{aMmG96}, ``is very promising for the future developments of applied dynamical system science and, in particular, for Celestial Mechanics.'' The fundamental work of Morbidelli, Guzzo, Froeschl\'{e}, and others, have brought to light the existence of specific structures in the phase space when Nekhoroshev's theorem is satisfied, which in turn imply this celebrated long-time stability result \citep{aM95,aMmG96,aMcF96,aM02}. Leaving aside mathematical intricacies, they essentially showed that even if chaos exists in the phase space (which is not precluded by the Nekhoroshev theorem), there always exists in a mesh of the resonant web a no-resonant domain, which is filled by many invariant tori. Conversely, if the system is not in Nekhoroshev form, such no-resonant domains cannot be defined: resonances overlap and invariant tori become rare \citep{aM95}.    

The divergent behavior of trajectories can be easily investigated numerically using the broad family of Lyapunov and affiliated indicators. The problem of the eventual overlapping of resonances is thus reduced to the numerical computation of indicators that distinguish between stable, resonant, and chaotic orbits. For such investigations, we chose from the chaos toolbox the fast Lyapunov indicator (FLI) \citep[q.v.][]{cF97}. Writing the $n$-dimensional dynamical system in first-order autonomous form, $\dot{x}~=~f(x)$, where $x \in \mathbb{R}^{n}$ and $f: \mathbb{R}^{n} \to \mathbb{R}^{n}$ represents the vector field, the variational system in $\mathbb{R}^{2n}$ can be stated as
\begin{align}
	\label{eq-varia}
	\left\{ 
	\begin{aligned} 
	\dot{x}  &= f(x) \\
	\dot{w} &= \Big( \frac{\partial f(x)}{\partial x} \Big) w,
	\end{aligned} 
	\right.
\end{align}
where $w \in \mathbb{R}^{n}$ stands for the deviation (or tangent) vector. Many first order stability indicators are based on the propagation of the variational system and on the monitoring of the stretching of $w$ with time \citep[see][and references therein for a nice survey]{chS10}. The FLI, introduced in \citet{cF97}, follows from the variational system and enables the discrimination between ordered and chaotic motions \citep{cF00_structure,mG02}. The indicator at time $t$ is defined by 
\begin{align}
	\textrm{FLI}(t) \equiv \sup_{\tau \le t} \log \vert \vert w(\tau)\vert \vert,
\end{align}
where $\vert \vert \bullet \vert \vert$ denotes the $L_{2}$-norm. More rigorously, we should note $\textrm{FLI}(t,x_{0},w_{0})$ instead of $\textrm{FLI}(t)$ since the FLI computation requires the initialization of the variational system (Eq.~\ref{eq-varia}). The choice of the initial tangent vector is a recurrent question when dealing with first order variational indicators, as discussed by \citet{rB09_spurious}. The FLI atlas---a collection of FLI maps---presented here depend on two parameters: (i) the initial semi-major axis, ranging from roughly 3 to 5 Earth radii (the prominent MEO region) and (ii) a hierarchy of physical models, in order to illustrate the different dynamical mechanisms and how resonances and chaos appear. The physical models considered are schematically summarized by Table \ref{tab:physical-model} and discussed in more detail in Appendix \ref{sec:numerical_setup}, along with a description of the numerical setup.

\begin{table}
	\captionsetup{justification=justified}
	\centering
	\caption{Perturbations added to the central part of the geopotential for the numerical stability analysis. We refer to Appendix \ref{sec:numerical_setup} for more details.}
	\label{tab:physical-model} 
	\begin{tabular}{c|cccc}
	\multicolumn{5}{c}{\textsc{Dynamical models}} \\\hline\hline
							& Zonal terms & Tesseral terms & Lunar perturbation & Solar perturbation \\ \hline
		model $1$ 	&  $J_{2}$ & not considered 	&  up to degree  $2$ & not considered   \\
		model $2$ 	&  $J_{2}$ &  not considered &  up to degree $2$	&  up to degree $2$ \\
		model $3$ 	& $J_{2}, \cdots ,J_{5}$ & not considered &  up to degree $4$ 	
			& up to degree  $3$ \\
		model $4$ 	& $J_{2}, \cdots ,J_{5}$ & up to degree and order $5$ &  up to degree $2$ 	
			& up to degree $2$ \\\hline
	\end{tabular}
\end{table} 

\subsection{FLI stability atlas: hyperbolicity and predictability}

\begin{figure}[htp!]
	\centering
    \subfigure[$a_{0}=19,000$ km.] 
    {\includegraphics[width=7.9cm,height=4.7cm]{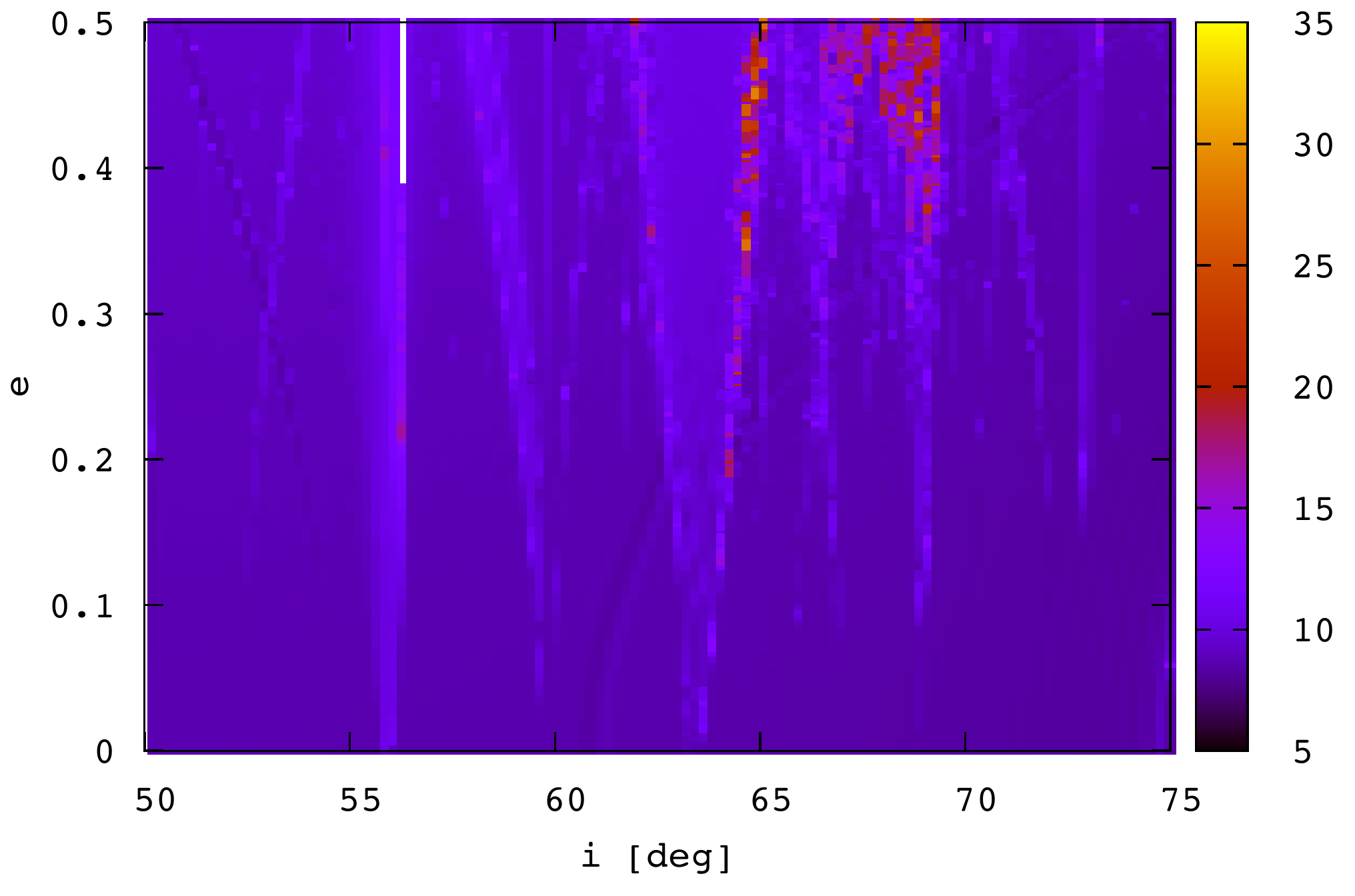}}
    \subfigure[$a_{0}=24,000$ km.]
    {\includegraphics[width=7.9cm,height=4.7cm]{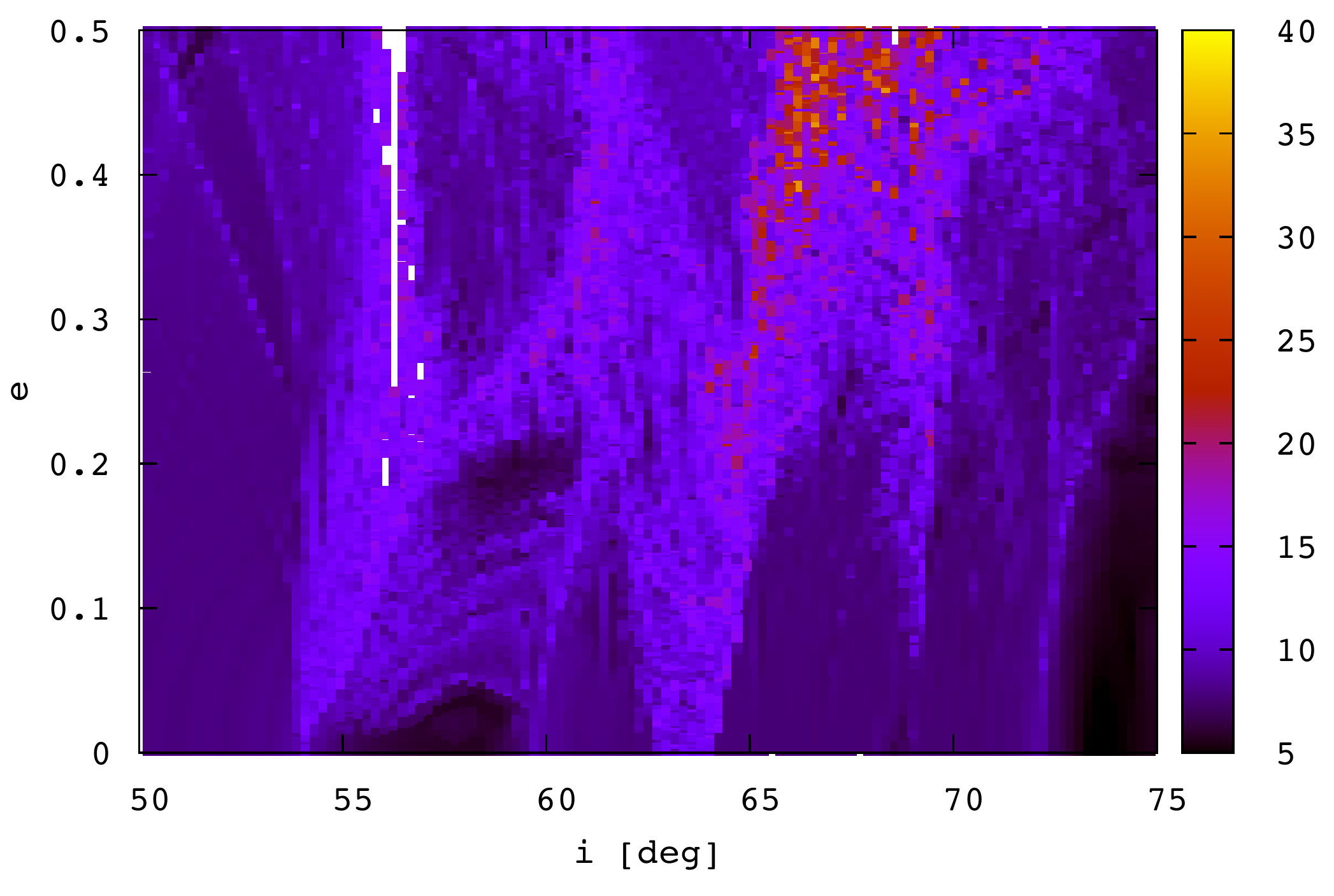}} 
    \subfigure[$a_{0}=25,500$ km.]
    {\includegraphics[width=7.9cm,height=4.7cm]{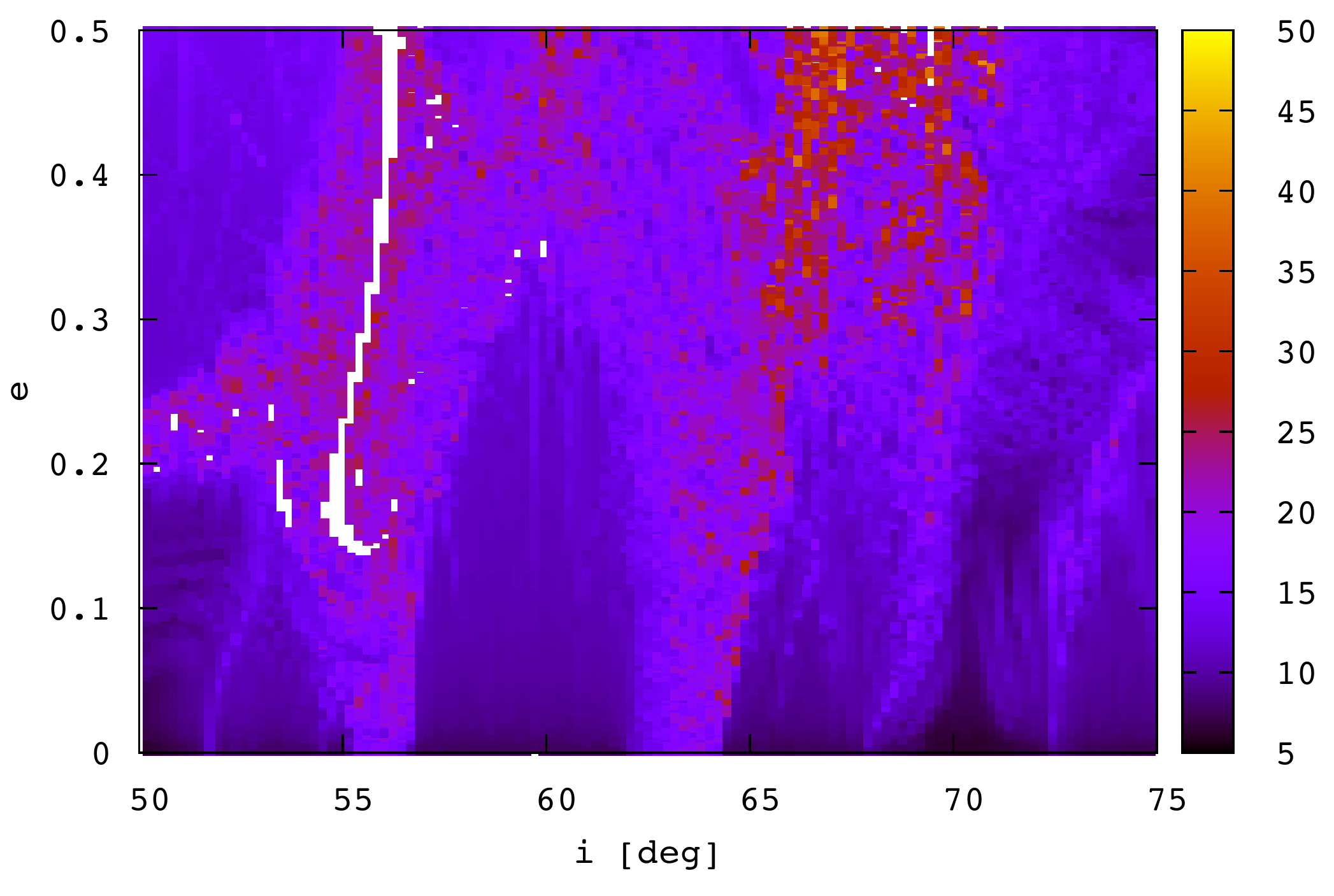}}        
    \subfigure[$a_{0}=29,600$ km.]
    {\includegraphics[width=7.9cm,height=4.7cm]{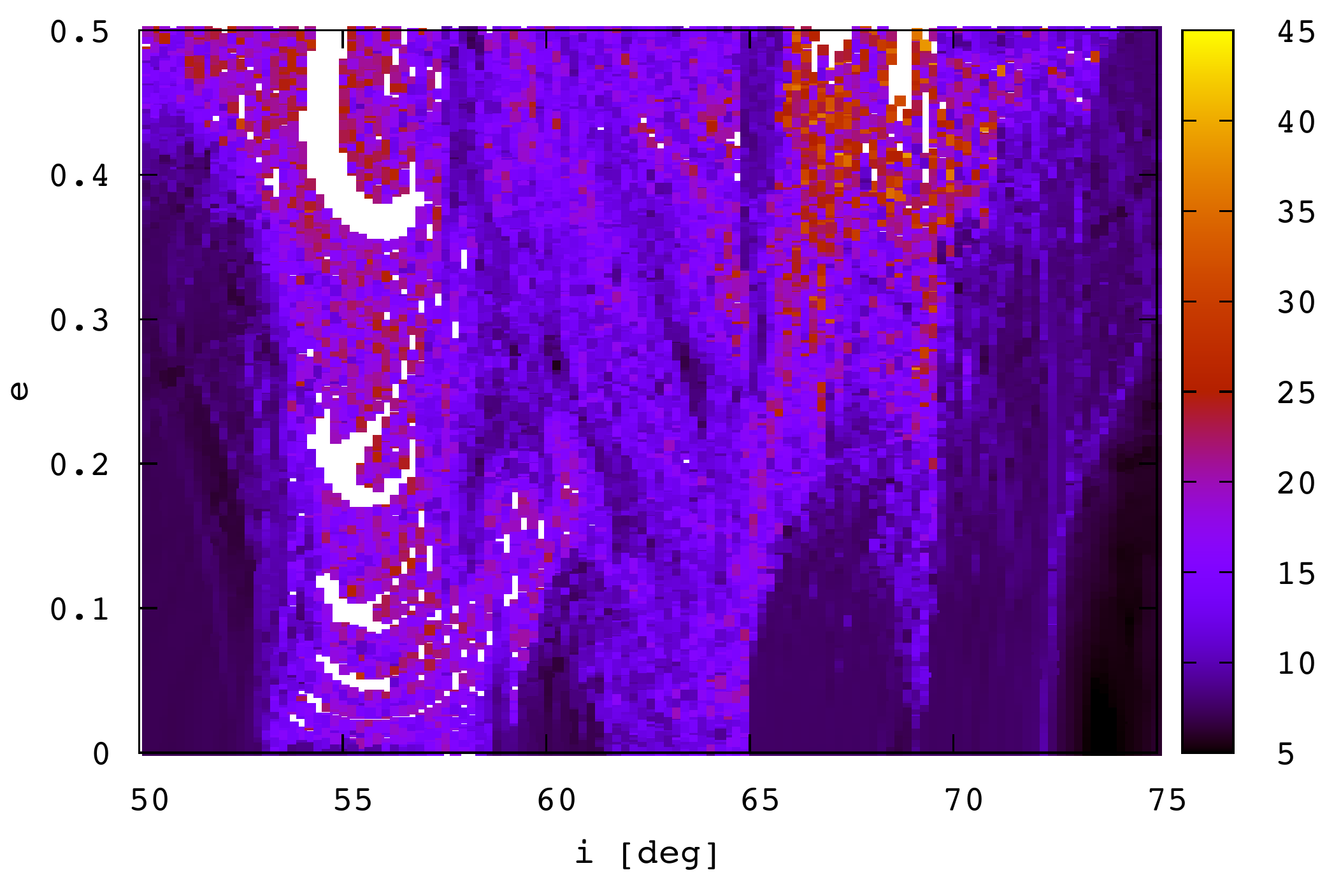}} 
    \caption{FLI stability maps for dynamical model 1.} 
    \label{atlas-m1}
\end{figure}

\begin{figure}[htp!]
	\centering
	\subfigure[$a_{0}=19,000$ km.]
    {\includegraphics[width=7.9cm,height=4.7cm]{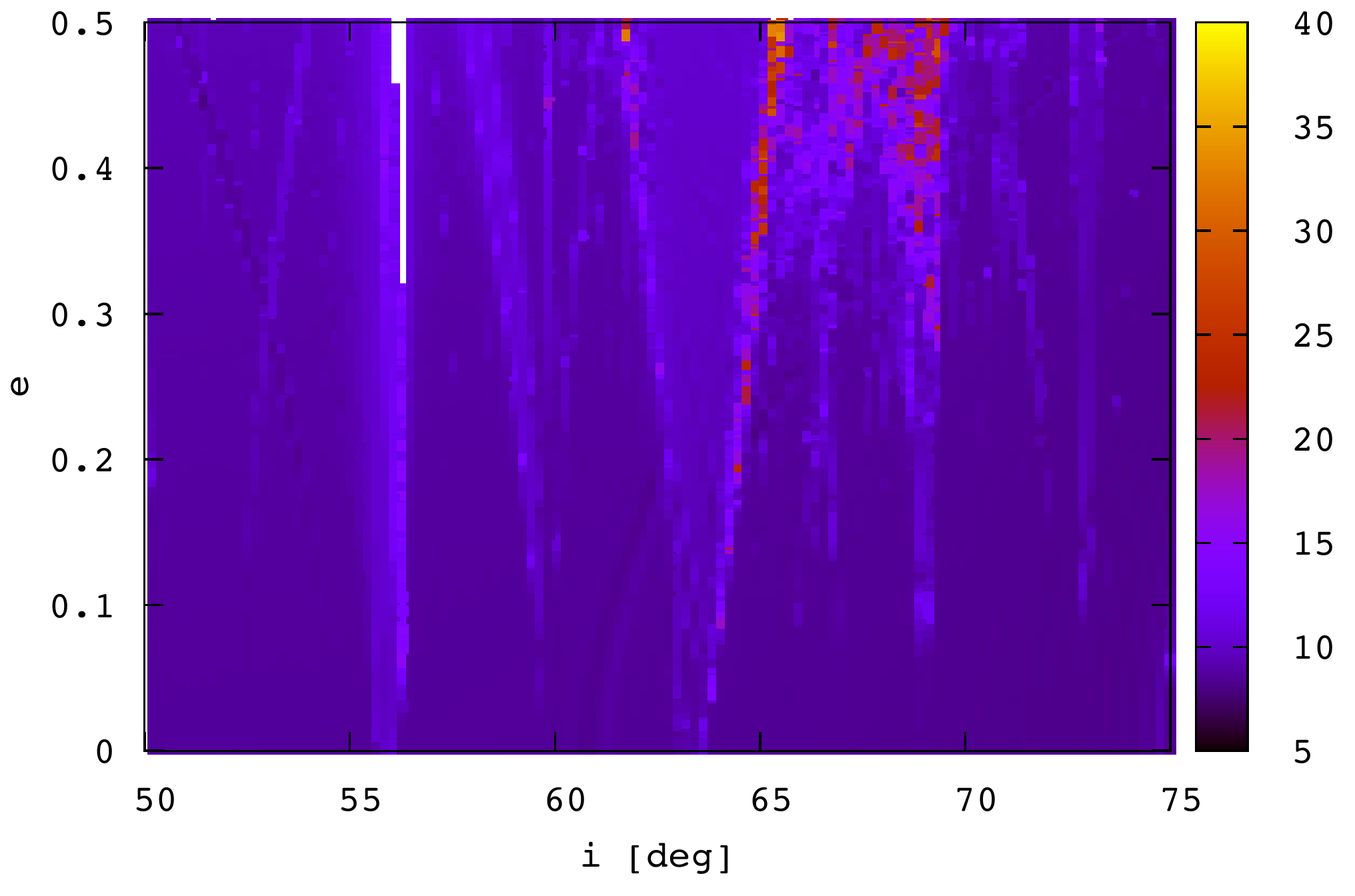}}
    \subfigure[$a_{0}=24,000$ km.]
    {\includegraphics[width=7.9cm,height=4.7cm]{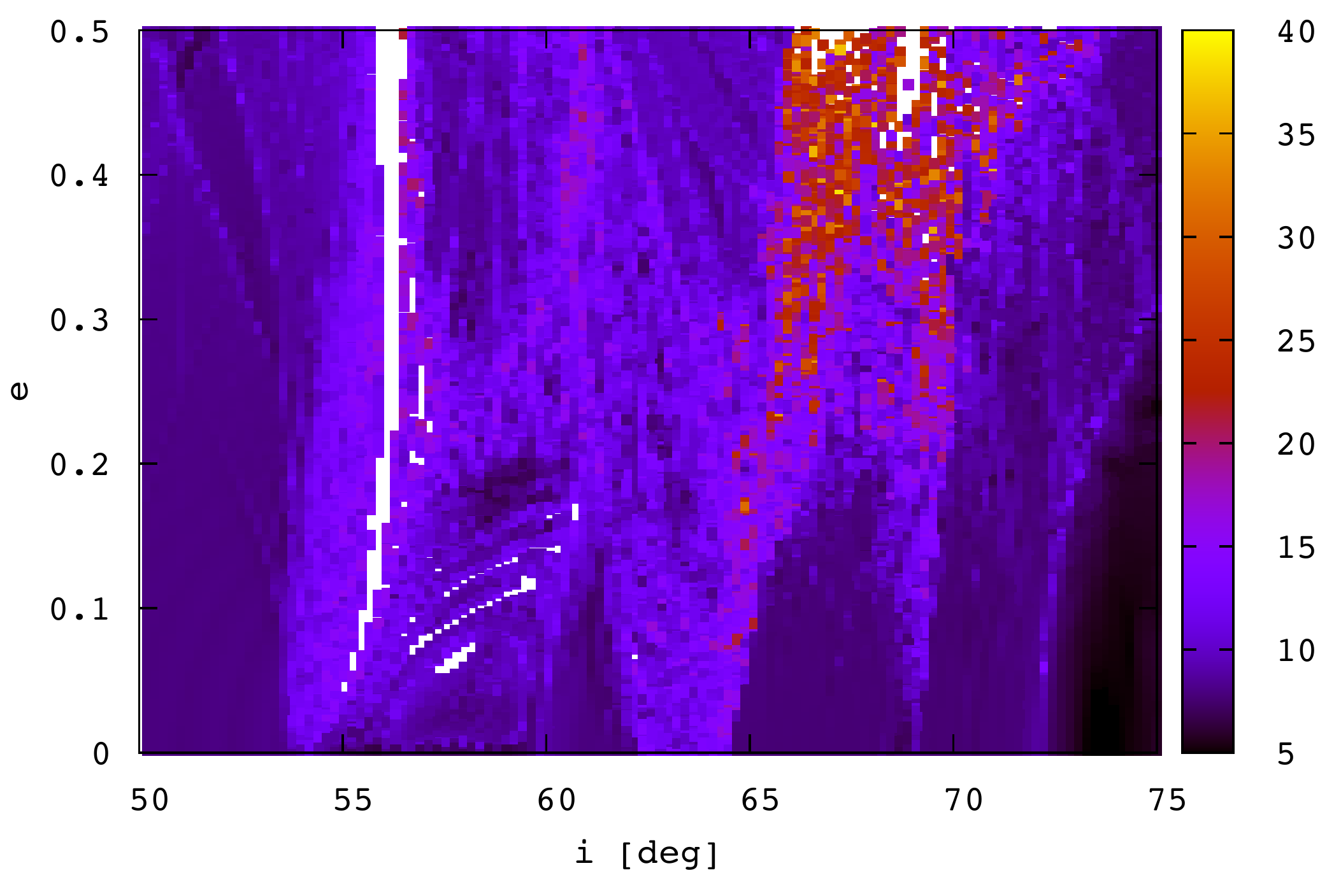}} 
    \subfigure[$a_{0}=25,500$ km.]
    {\includegraphics[width=7.9cm,height=4.7cm]{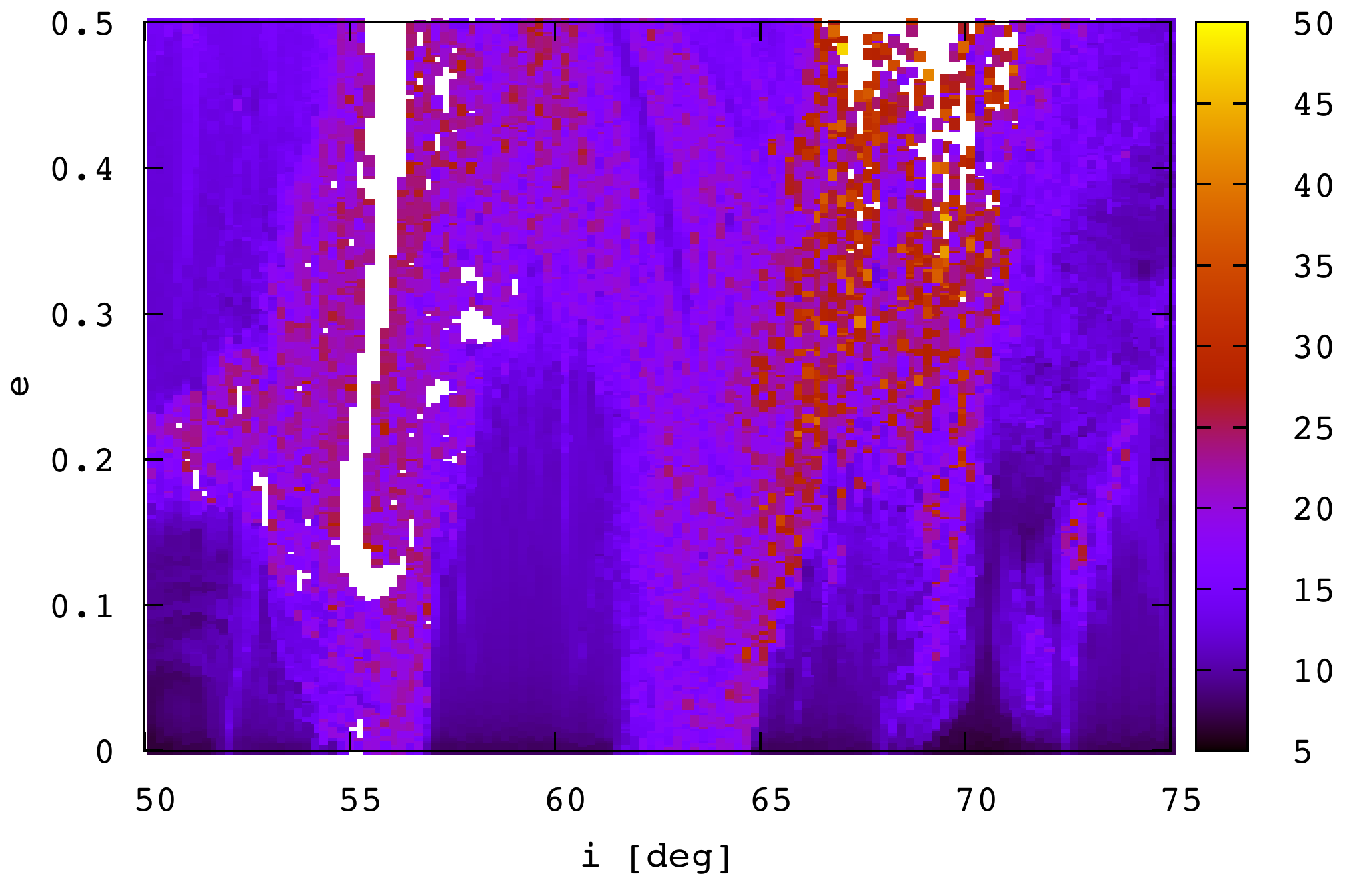}}        
    \subfigure[$a_{0}=29,600$ km.]
    {\includegraphics[width=7.9cm,height=4.7cm]{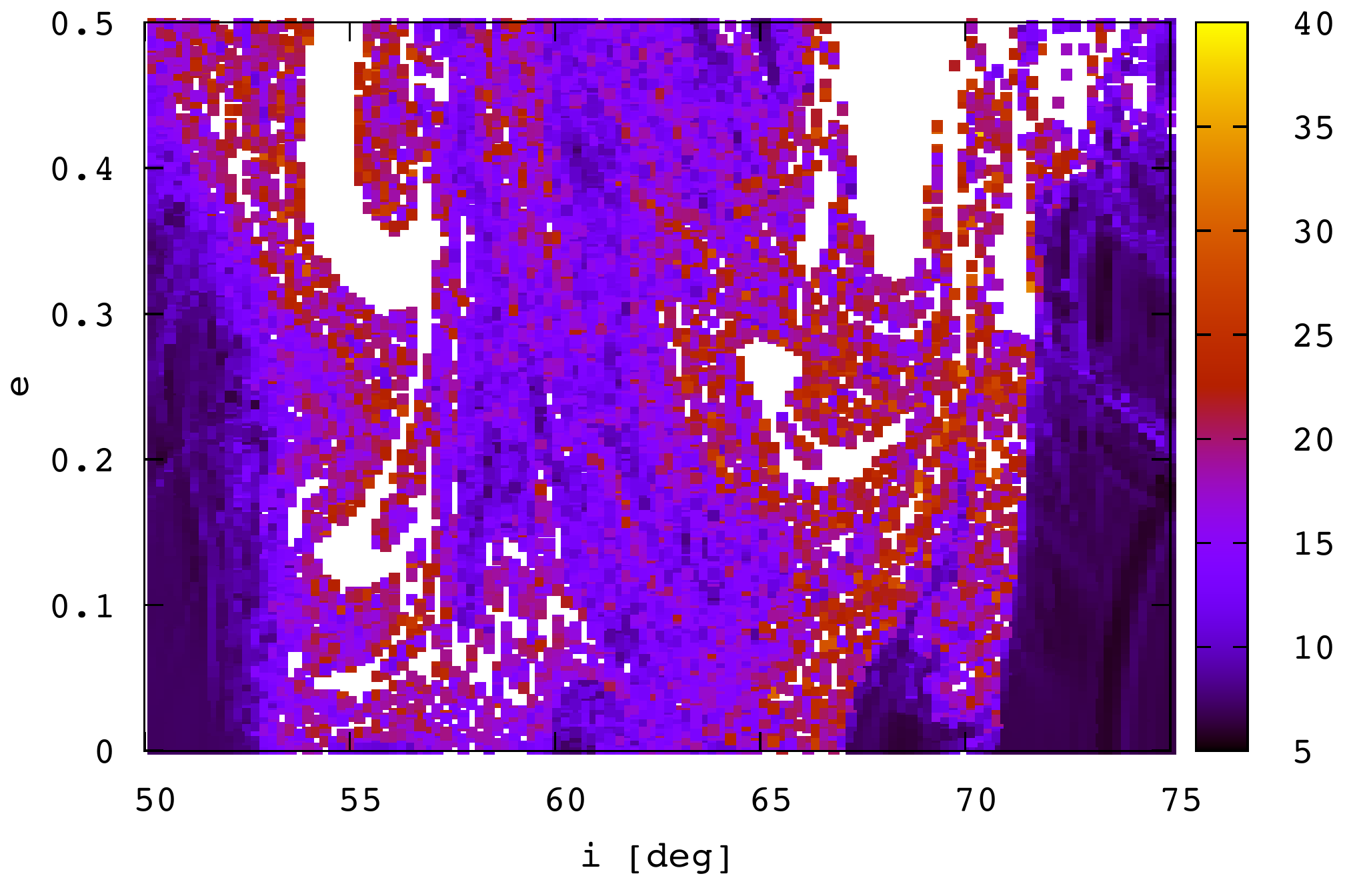}} 
    \caption{FLI stability maps for dynamical model 2.} 
    \label{atlas-m2}
\end{figure}

\begin{figure}[htp!]
	\centering
	\subfigure[$a_{0}=19,000$ km.]
    {\includegraphics[width=7.9cm,height=4.7cm]{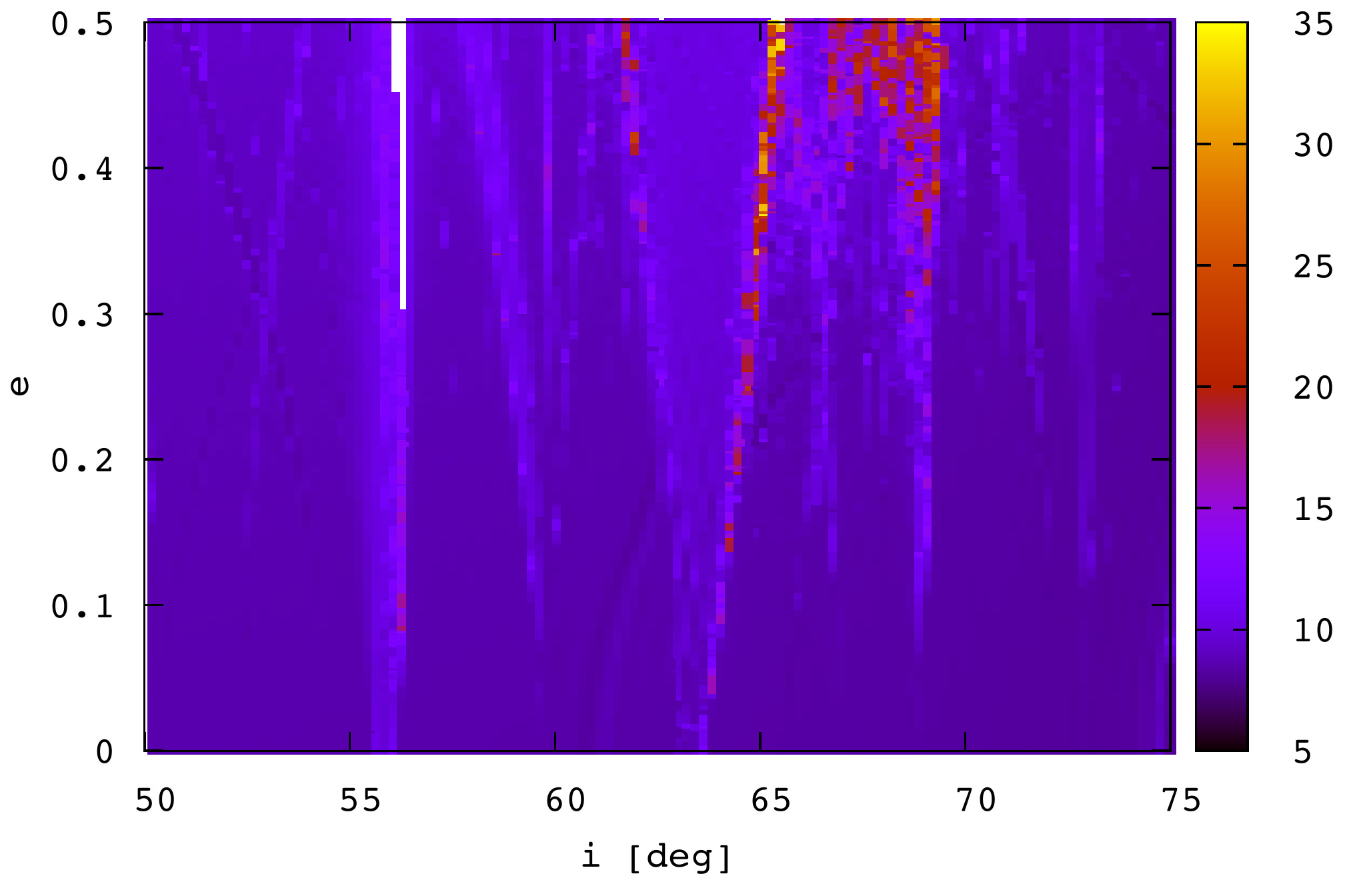}}
    \subfigure[$a_{0}=24,000$ km.]
    {\includegraphics[width=7.9cm,height=4.7cm]{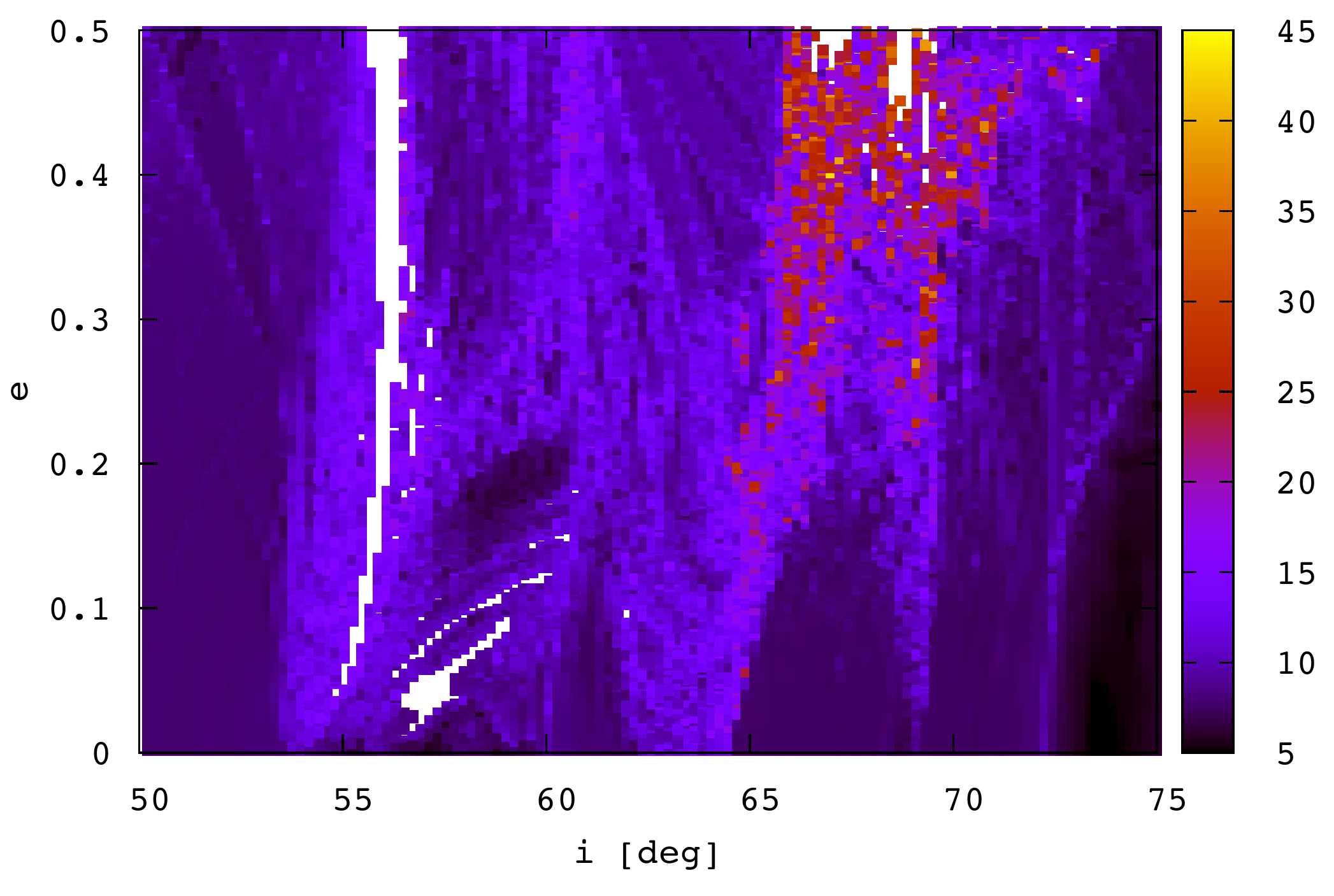}} 
    \subfigure[$a_{0}=25,500$ km.]
    {\includegraphics[width=7.9cm,height=4.7cm]{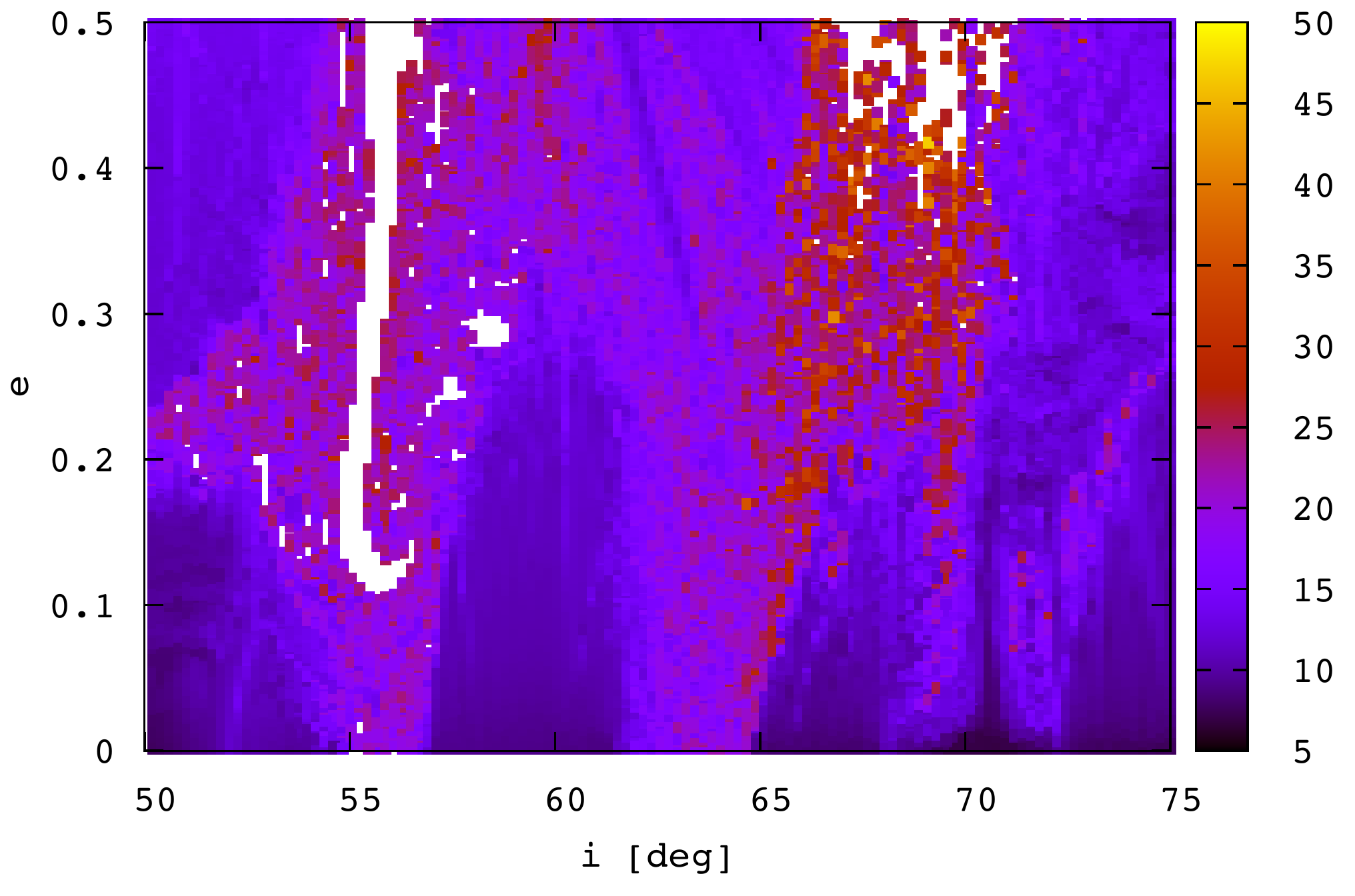}}        
    \subfigure[$a_{0}=29,600$ km.]
    {\includegraphics[width=7.9cm,height=4.7cm]{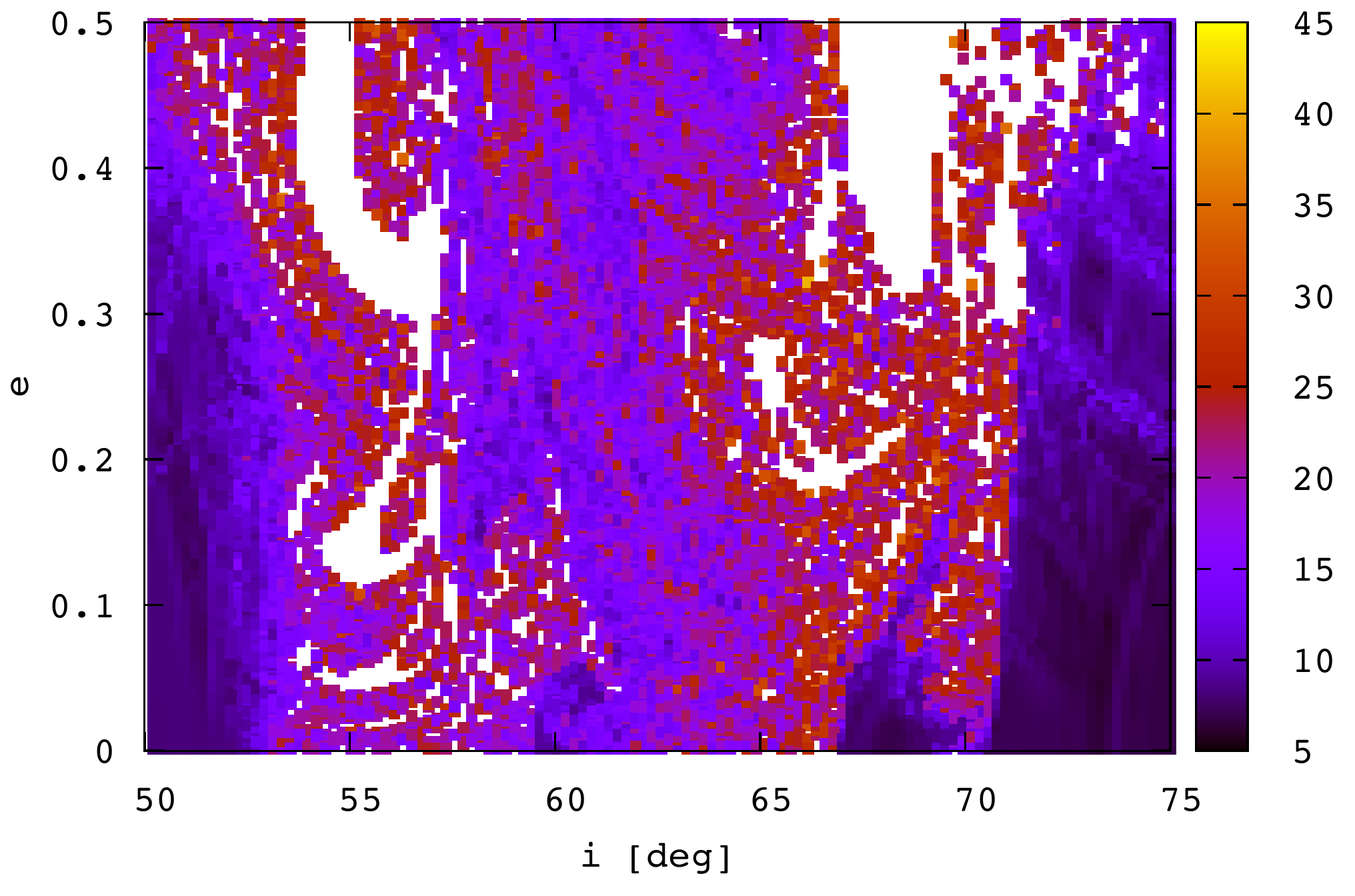}} 
    \caption{FLI stability maps for dynamical model 3.} 
    \label{atlas-m3}
\end{figure}

\begin{figure}[htp!]
  \centering
    \subfigure[$a_{0}=19,000$ km.]
    {\includegraphics[width=7.9cm,height=4.7cm]{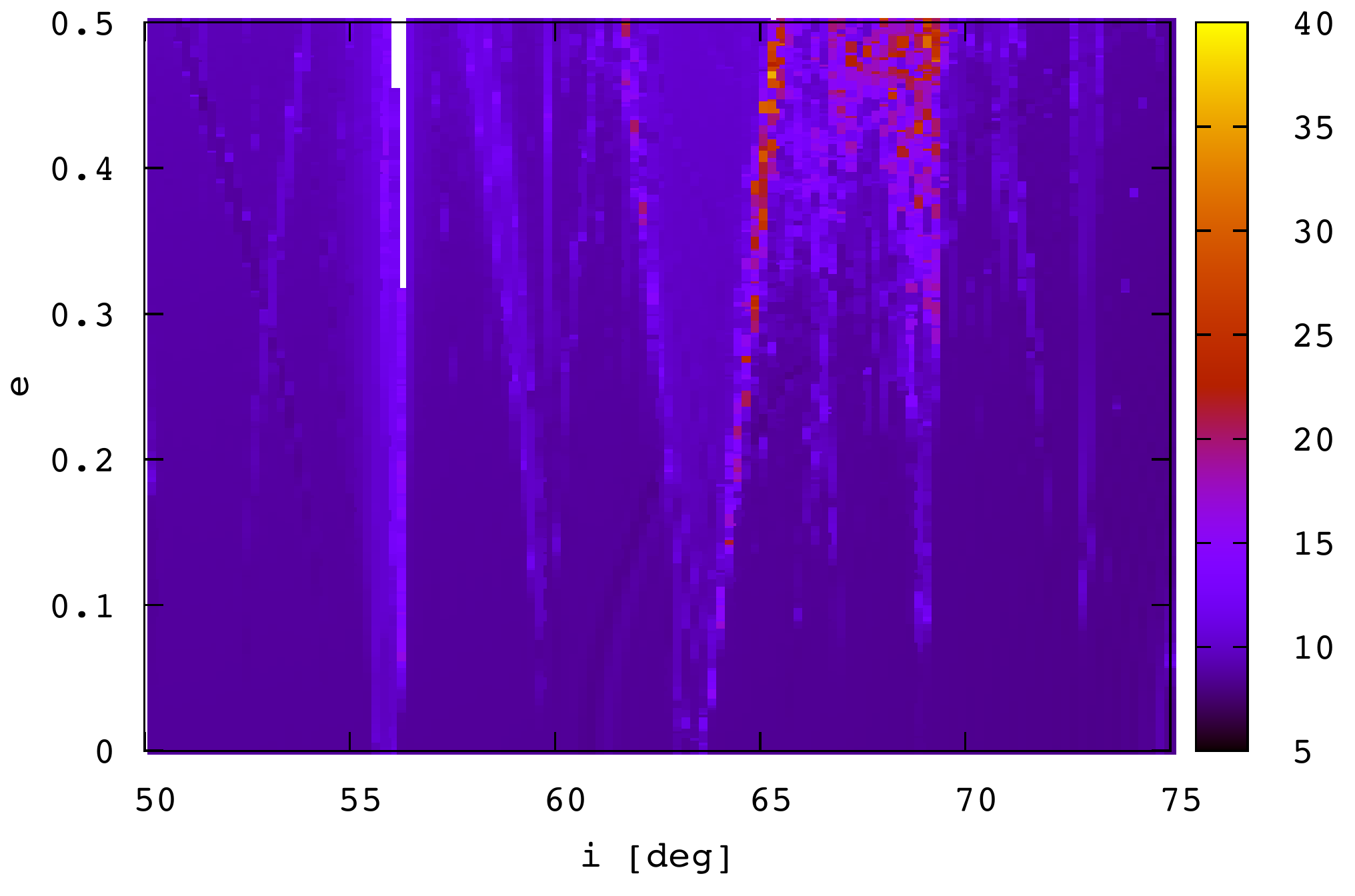}}
    \subfigure[$a_{0}=24,000$ km.]
    {\includegraphics[width=7.9cm,height=4.7cm]{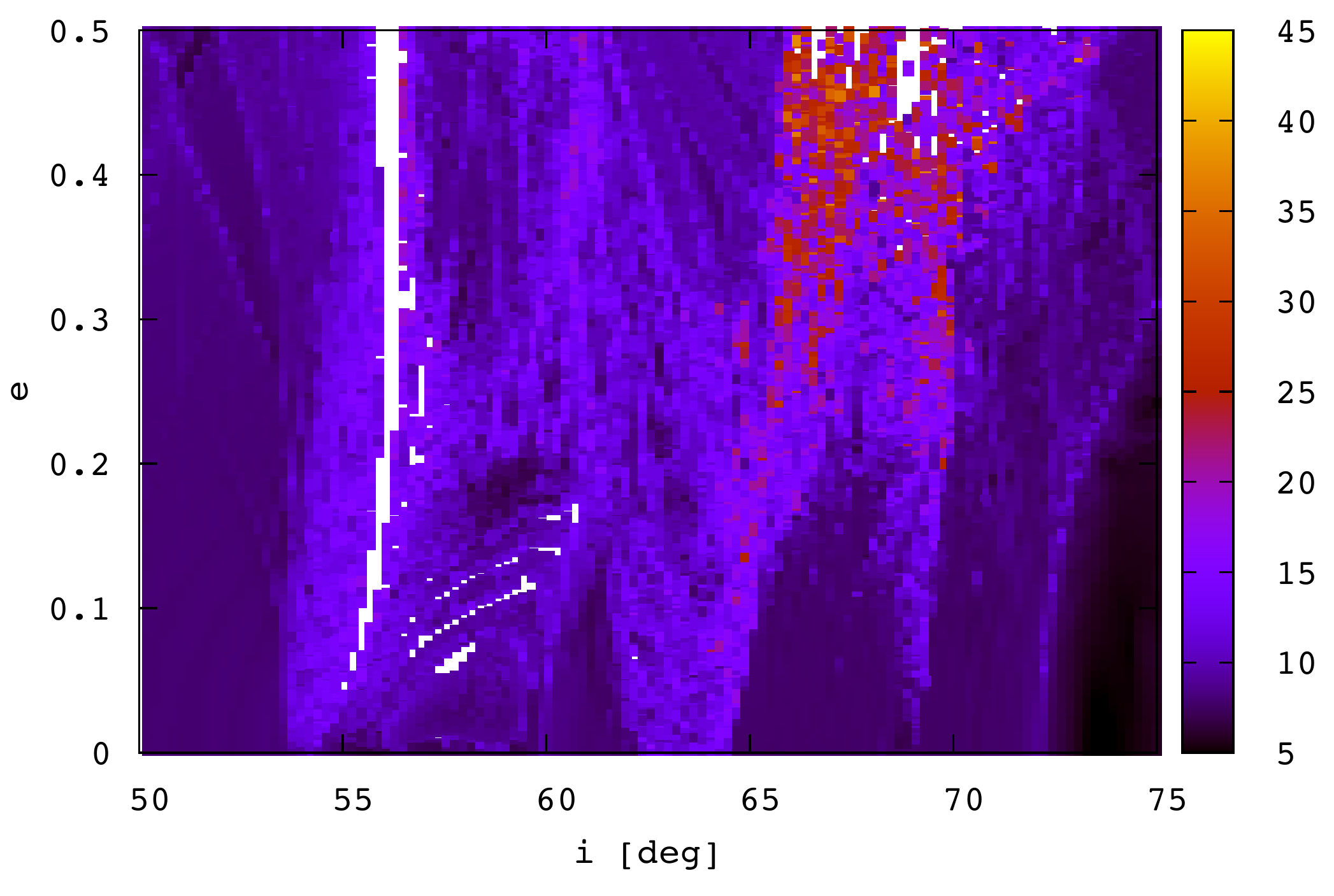}} 
    \subfigure[$a_{0}=25,500$ km.]
    {\includegraphics[width=7.9cm,height=4.7cm]{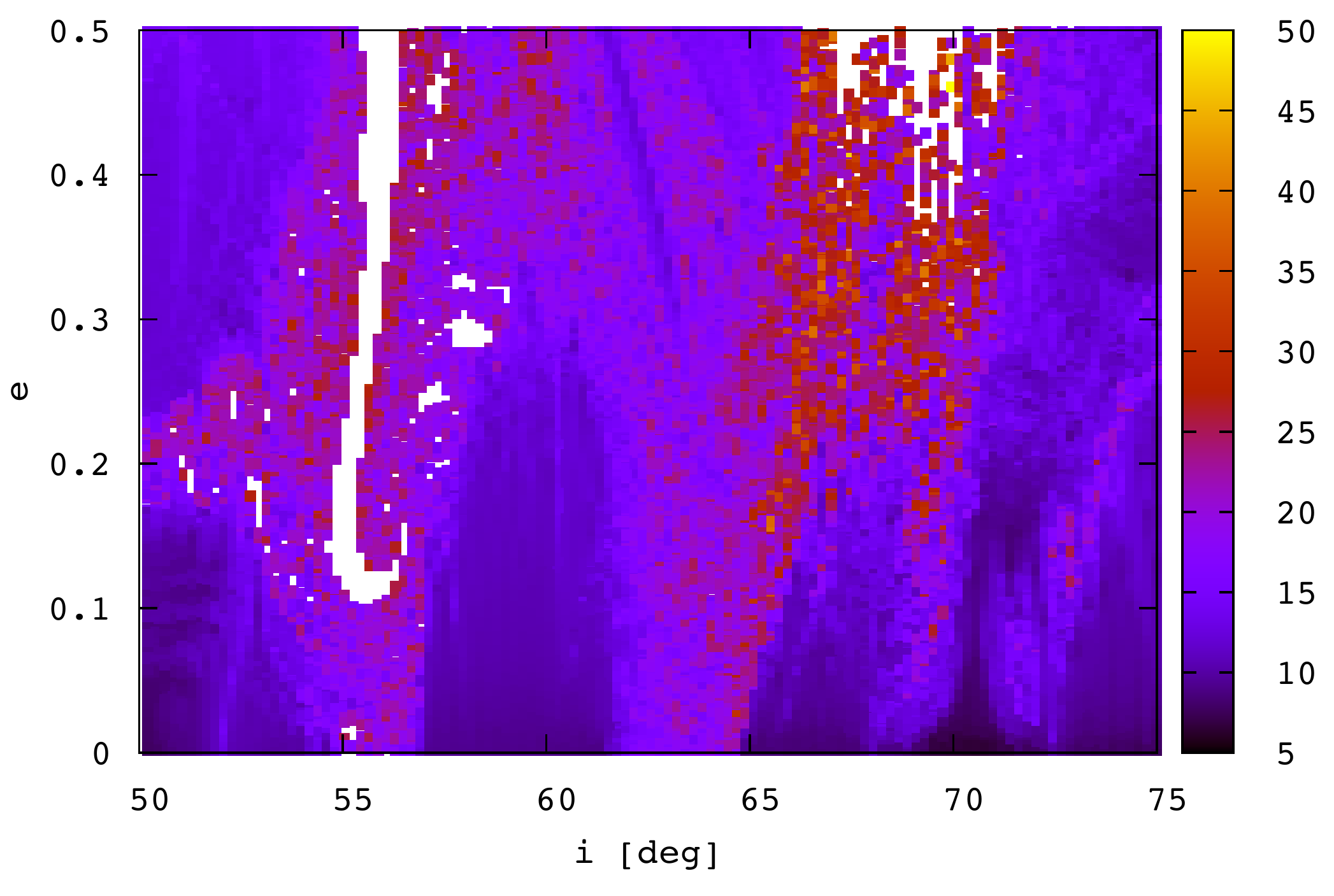}}        
    \subfigure[$a_{0}=29,600$ km.]
    {\includegraphics[width=7.9cm,height=4.7cm]{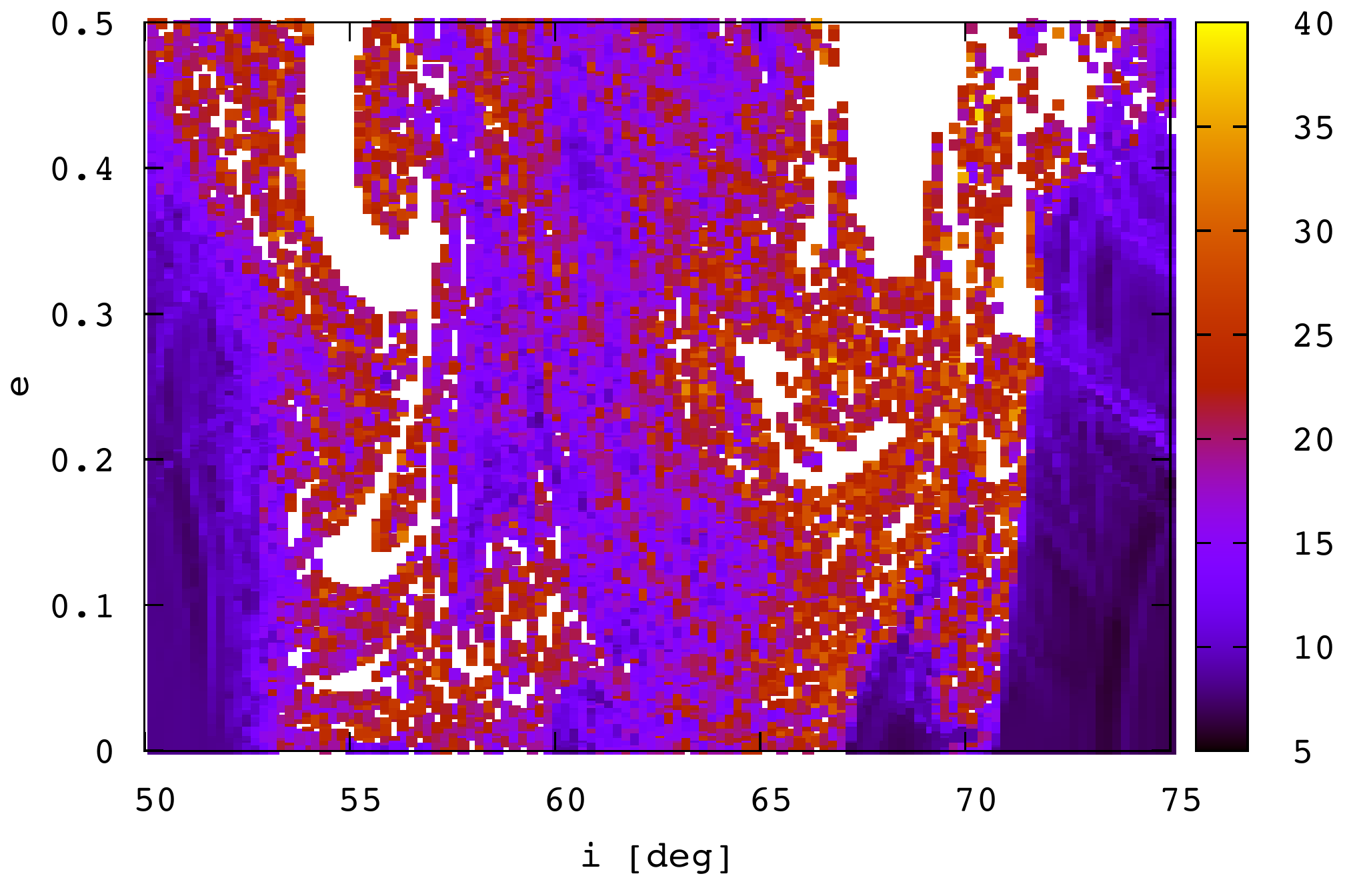}} 
  \caption{FLI stability maps for dynamical model 4.} 
  \label{atlas-m4}
\end{figure}

Figures~\ref{atlas-m1} through \ref{atlas-m4} shows the results of the FLI analysis corresponding to the zoomed-in portion of the action space presented in Fig.~\ref{fig:apertures_zoom} for the four dynamical models of Table \ref{tab:physical-model}. On each of the pictures, the initial conditions $i_0$, $e_0$ are associated to the corresponding FLI value using a black-blue-red-yellow color scale. The FLI of all regular orbits have approximately the same value and appear with the same dark blue color. Light blue corresponds to invariant tori, red and yellow to chaotic regions, and white to collision orbits. The striking feature of Figs.~\ref{atlas-m1}--\ref{atlas-m4} is the transition from ordered to chaotic motion as the perturbation parameter increases. For $a_0 = 19,000$ km, ordered motions dominate in the phase space. Resonant structures are mostly isolated or emanate in a ``V'' shape at the birth of inclination-dependently-only resonances at zero eccentricity. Chaotic orbits are mostly confined along the principal V shape of the critical inclination resonance ($63.4^\circ$) and to small pockets of instability at higher eccentricities. As the semi-major axis increases to $24,000$ km, the geometrical organization and coexistence of regular and chaotic orbits becomes more complex. At $25,500$ km, more and more of the resonant tori are rendered unstable, and the width of the stochastic layers and their degree of overlap are increased. A common characteristic between these maps and the $24,000$ km case is the location of re-entry orbits along the $56.1^\circ$ resonance. Finally, at $a_0 = 29,600$ km (the nominal semi-major axis of the Galileo constellation), roughly speaking, all orbits belonging to the domain delimited by $(i_0, e_0) \in [52^{\circ}, 71^{\circ}] \times [0, 0,5]$ are resonant or chaotic. Re-entry orbits display a very complicated geometry and may now concern quasi-circular orbits near $i_0 = 55^\circ$; that is, strong instability exist that can potentially affect Galileo-like orbits.                

\begin{figure}
	\captionsetup{justification=justified}
	\centering
    \subfigure[dynamical model $1$]
    {\includegraphics[width=7.9cm,height=4.7cm]{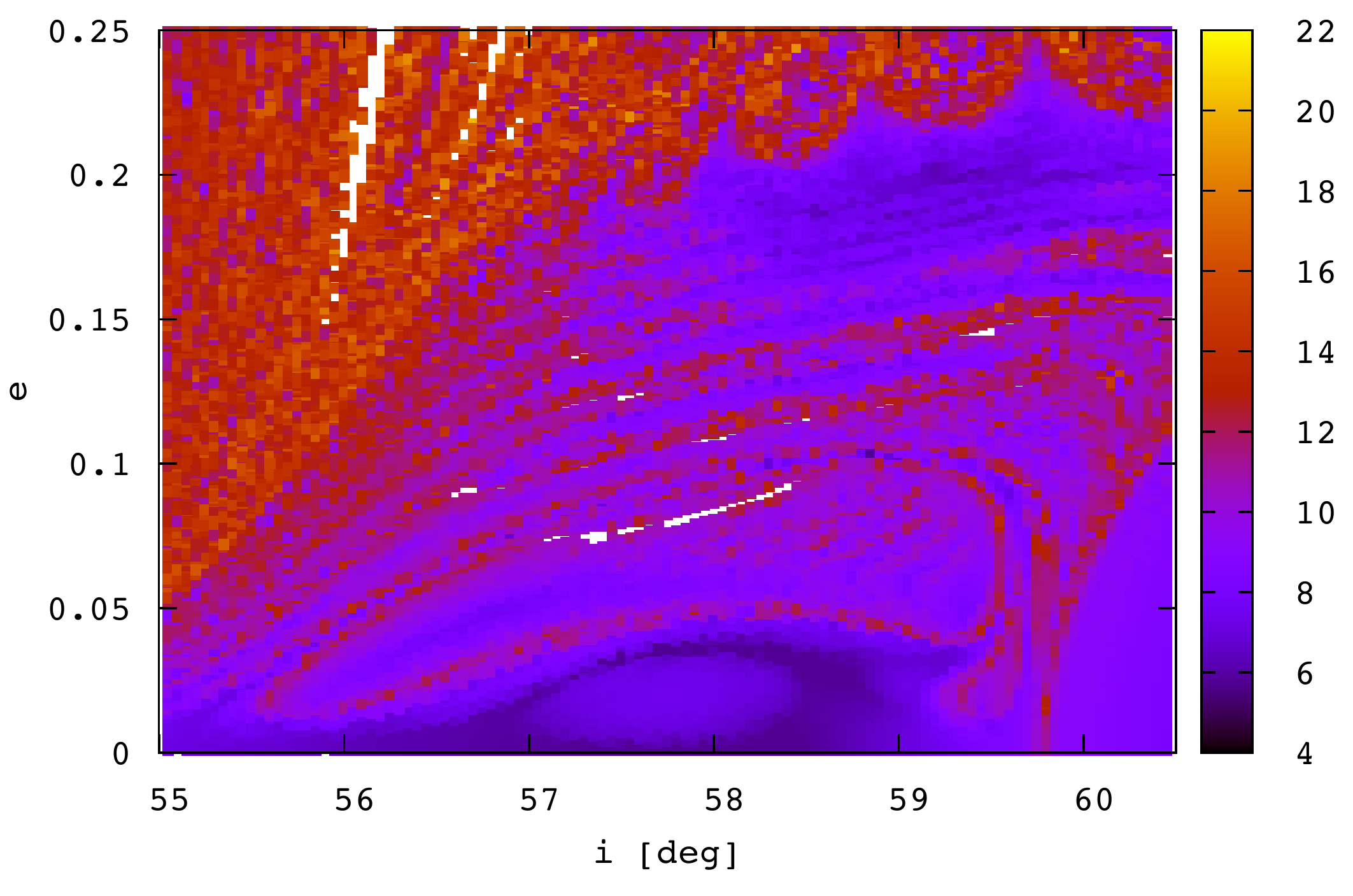}}
    \subfigure[dynamical model $2$]
    {\includegraphics[width=7.9cm,height=4.7cm]{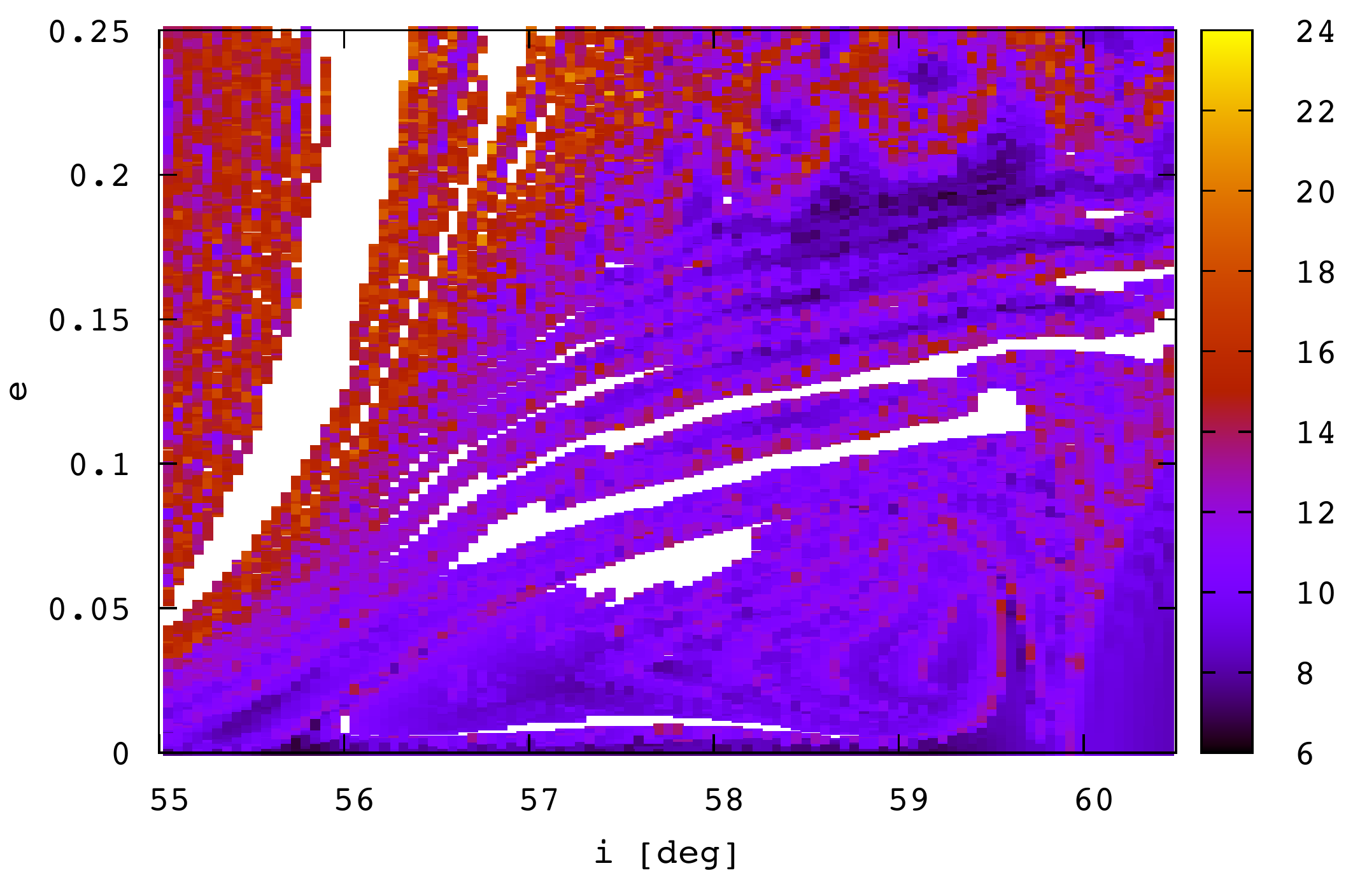}} 
    \subfigure[dynamical model $3$]
    {\includegraphics[width=7.9cm,height=4.7cm]{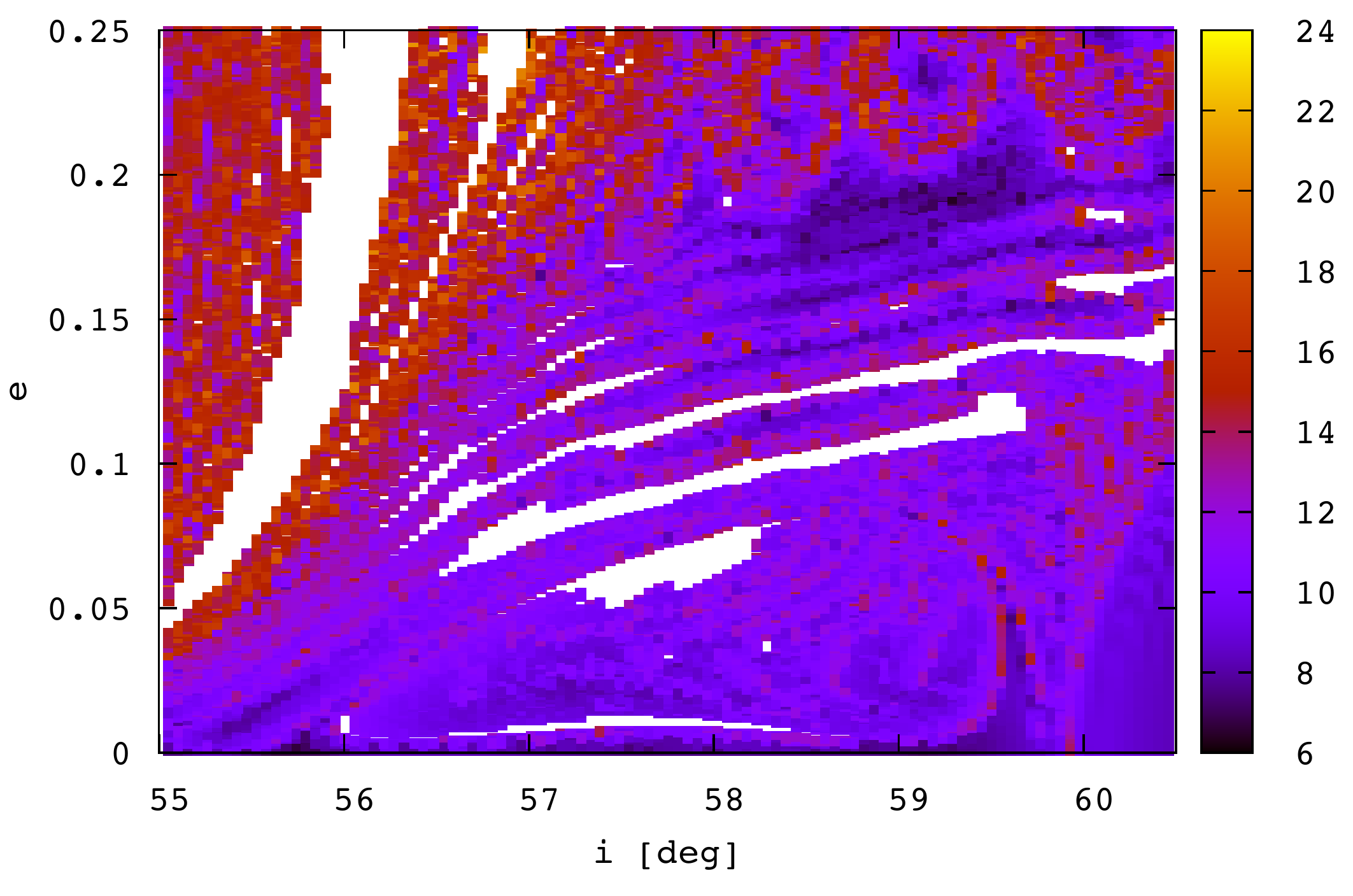}}        
    \subfigure[dynamical model $4$]
    {\includegraphics[width=7.9cm,height=4.7cm]{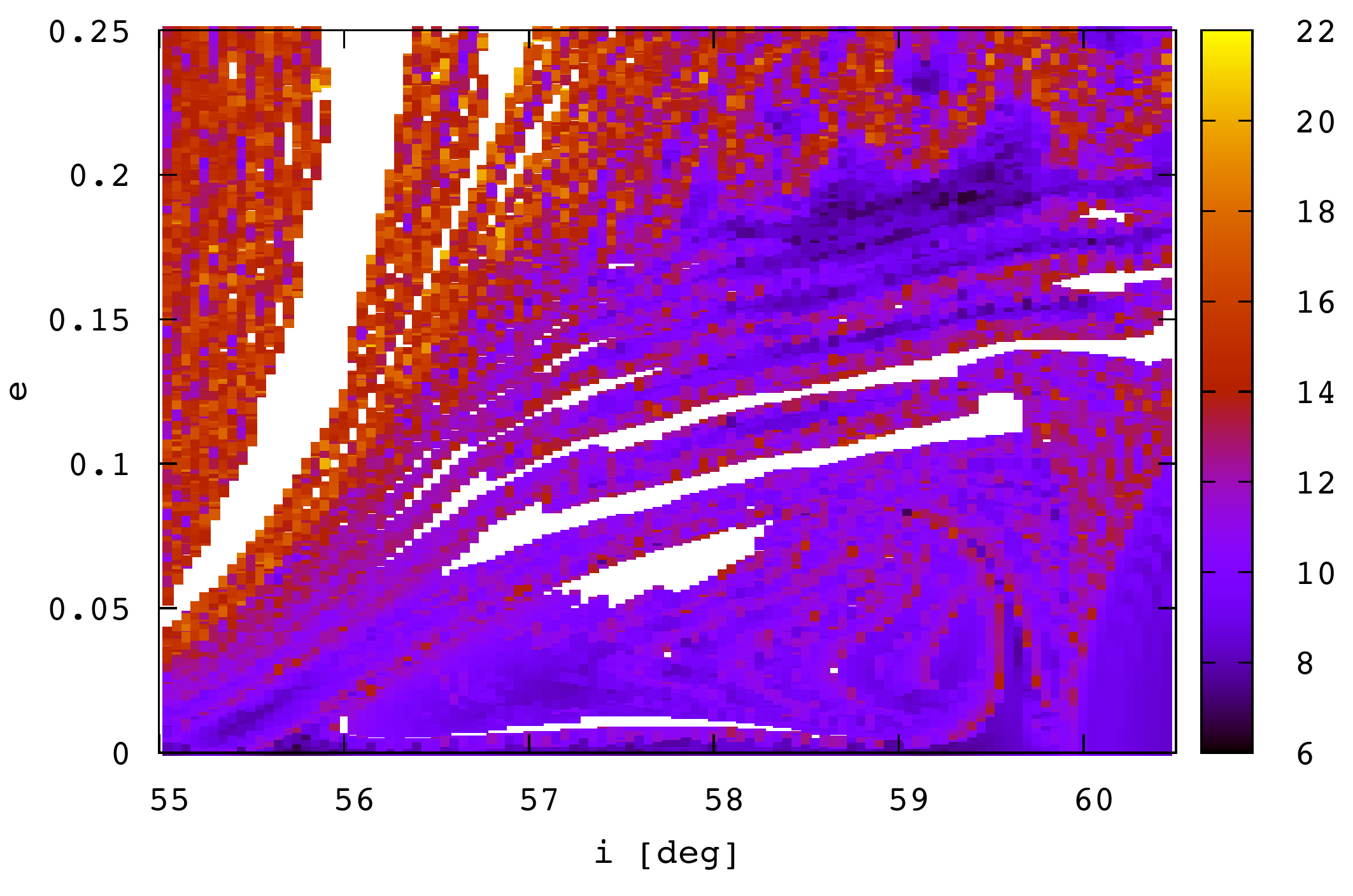}} 
	\caption{Zoomed-in portion for $a_{0}=24,000$ km near the $2\dot{\omega}+ \dot{\Omega}$ only-dependent-inclination resonance under the various dynamical models. Initial conditions have been propagated from the initial epoch $2$  March $1969$ until the final date set to $15$ November  2598. The precise detection of the stable manifolds allows to predict the set of re-entry orbits. These maps also corroborate model 2 as the basic physical model.}
	\label{magnification-atlas2}	
\end{figure}

A fundamental conclusion reached via this parametric approach, using a hierarchy of dynamical models, is that model 2 can be legitimately declared as the basic force model; i.e., the simplest physical model that can capture nearly all of the qualitative and quantitate features (dynamical structures in the maps, degree of hyperbolicity, domains of collision orbits, etc.) of more complicated and realistic force models. In particular, there are no significant changes in the FLI maps of Figs.~\ref{atlas-m3} and \ref{atlas-m4}, when compared to that of Fig.~\ref{atlas-m2}. Force model 2 differs from model 1 only by the presence of the solar third-body perturbation, developed at the quadrupole order. Comparing Figs.~\ref{atlas-m2} and \ref{atlas-m1}, it is clear that that solar perturbations play a non-negligible role on the long-term dynamics. We note, specifically, the manifest widening of the resonant regions and the increase of chaotic orbits near the inclination-dependent-only resonances, principally near $i=56.1^\circ$ and $63.4^\circ$ for all eccentricity values. Moreover, the volume of collision (re-entry) orbits is larger when solar perturbation are taken into account, as is well illustrated in panels (b), (c), and (d) for highly eccentricity orbits along $i = 69^\circ$. This numerical finding confirms, a posteriori, the analytical refinement given by Eq.~\ref{eq:magnitude-harmonic}. The robustness of the physical model 2 was further tested at a smaller phase-space scale, as shown in Fig.~\ref{magnification-atlas2}, giving a magnification of a zone containing many secondary and transverse structures for $a_0 = 24,000$ km. To show finer details and sharper contrast, the resolution and propagation time were enhanced for an objective comparison between the various force models. We can see that, even at this scale, force model 2 can unquestionably be considered the basic physical model: changes in the dynamical structures, and the like, with models 3 and 4 are nearly undetectable. These maps (Fig.~\ref{magnification-atlas2}) also reveal the extraordinary richness of the phase space and the apparition of transverse and thin hyperbolic manifolds (structures with high FLI values). It is worth nothing that re-entry orbits, appearing when the solar contribution is included, are located precisely along those thin manifolds, already present with force model 1. Furthermore, the main collision region in the vicinity of the $56^\circ$ inclination-dependent-only resonance widens significantly with model 2. The identification of the relevant dynamical model is undoubtedly the first question to be addressed when dealing with real (physical) problems. Strangely enough, the isolation of the basic force model for the MEO problem, was largely missing from the literature, though was speculated on in \citet{aR15}. Thus, all future work in this area can be performed under this basic model of oblateness and lunisolar perturbations, and, without loss of generality, the obtained results may be considered representative of the other more refined, and complicated, models.          

\begin{figure}[htp!]
	\captionsetup{justification=justified}
	\centering
    \subfigure[$\Omega_{0}=120^{\circ}$, $\omega_{0}=30^{\circ}$.]
    {\includegraphics[width=7.9cm,height=4.7cm]{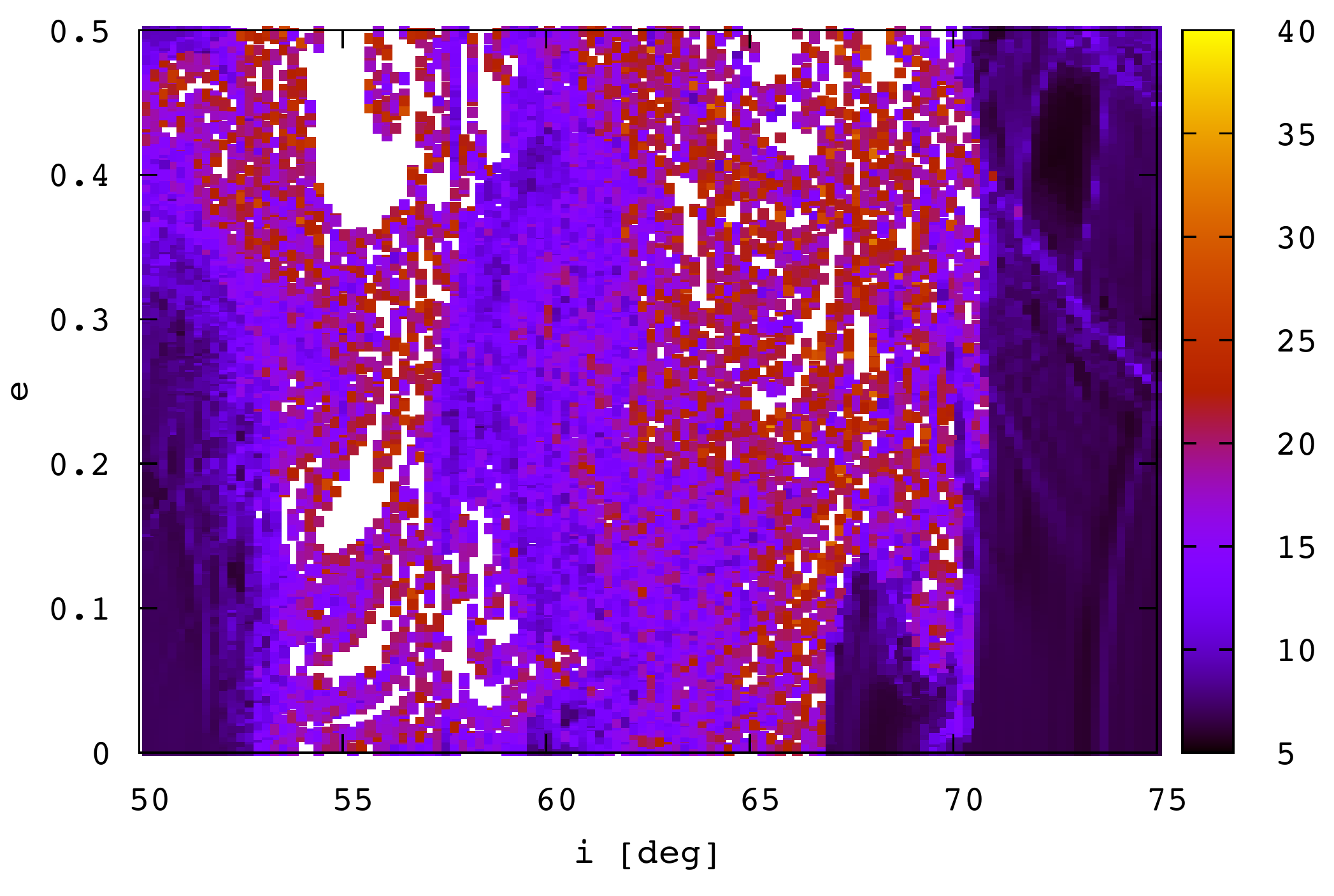}}
    \subfigure[$\Omega_{0}=240^{\circ}$, $\omega_{0}=30^{\circ}$.]
    {\includegraphics[width=7.9cm,height=4.7cm]{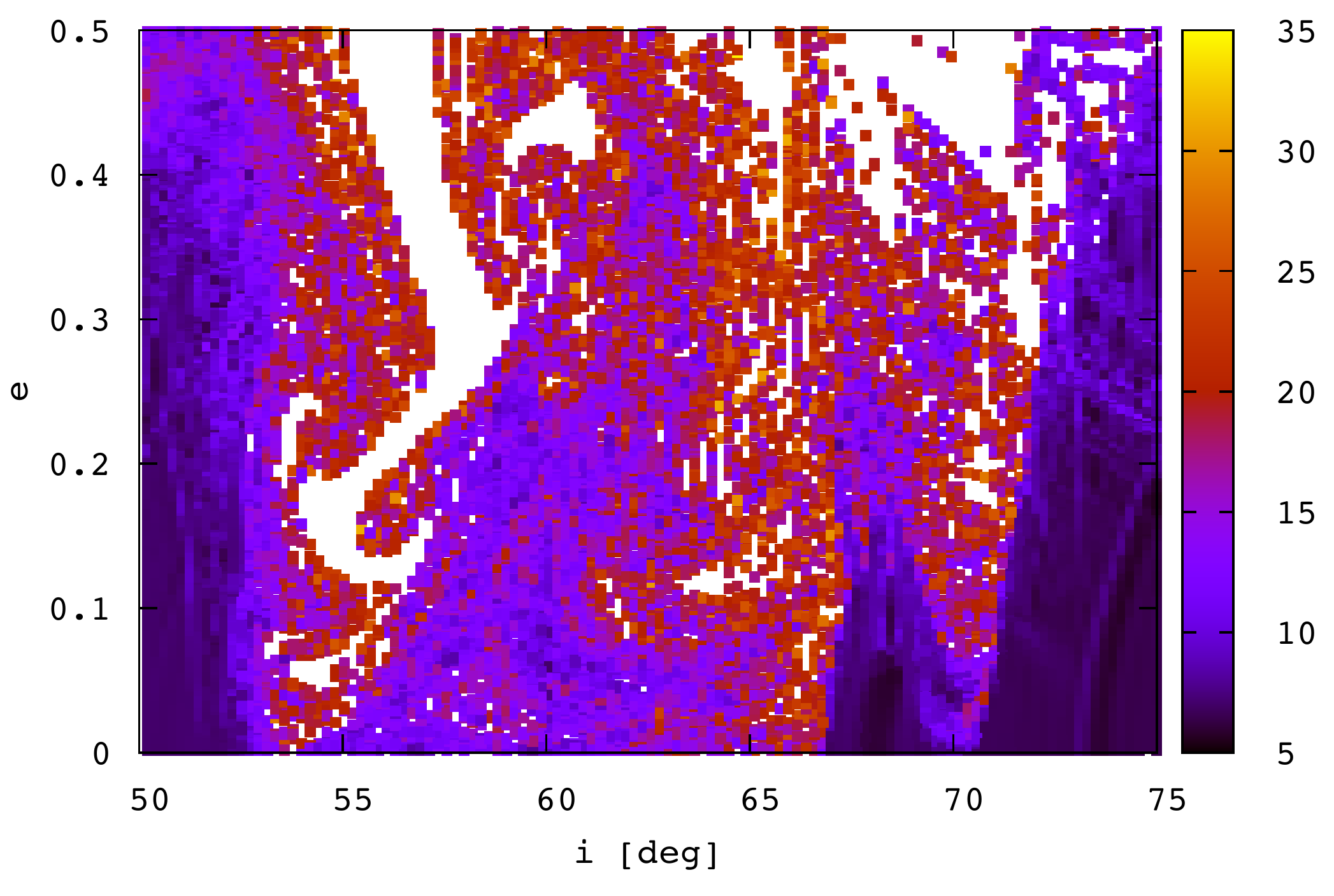}} 
    \subfigure[$\Omega_{0}=120^{\circ}$, $\omega_{0}=75^{\circ}$.]
    {\includegraphics[width=7.9cm,height=4.7cm]{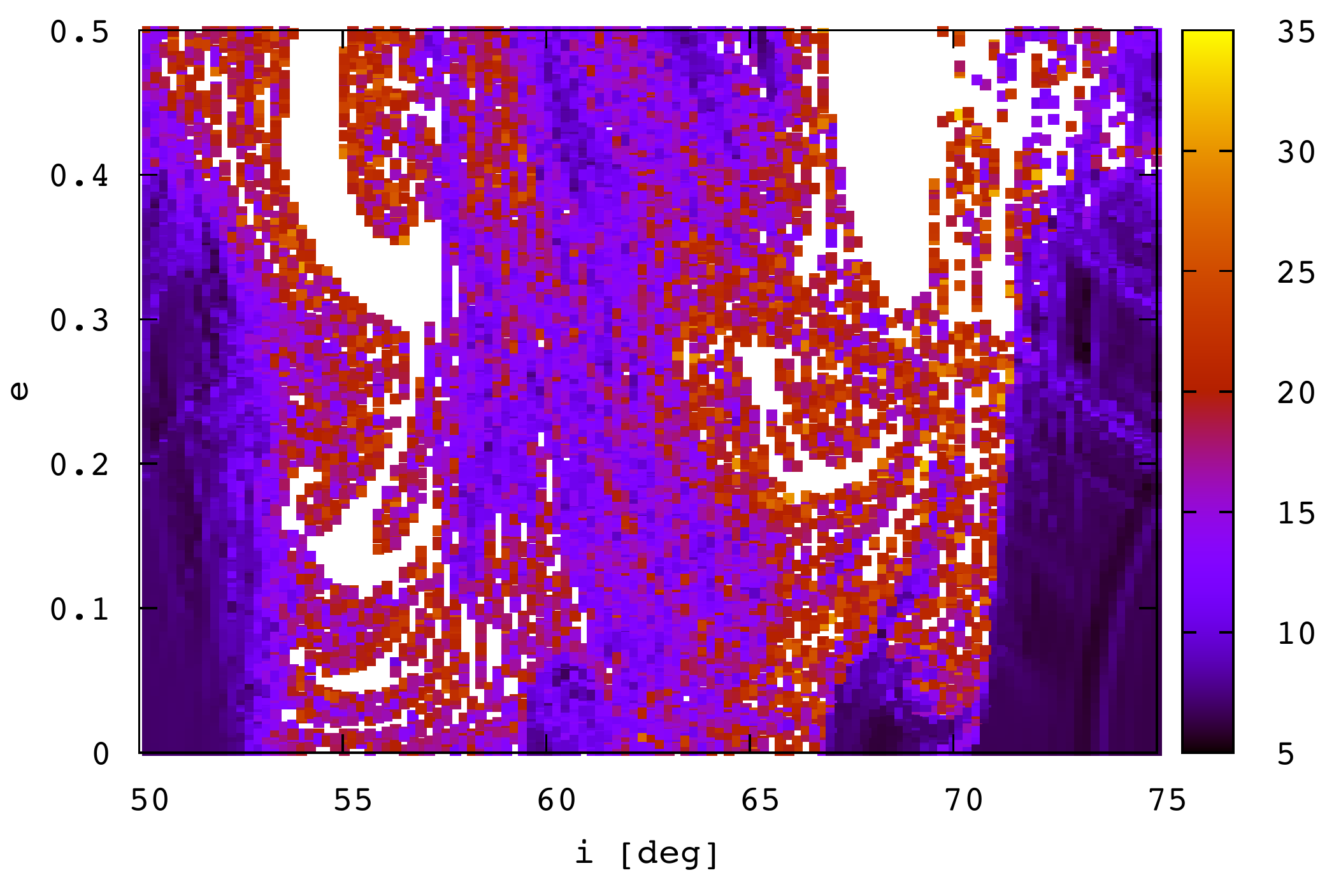}}        
    \subfigure[$\Omega_{0}=120^{\circ}$, $\omega_{0}=30^{\circ}$, initial epoch is 11 March 1975.]
    {\includegraphics[width=7.9cm,height=4.7cm]{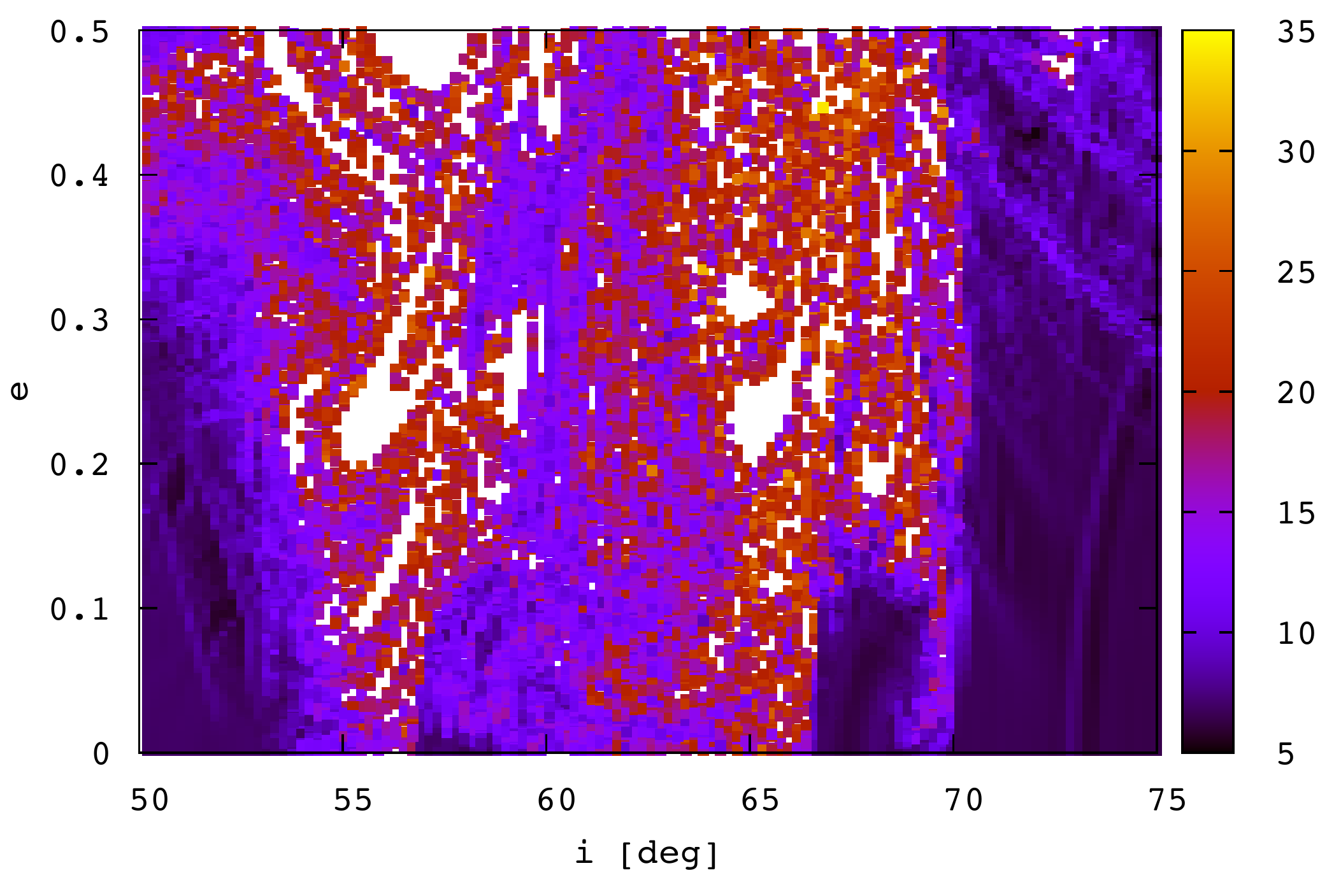}} 
    \caption{FLI stability maps for $a_{0}$ set to $29,600$ km under force model $2$. Unless otherwise stated, the initial epoch of the simulations, which determines the initial dynamical configuration of the Earth-Moon-Sun system, is $02$ March $1969$.}
    \label{fig:galileo}    
\end{figure}

\begin{figure}[htp!]
  \centering
    \subfigure[$\Omega_{0}=120^{\circ}$, $\omega_{0}=30^{\circ}$.] 
    {\includegraphics[width=7.85cm,height=4.7cm]{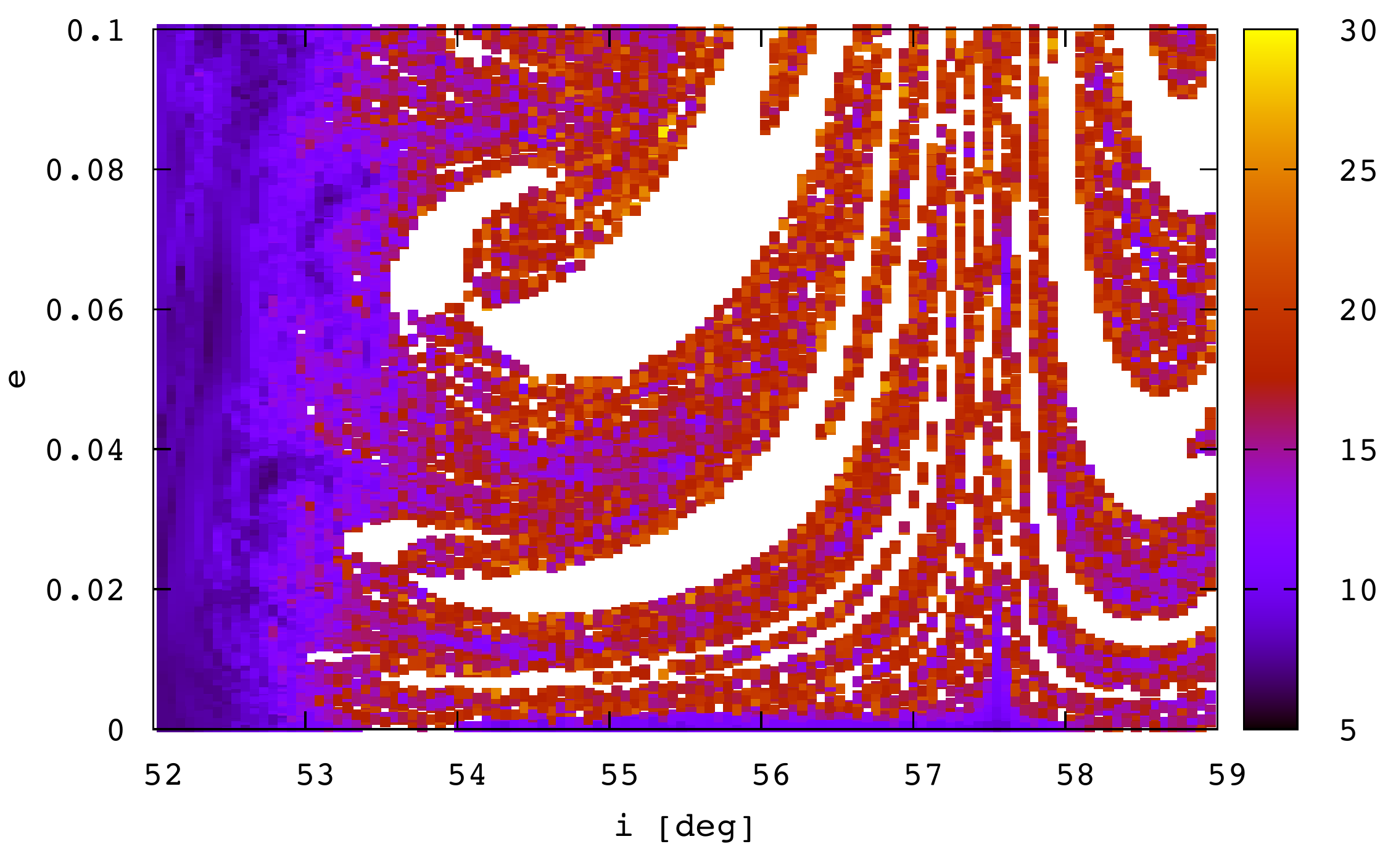}
    \label{sfig:gal_zoom_a}}
    \subfigure[$\Omega_{0}=240^{\circ}$, $\omega_{0}=30^{\circ}$.]
    {\includegraphics[width=7.85cm,height=4.7cm]{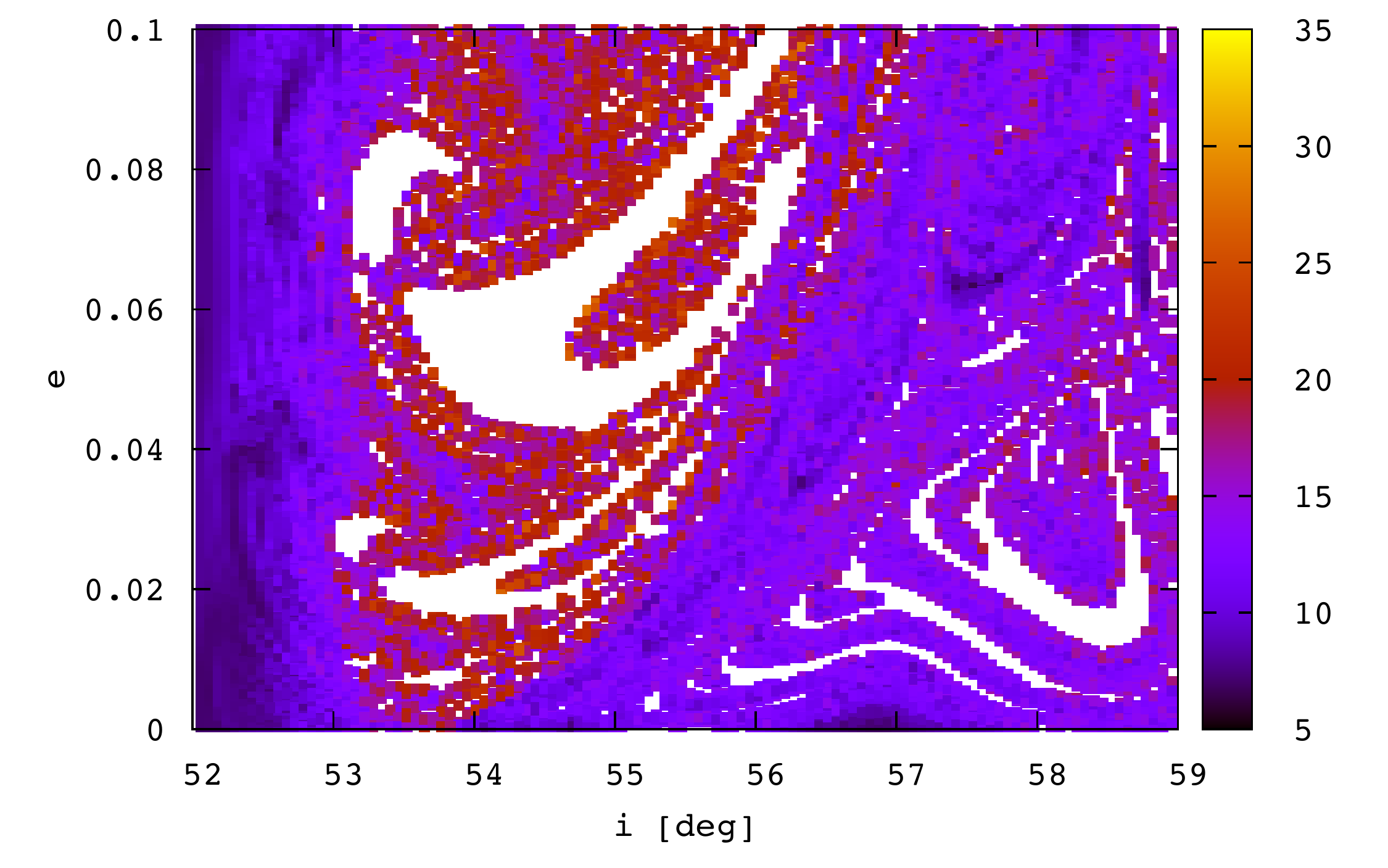}} 
    \subfigure[$\Omega_{0}=120^{\circ}$, $\omega_{0}=75^{\circ}$.]
    {\includegraphics[width=7.85cm,height=4.7cm]{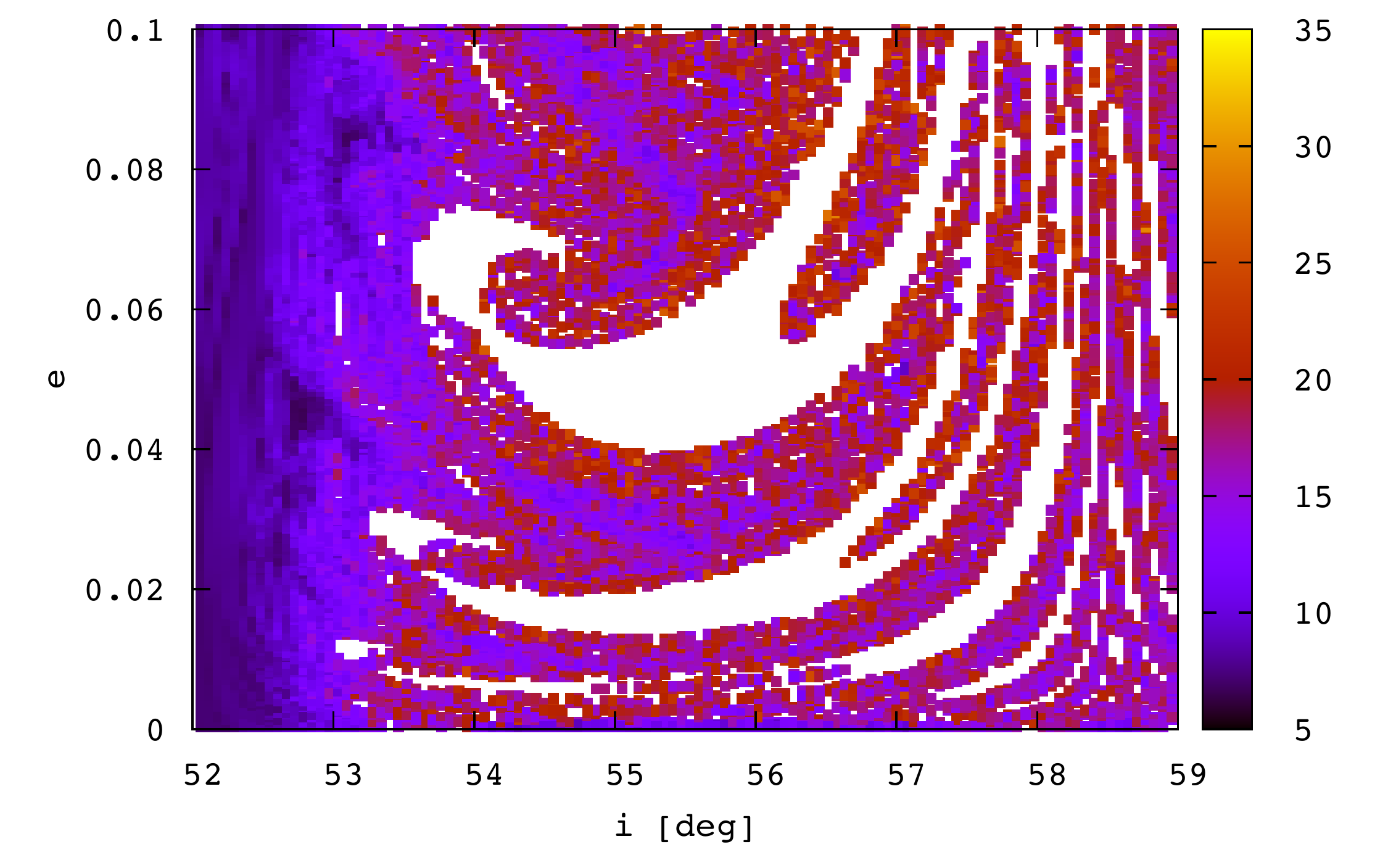}}        
    \subfigure[$\Omega_{0}=120^{\circ}$, $\omega_{0}=30^{\circ}$, initial epoch is 11 March 1975.]
    {\includegraphics[width=7.85cm,height=4.7cm]{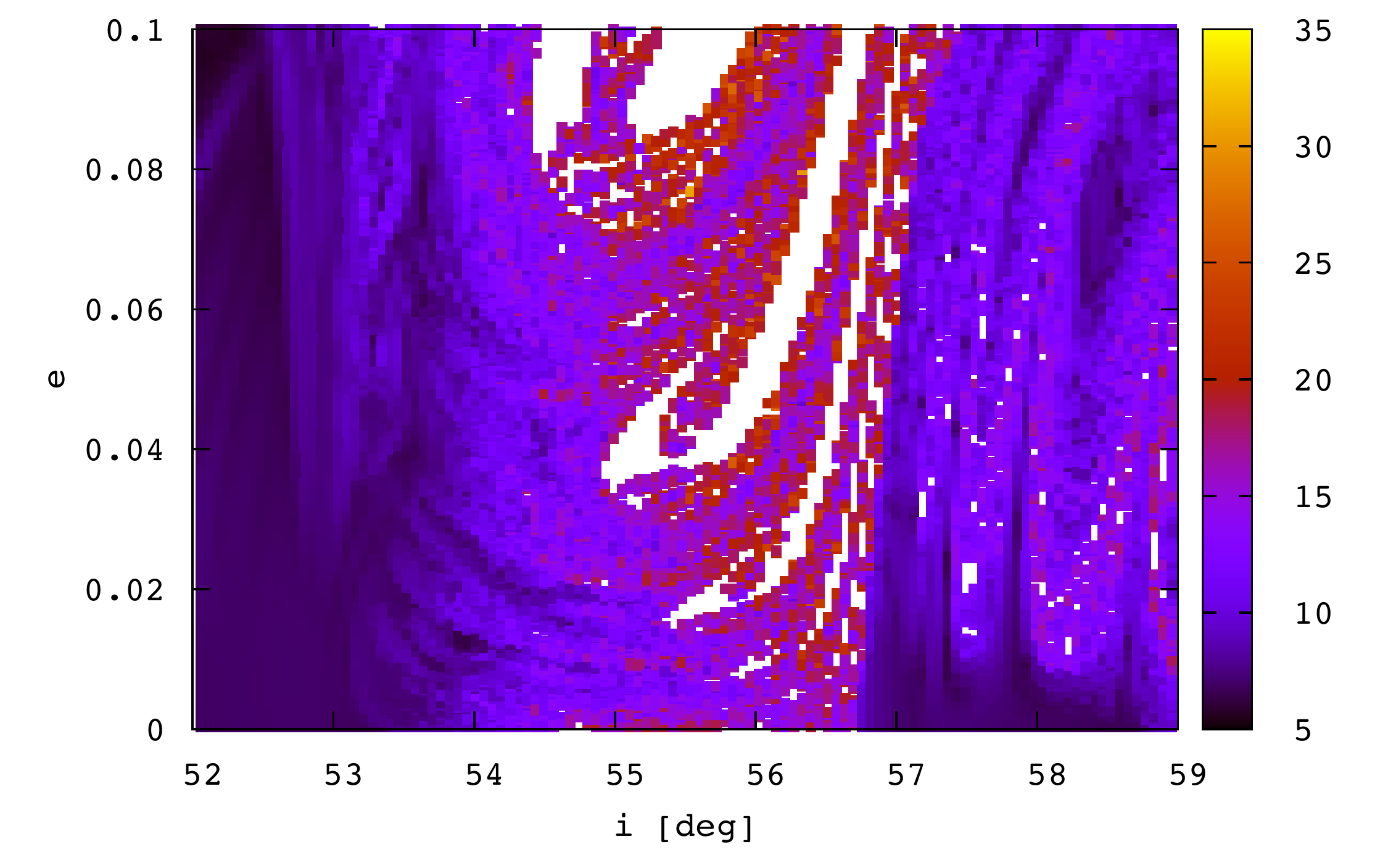}
     \label{sfig:gal_zoom_d}} 
  	\caption{Zoomed-in portion of Fig.~\ref{fig:galileo}, representative of Galileo-like orbits.}
	\label{fig:galileo-zoom}	
\label{magnification-atlas-gal}
\end{figure}

It should be emphasized that all of these charts have been obtained by varying only the initial inclination and eccentricity, with no regard for the initial phases, themselves being fixed for all computed FLI. Figures~\ref{fig:galileo} and \ref{fig:galileo-zoom} illustrate the inherent difficulty to capture the dynamics of a six-dimensional phase space in a plane of dimension two (recall that Hamiltonian, Eq.~\ref{eq:Hamiltonian_3DOF}, is three DOF) \citep{mR14}. They present the FLI maps for $a_0 = 29,600$ km under model 2, at two different phase-space scales, in which the results of the latter may be considered representative of Galileo-like orbits. We can observe how the dynamical structures (stable, resonant, chaotic, or collision orbits) evolve by changing the initial angles $\omega$, $\Omega$, or even the initial configuration of the Earth-Moon system (equivalent to changing the initial epoch of the simulation). Despite this dependence on the initial phases, however, the regime of the global phase space still persists: the region is dominated by chaotic orbits. This does not appear to be the case for the zoomed-in portion of the phase space, where it is clear from Figs.~\ref{sfig:gal_zoom_a} and \ref{sfig:gal_zoom_d} how the initial epoch may strongly influence the stability analysis. Certainly, the dependence of the dynamical structures on the initial phases $(\omega, \Omega, \Omega_\M)$ is a subject that requires further studies, and the associated difficulties are, in essence, due to the representation of the dynamics in a reduced dimensional phase space \citep{mR14}.

\begin{figure}[h!]
	\captionsetup{justification=justified}
	\centering
	\includegraphics[width=7.9cm,height=4.7cm]{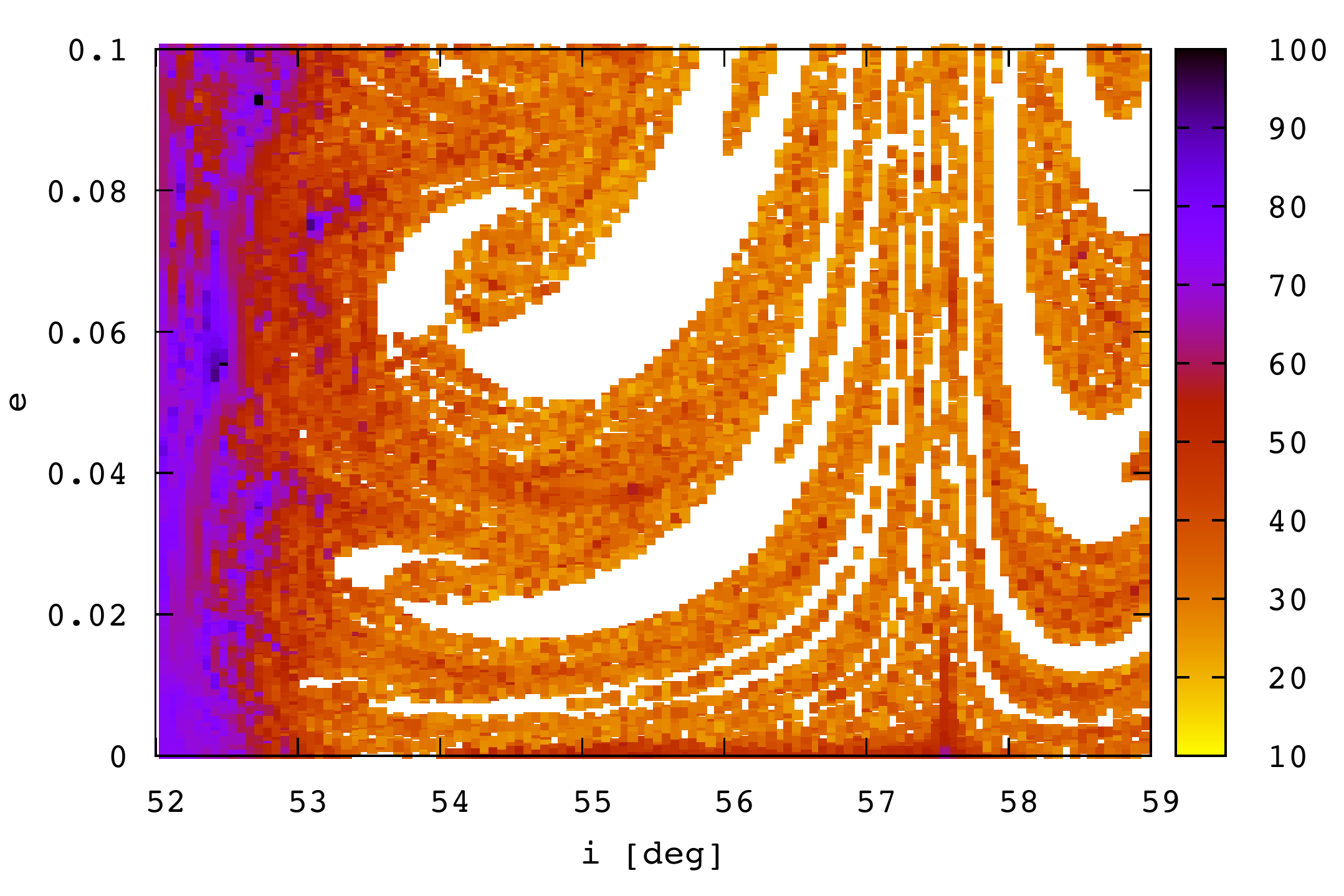} 
	\caption{Map of Lyapunov-time expressed in years. The Lyapunov time represents the barrier of predictability of the dynamical system.  Far beyond this time, the system loses the memory of his initial conditions and statistical properties are more suitable to characterize the evolutionary properties of the motion. Initial conditions are those of Fig.~\ref{sfig:gal_zoom_a}.}
	\label{TL-gal1-m2}
\end{figure}

Besides the quantification of the local hyperbolicity, the predictability is an another important aspect of dynamical interest that we would like to discuss here. The Lyapunov time $\textrm{T}_{\mathcal{L}}$, defined as as the inverse of the maximal Lyapunov characteristic exponent (mLCE), represents physically the barrier of predictability in the dynamical system. Recall that this time, defined by
\begin{align}
	\label{LyapunovTime}
	\textrm{T}_{\mathcal{L}} \equiv \left( \lim_{t \to \infty} \frac{\log \vert \vert w(\tau)\vert \vert}{t} \right)^{-1}, 
\end{align}
goes to infinity for regular orbits (because the mLCE decreases to zero), and, as a result, the dynamical system is predictable for all future time $t$. However, for chaotic systems, the mLCE converges to a positive value and $\textrm{T}_{\mathcal{L}}$ converges to a finite value. We have estimated Eq. \ref{LyapunovTime} numerically for the charts of Fig.~\ref{magnification-atlas-gal} and we have found that many of the orbits have Lyapunov times on the order of decades, as shown by Fig.~\ref{TL-gal1-m2} for a specific case. For the propagation time chosen, slightly larger than $530$ years, the most stable orbits in the maps exhibit, roughly speaking, a Lyapunov time of between $100$ and $120$ years. Thus, in Fig.~\ref{TL-gal1-m2}, the orbits with a Lyapunov time bigger than $100$ years have been assigned to an orbit with a Lyapunov time equals to $100$ years. This does not alter the stability results, but only provide more visual contrast in the maps. This order of predictability, found to be very small indeed for Galileo-like orbits, is a nice invitation to the general reflection of the practical meaning of long-time simulations, so often preformed in astrodynamics. In fact, the problem should be considered as probabilistic in nature and the evolutions of orbits, far beyond the Lyapunov time, have to be considered as only a possible future. In this sense, the long time tracking of individual orbits appears to be inappropriate. The problem requires more fundamental statistical studies that describe the properties of an ensemble of trajectories: this approach is far from common in our community. 

\subsection{Transport in phase space}

The local hyperbolicity and predictability in the MEO region, both first-order variational quantities, have been quantified via FLIs and numerical estimation of Lyapunov times, respectively. Another important feature associated with chaos, which has yet to be addressed, is the transport properties \citep{AstroFica}; that is, the possibility for the dynamical system to explore its phase-space domain. As typified by the asteroid Helga \citep{aM92}, stable chaos may occur in actual (physical) solar system dynamics, for which orbits with short Lyapunov times display no signifiant changes on very long timescales (even several orders of magnitude larger than $\textrm{T}_{\mathcal{L}}$). For the MEO problem, stable chaos is evidently not at the heart of the instabilities and, on the countrary, the apparent growth of the eccentricity on short timescales was how chaos was initially suspected \citep{aR15}. The most rewarding suggestions of the transport properties concerning MEO dynamics are due to \citet{tE02}, who showed the tendency for typical GPS orbits to follow the resonant skeleton give by Fig.~\ref{fig:skeleton}. This explains how quasi-circular orbits may become nearly hyperbolic on millennia timescales. Similar results were obtained recently by \citet{aR15}, using stroboscopic techniques, who demonstrated this phenomenon on much short timescales for the various navigation constellations. Here, we use our FLI analysis, which more clearly reveals the extent and geometry of the chaotic sea in the phase space, to further enliven this idea, by showing numerically how the dynamical structures in the FLI maps affect the long-term evolutionary properties of the motion. 

\begin{figure}[htp!]
	\captionsetup{justification=justified}
	\centering
	\subfigure[transport along the isolated $2\dot{\omega} + \dot{\Omega} = 0$ resonance.]
	{\includegraphics[width=7.8cm,height=4.7cm]{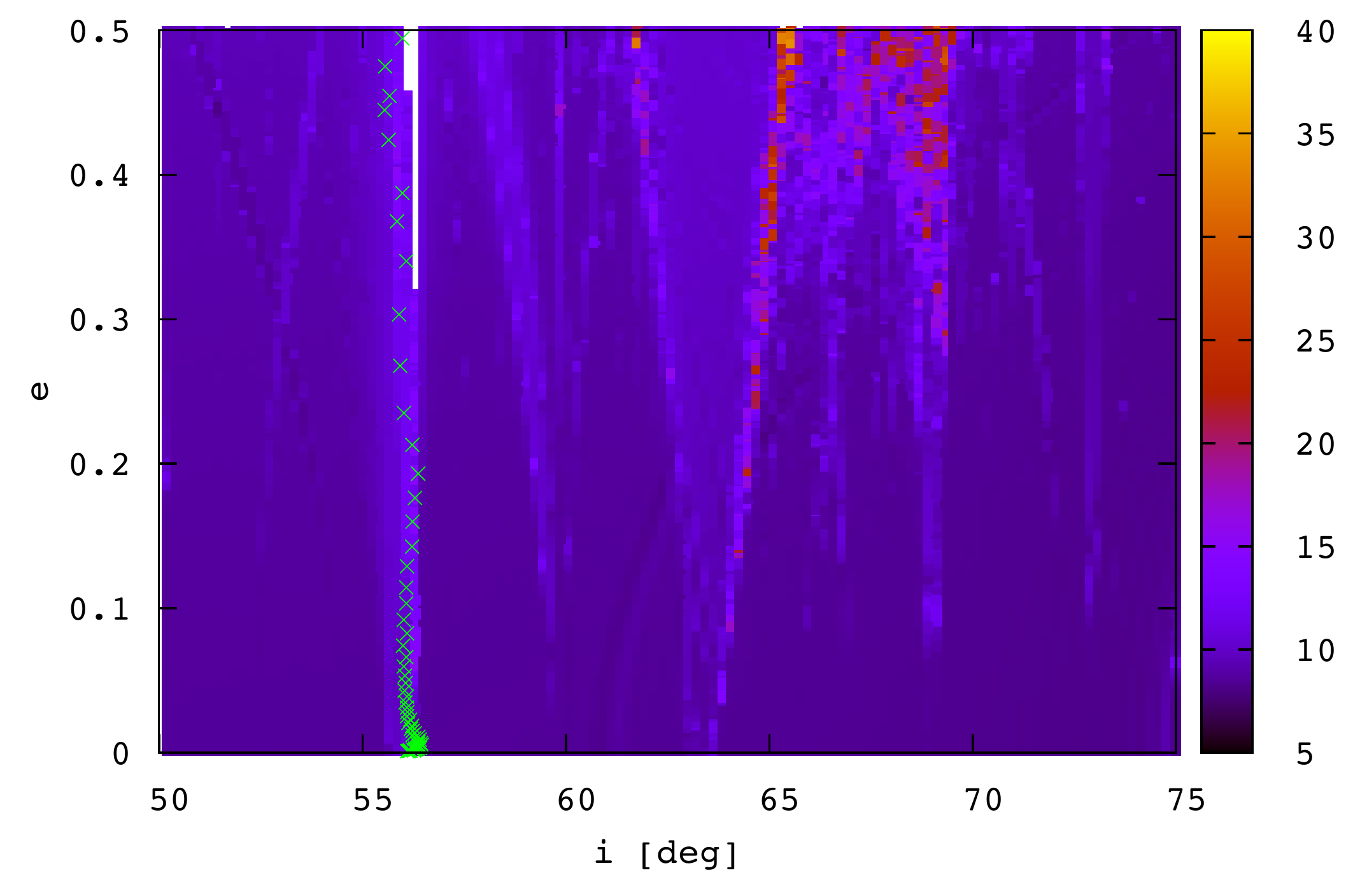}
    \label{sfig:trans_atlas1_i56}}
    \subfigure[transport along the isolated $2\dot{\omega} + \dot{\Omega} =0$ resonance; 
    zoomed-in portion.]
    {\includegraphics[width=7.8cm,height=4.7cm]{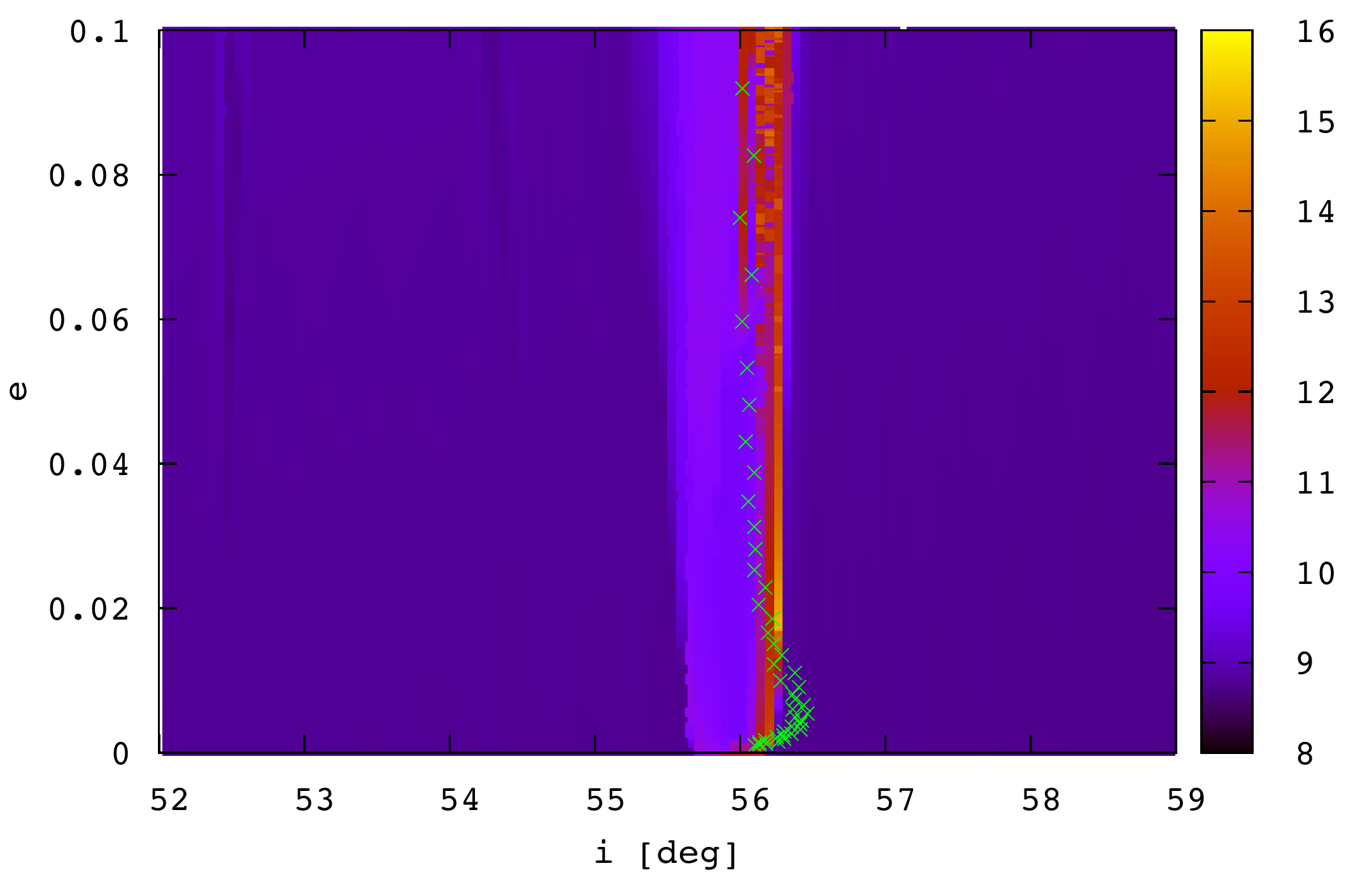}
    \label{sfig:trans_atlas1_i56_zoom}} 
    \subfigure[Transport along the isolated critical inclination resonance.]
    {\includegraphics[width=7.8cm,height=4.7cm]{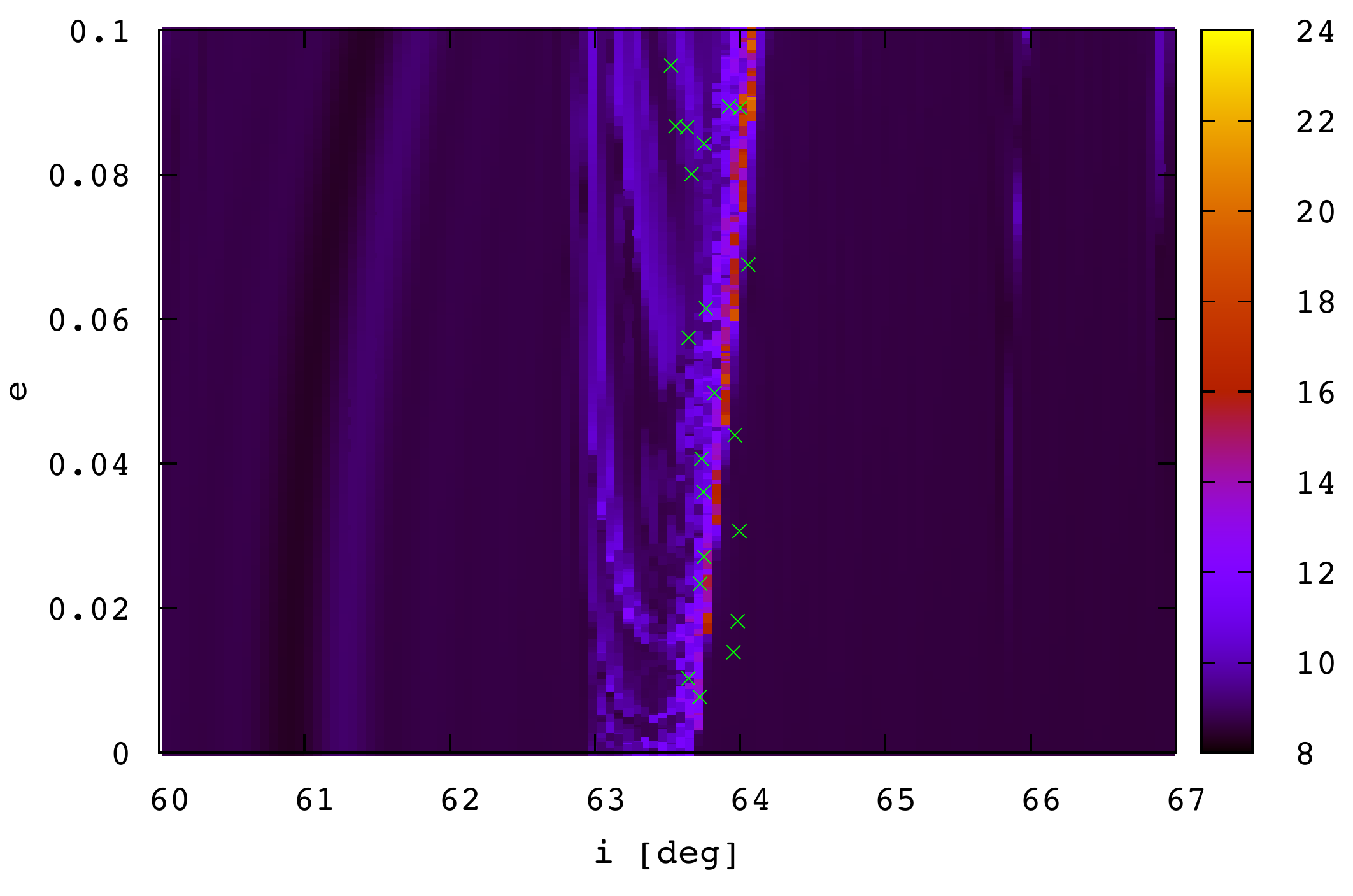}
    \label{sfig:trans_atlas1_i63_zoom}}        
    \subfigure[Macroscopic transport and confinement.]
    {\includegraphics[width=7.8cm,height=4.7cm]{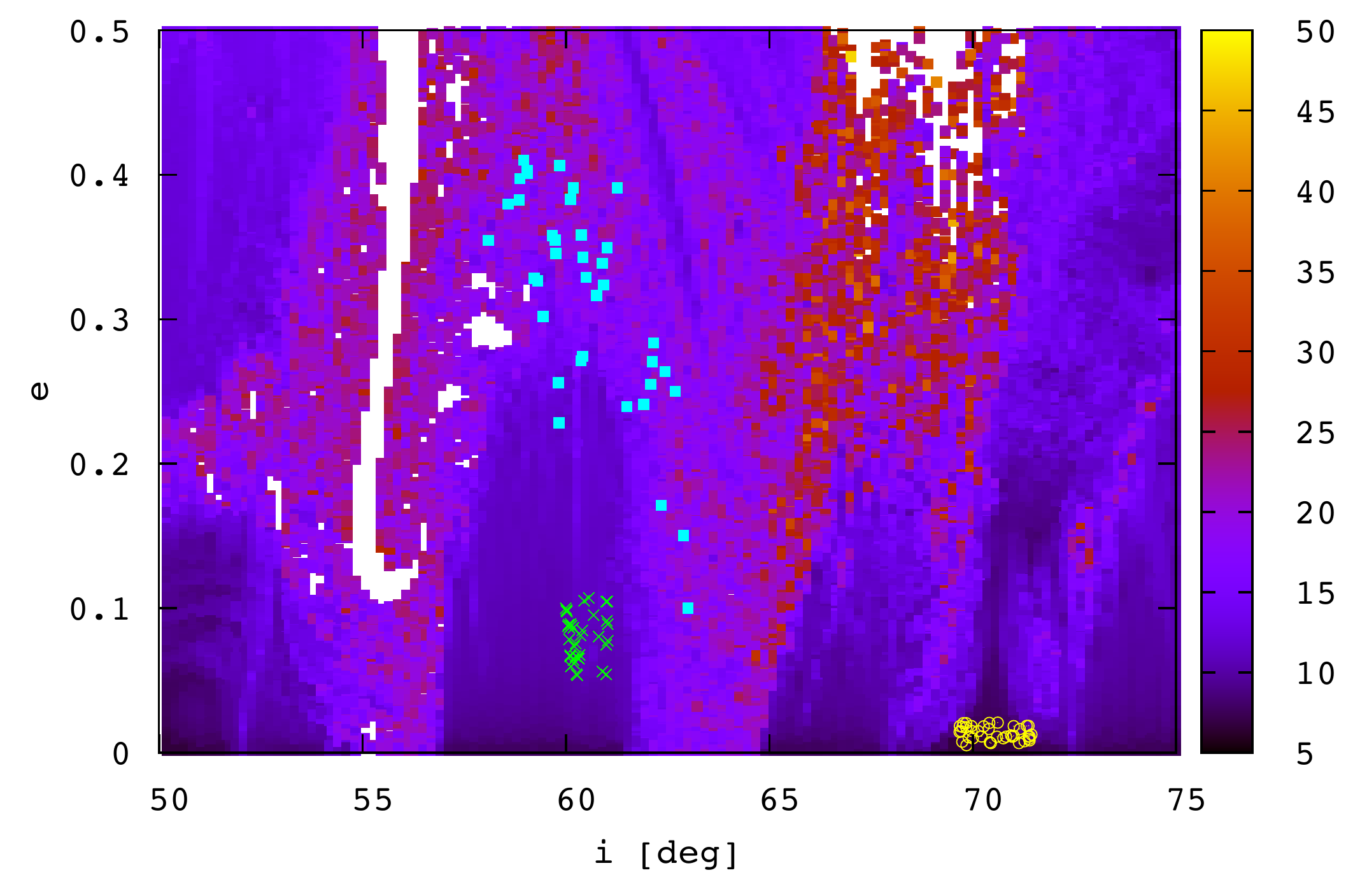}
    \label{sfig:trans_atlas3_transport_confinement}} 
    \caption{Illustration of the transport properties for orbits with initial condition near isolated resonances or for orbits with initial condition in the resonance overlapping regime or far from them. All orbits are semi-analytically propagated using the basic force model $2$ and are stroboscopically overlaid onto the FLI maps obtained with the same force model. The nature of the resonances compel the transport in the phase space. The structures in the FLI maps, in addition to quantifying the hyperbolicity, also reveal the preferential and possible routes for transport in phase space.}
    \label{fig:transport}
\end{figure}

Figure~\ref{fig:transport} shows the long-term evolution (plotted stroboscopically once per lunar nodal period) of orbits with initial inclinations and eccentricities in particular regions of the phase space superimposed on the background dynamical structures obtained from the FLI computations. For orbits initial located far from the resonances, nothing dynamically interesting happens in terms of transport: the motion is quasi-periodic and the excursions in the action space are modest, bounded, and confined by the invariant tori. Figures~\ref{sfig:trans_atlas1_i56}--\ref{sfig:trans_atlas1_i63_zoom} show that in the vicinity of isolated resonances (where no adjacent resonances exist), the orbits have the tendency to explore a larger portion of their phase space, evolving along this resonance. The evolutionary properties of the eccentricity and inclination are mainly determined by the shape of the resonance, as revealed by the FLI analysis. Because of the geometry of the dominate isolated resonances (the V shapes clearly distinguishable for $a_0 = 19,000$ km), we see how some orbits can have a very stable inclination despite significant growth in eccentricity (evolving along the resonant line). For orbits with initial conditions in the resonance overlapping regime, where invariant tori are destroyed, we find that excursions of the eccentricity and inclination are fast and macroscopic (Fig.~\ref{sfig:trans_atlas3_transport_confinement}). The exploration of a large phase-space volume is permitted, and, despite the erratic appearance of the motion, when sampled stroboscopically we see a tendency for the orbits to jump from one resonance domain to another, being confined to the chaotic sea. Figure~\ref{sfig:trans_atlas3_transport_confinement} also shows the evolution of orbits in the stable domains (marked green and yellow), where the variations are quasi-periodic. Such results were further confirmed by a large amount of simulations, and thus we can conclude that the FLI maps clearly reveal how transport is mediated in the phase space. 

\section{Testing the Chirikov criterion} 
The dynamical structure of the phase space of the MEO region was enlightened using two different, but complementary, approaches: the heuristic Chirikov criterion and a CPU-expansive numerical stability survey. We would like to stress at the outset that, to the best of our knowledge, the Chirikov principle has never been applied to a real (physical) system of more than 2 degrees of freedom; recall that in our study the autonomous version of the Hamiltonian is 3 DOF, leading to a 6 dimensional phase space, whereby the resonance widths involve two actions. The realistic nature of Chirikov's analytical results is then of primary importance, and so a synthesis of the theoretical and numerical aspects of this work is presented here. 

\begin{figure}[htp!]
	\centering
	\subfigure[$a_{0}=19,000$ km.]
	{\includegraphics[scale=0.725]{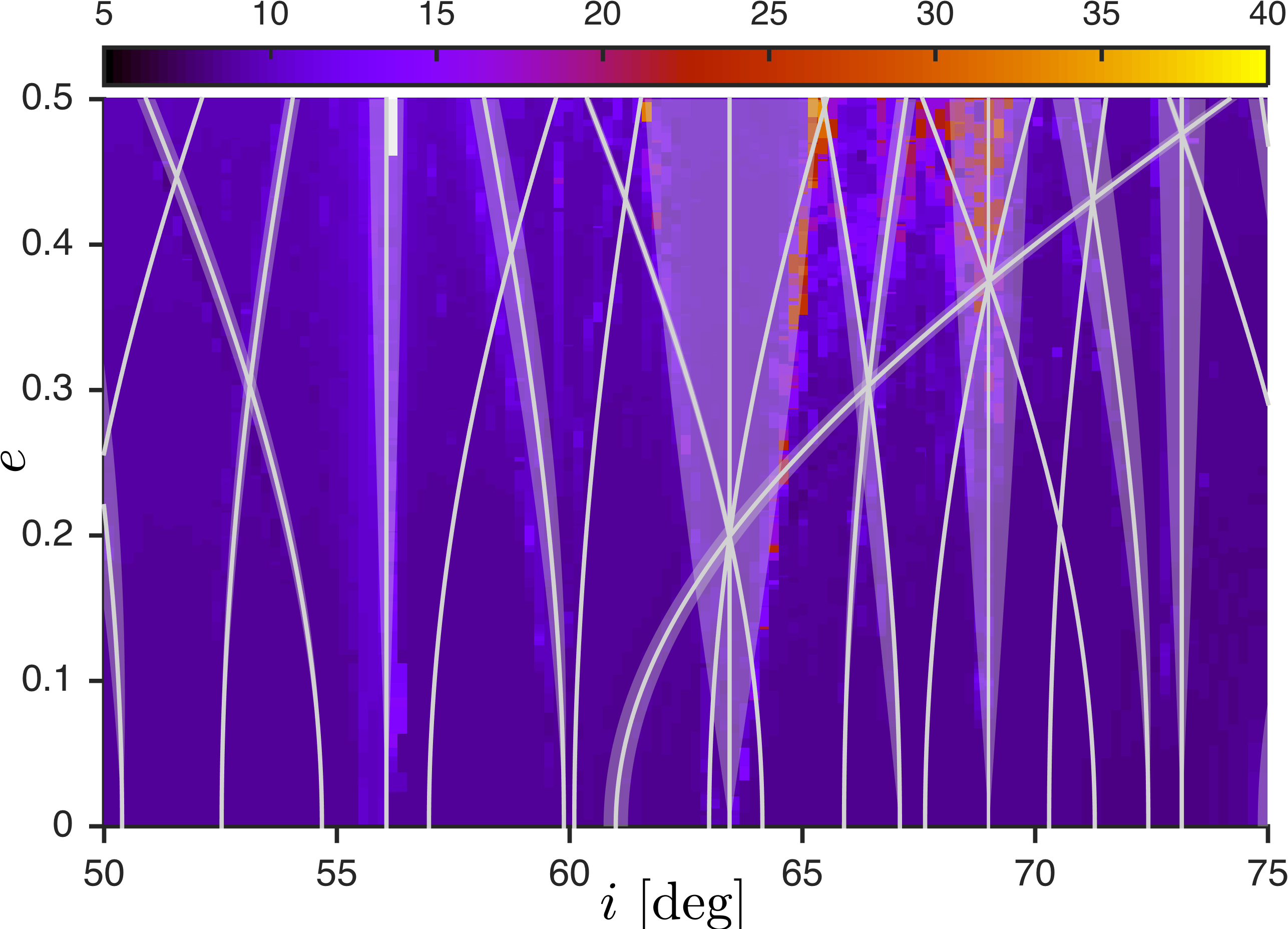}}
    \subfigure[$a_{0}=24,000$ km.]
    {\includegraphics[scale=0.725]{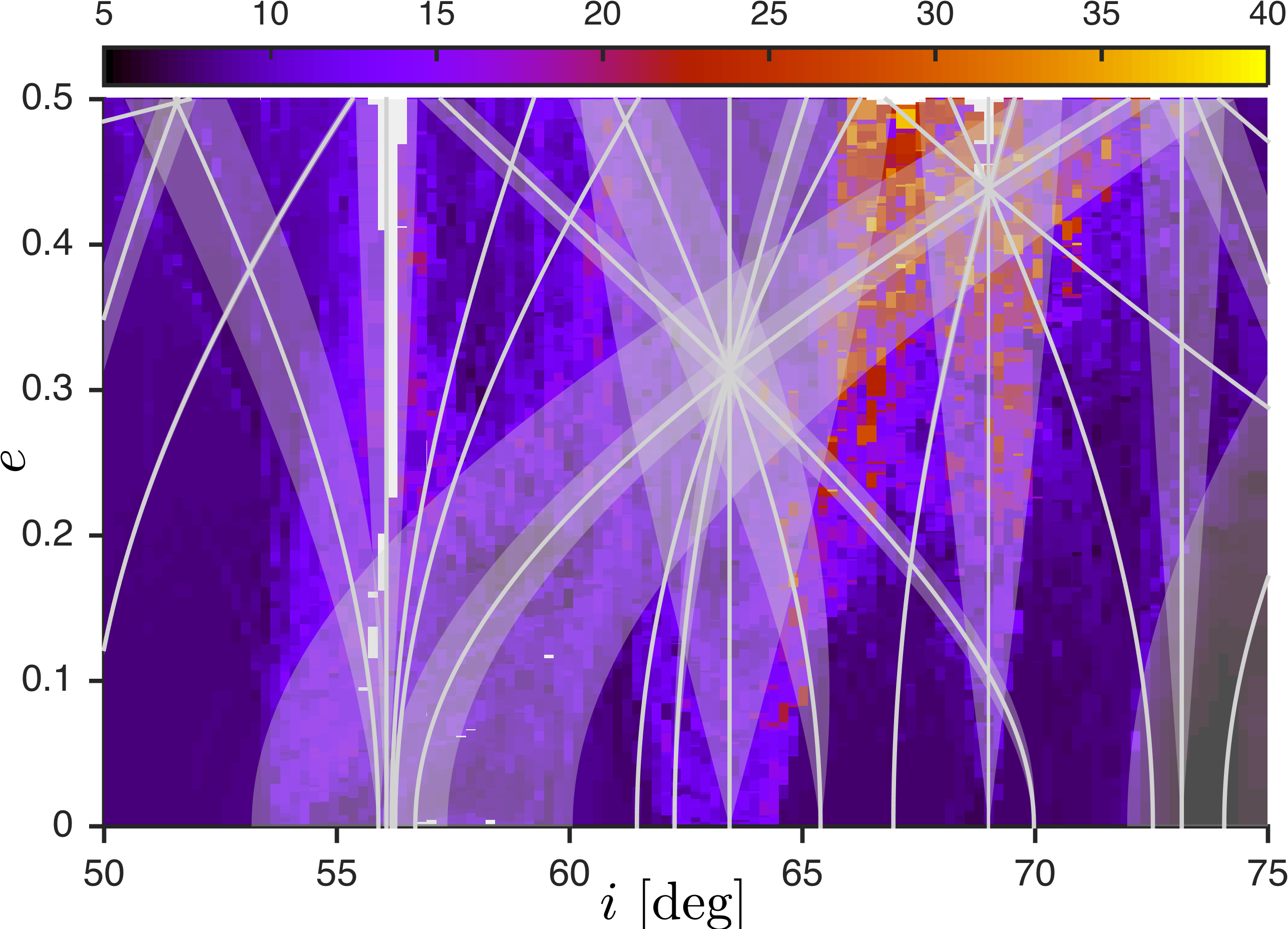}} 
    \subfigure[$a_{0}=25,500$ km.]
    {\includegraphics[scale=0.725]{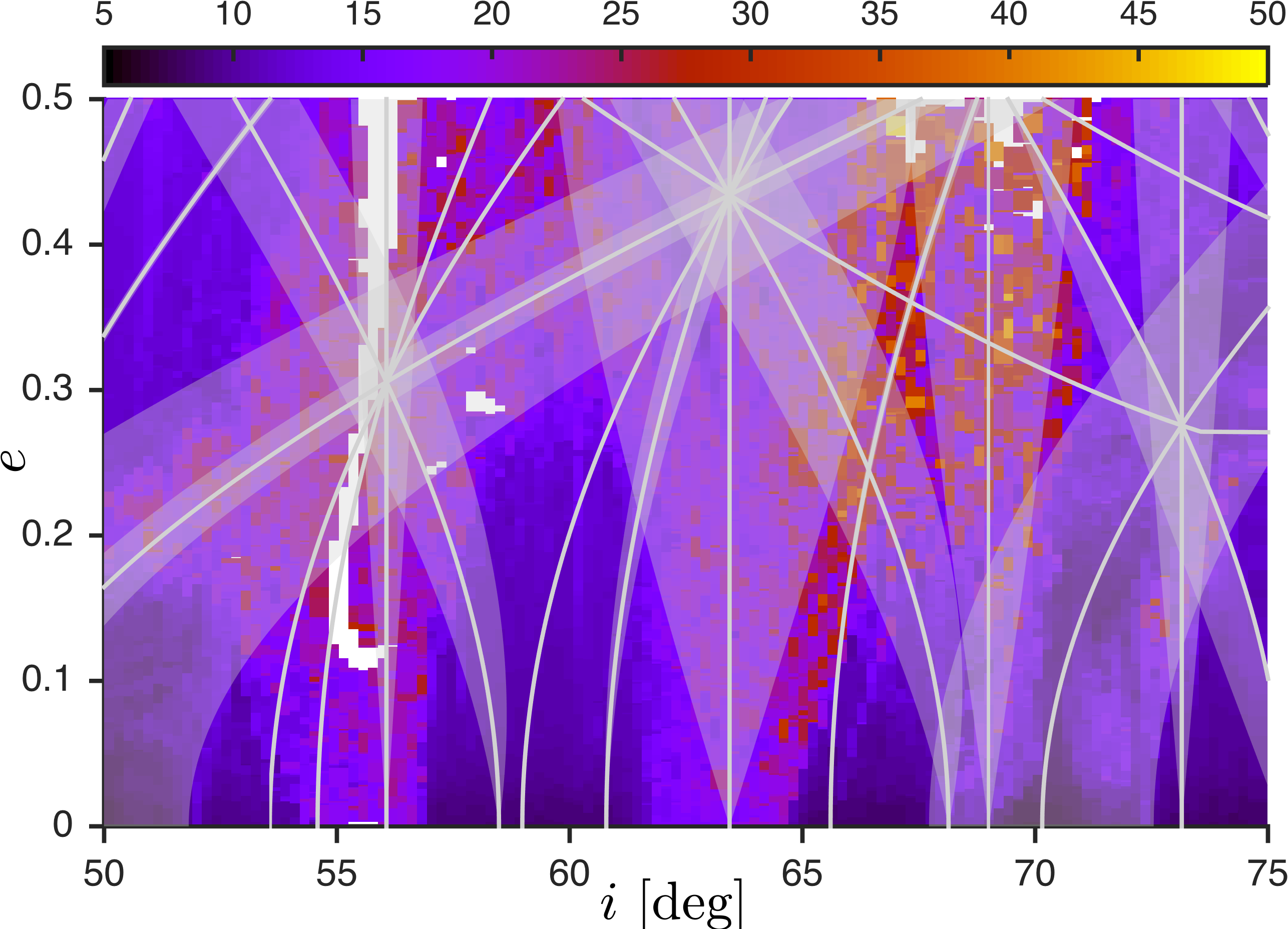}}        
    \subfigure[$a_{0}=29,600$ km.]
    {\includegraphics[scale=0.725]{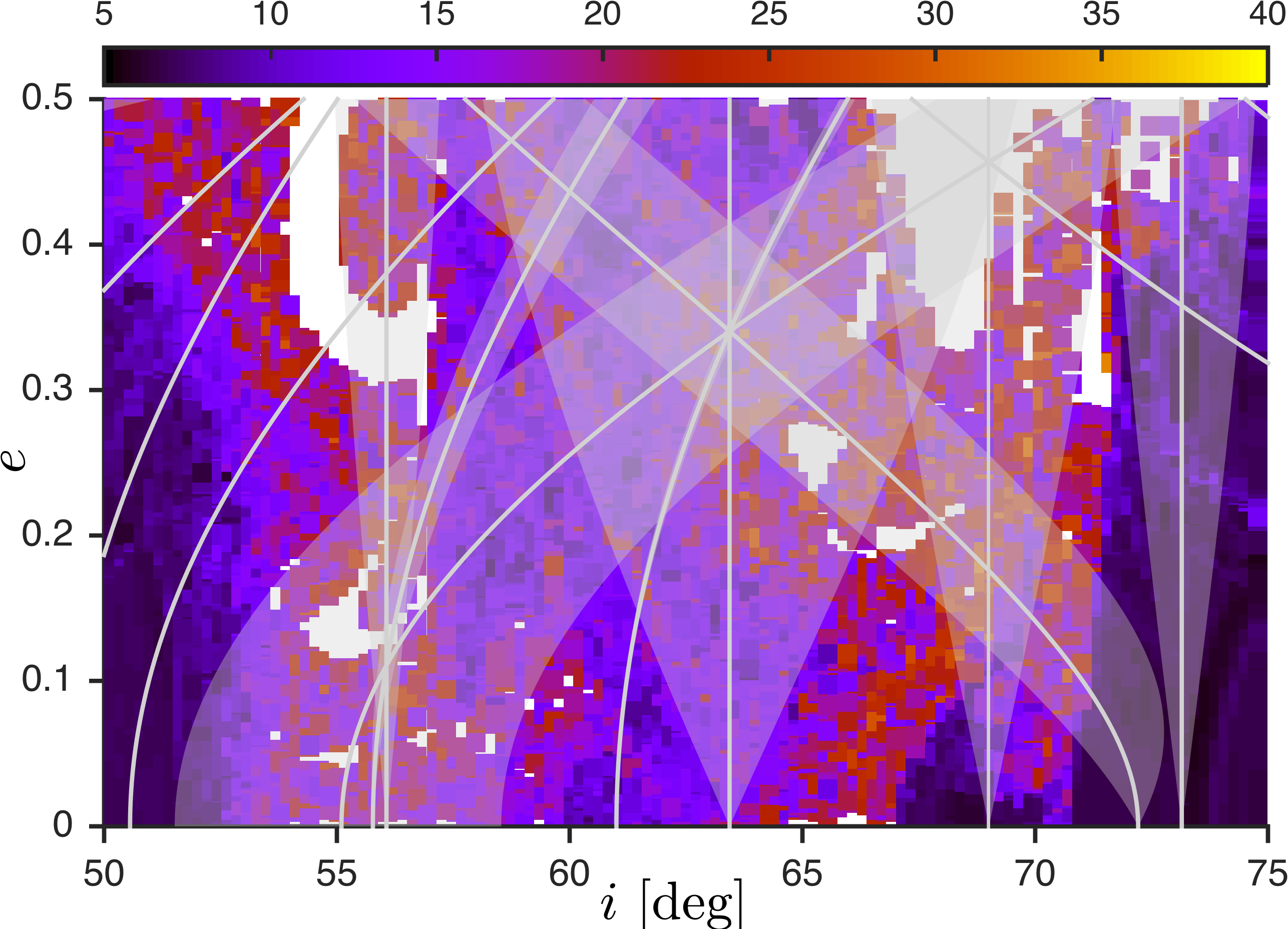}} 
    \caption{FLI stability maps versus Chirikov's resonance overlap for force model $2$.} 
  \label{fig:FLI_vs_Chirikov}  
\end{figure}

Figure~\ref{fig:FLI_vs_Chirikov} presents the lunisolar secular resonance centers and widths of Fig.~\ref{fig:apertures_zoom}, now appearing as light gray, superimposed on the corresponding FLI stability maps of Fig.~\ref{atlas-m2}, both obtained under the same basic force model of oblateness and lunisolar perturbations. For $a_0 = 19,000$ km, where the resonances are mostly isolated (the fundamental assumption made in our analytical treatment), the agreement is both qualitatively and quantitatively very satisfactory. The FLI numerically detected V shapes, emanating from $56.1^\circ$, $60^\circ$, and $63.4^\circ$ at zero eccentricity, are well captured analytically. As the semi-major axis increases, however, the discrepancy with the isolated resonance hypothesis becomes manifest, as was noted from the FLI analysis. Accordingly, the comparison between the two approaches becomes more and more difficult to quantify, though we note that the most important qualitative feature---the fact that the systems tends towards the overlapping chaotic regime---can be seen from both. The analytical development captures indubitably the tendency of global overlap, even if a $1-1$ correspondence between the overlaps and the numerical detection of chaos is hard to formulate. Nevertheless, we note that some numerical structures are nicely captured analytically, such as the widths of the resonances for moderately eccentric orbits near the inclination of $54^{\circ}$--$60^{\circ}$ and $62^{\circ}$--$64^{\circ}$ for $a_0 = 24,000$ km, or near the critical inclination for $a_0 = 25,500$ km. Also apparent for all semi-major axes is the large number of chaotic or escaping orbits along the $69^\circ$ resonance at high eccentricities at the intersection of multiple resonant curves.

Reasons for any discrepancies surely arise from the fact that the Chirikov criterion is imperfect by nature: resonances are treated solely in isolation. Certainly, more sophisticated and rigorous analytical methods that treat resonance interactions, such as those detailed in \citet{gH99}, should be pursued in the future. But despite the approximations of the Chirikov criterion, it has permitted us to gain a formidable intuition into the nature of this physical problem. The application of this analytical criterion is a strong testimony that despite the great power and scope of modern computers, there is still a place in celestial mechanics for pen-and-paper calculations in the style of our forebears. 

\section{Discussion and future work} 

We present below several interesting aspects of the MEO problem that have not been adequately addressed here; points on which to concentrate in future works.  
\begin{enumerate}

	\item We have pointed out the critical role of the initial phases $(\omega, \Omega, \Omega_\M)$ on the dynamical structures in the FLI maps, leading to different chaotic pictures in the phase space. Yet, we gave no information about which angles will ensure or avoid chaos for a given initial inclination and eccentricity. This question is clearly of significant practical interest and needs further investigation. 
	
	\item Even if we demonstrated a $1$--$1$ correspondence between chaos and large-volume exploration of phase space, the precise nature of the transport has not been specified. In particular, no distribution of the increments $\Delta e$ with time was given. Such analytical modeling of $f (\Delta e, t)$ is  currently being studied, based on considerations in \citet{nM97} or \citet{hV04}.  
		
	\item The general link between local hyperbolicity (chaos) and stochasticity \citep{bC79_universal} needs to be studied, a question which has hitherto remained seriously underrated in celestial mechanics. Statistical descriptions of the action evolution, generally through adequate averaged quantitates, are at the foundations of statistical mechanics and ergodic theory---the theory of the long-term statistical behavior of dynamical systems. The injection of the concepts and tools of these fields to the applied (physical) MEO problem must still be performed. This need, moreover, is deeply stimulated by the estimated short Lyapunov times, implying that predictions must be statistical in nature.    

	\item In the Chirikov overlapping regime ($a_0 = 29,600$ km), re-entry orbits represented a significant portion of the phase space. The exact period of time after which an orbit collides, however, is not illustrated in the FLI maps, so that there might exist orbits that collide after 100 years or after 500 years, both represented in the maps by the white color. We preferred here to keep our study succinct and focused only on dynamical principles, but such considerations are obviously relevant for the analysis of disposal strategies for the four constellations located in this precarious region of space.  

	\item Though averaging methods have proven their use in celestial mechanics, given the criticism of Arnold to the so-called ``averaging principle" that was in vogue 50 years ago, we are currently testing the simulation of the whole system (the non-averaged system) to test the validity of the averaging procedure with respect to chaotic dynamics
\citep{vA06,jD15}	

\end{enumerate}
 
\section{Conclusions}

We have presented a $2.5$-DOF secular Hamiltonian governing the long-term dynamics of the MEO region under oblateness and lunisolar perturbations. Using this analytically tractable Hamiltonian, written in terms of the Delaunay elements and reduced to a $1$-DOF pendulum system near specified resonances, we have estimated the widths of the dense network of secular lunisolar resonances permeating the $i$--$e$ phase space. As the semi-major axis recedes from Earth, scanning the MEO region (from $3$ to $5$ Earth radii), we found a transition from a globally stable regime, where resonances are thin and well separated, to a globally unstable one, where resonances widen and overlap. Surprisingly, the application of the well-known analytical Chirikov criterion, frequently used for architectural celestial mechanical studies, was never fully applied to the MEO problem, because the fundamental work of \citet{tEkH97} was overlooked by the community for nearly twenty years. This analytical methodology was completed here, and tested against predictions from a semi-numerical stability analysis, which focused on the region of MEO populated by the navigation constellations. In essence, both approaches agree and confirm the transition to chaos; yet it is well known that the approximation of treating the resonances in isolation breaks down in the overlapping regime, implying that the quantitative extent and geometry of the chaotic domains can only be obtained numerically. Our parametric and extensive numerical simulations, moreover, have permitted the isolation of the basic physical model governing the dynamics in this region. Specifically, we have shown that the grandiloquent force models, often employed in such studies, are not required to capture the stable or chaotic features of the phase space. Future analytical or numerical effects can be made with the basic 2.5-DOF averaged model of oblateness and lunisolar perturbations, developed to quadrupole order. The semi-analytical investigations have also illustrated the extraordinary richness of the dynamical structures in the $i$--$e$ phase space, whose full understanding is far from complete. The predictably of typical navigation satellites has also been quantified through the estimation of Lyapunov times, finding a barrier of predictability of only decades. It is hoped that such results will stimulate and guide the community towards systematic and fundamental statistical approaches, more suitable for describing the distribution of the actions. We have also investigated the transport characteristics in the phase space, and how resonances and chaos influence the evolutionary behavior of the inclination and eccentricity. The nature of this transport will be discussed in a forthcoming paper, as with some general stochastic properties of the motion induced by the hyperbolic (chaotic) nature of the dynamical system. The effective application of these results towards the management of the navigation satellite systems is a challenging task, the goal of which is to exploit the instabilities to identify suitable collision orbits or to use the associate transport routes to reach stable regions in phase space (parking orbits). In this regard, we stress that analytical and numerical techniques cannot be separated, and a complete, logically ordered picture can be obtained only by the application of both methods jointly.

\begin{acknowledgements}
The present form of the manuscript owes much to the critical comments and helpful suggestions of many colleagues and friends. J.D. would like to thank M. Fouchard for discussions on the FLI computations, E. Bignon and P. Mercier for support with the Stela software, as well as the ``Calcul Intensif'' team from CNES where numerical simulations were hosted. A.R. owes a special thanks to K. Tsiganis for hosting him at the Aristotle University of Thessaloniki in March, and for the numerous insightful conversations that ensued. Discussions with A. B\"{a}cker, A. Celletti, R. de la Llave, G. Haller, and J.D. Meiss at the Global Dynamics in Hamiltonian Systems conference in Santuari de N\'{u}ria, Girona, 28 June -- 4 July 2015, have been instrumental in shaping the analytical component of this work. This work is partially funded by the European CommissionÕs Framework Programme 7, through the Stardust Marie Curie Initial Training Network, FP7-PEOPLE-2012-ITN, Grant Agreement 317185. Part of this work was performed in the framework of the ESA Contract No. 4000107201/12/F/MOS ``Disposal Strategies Analysis for MEO Orbits''.
\end{acknowledgements}

\begin{appendices}

\section{Special functions in the lunisolar disturbing potential expansions} \label{sec:harmonic_coefficients}

Recall that the lunar disturbing function can be written in the compact form
\begin{align*}
	\R_\M
	= \ds \sum_{m = 0}^2 \sum_{s = 0}^2 \sum_{p = 0}^2 h_{2-2p, m, \pm s}^\M  
		\cos \Phi_{2-2p, m, \pm s}^\M, 
\end{align*}
with harmonic angle and associated harmonic coefficient
\begin{align*}
	\Phi_{2-2p, m, \pm s}^\M 
	& = (2 - 2 p) \omega + m \Omega \pm s (\Omega_\M - \pi/2) - y_s \pi, \\
	h_{2-2p, m, \pm s}^\M
	& = \frac{\mu_\M a^2}{a_\M^3 (1 - e_\M^2)^{3/2}}
		(-1)^{\floor*{m/2}} \epsilon_m \frac{(2 - s)!}{(2 + m)!}
		F_{2,m,p} (i) F_{2,s,1} (i_\M) H_{2,p,2p-2} (e) (-1)^{m + s} U_2^{m, \mp s} (\epsilon),
\end{align*}
so that a lunar resonance occurs when
\begin{align*}
	\dot\psi_{2 - 2 p, m, \pm s} = (2 - 2 p) \dot\omega + m \dot\Omega \pm s \dot\Omega_\M \approx 0.
\end{align*}
The solar disturbing function can be written in the compact form
\begin{equation*}
\begin{array}{l}
	\ds \R_\S
	= \ds \sum_{m = 0}^2 \sum_{p = 0}^2 h_{2-2p, m}^\S  \cos \Phi_{2-2p, m}^\S,
\end{array}
\end{equation*}
with harmonic angle and coefficient
\begin{align*}
	\Phi_{2-2p, m}^\S 
	& = (2 - 2 p) \omega + m (\Omega - \Omega_\S), \\
	h_{2-2p, m}^\S
	& = \frac{\mu_\S a^2}{a_\S^3 (1 - e_\S^2)^{3/2}}
		\epsilon_m \frac{(2 - m)!}{(2 + m)!}
		F_{2,m,p} (i) F_{2,m,1} (i_S) H_{2,p,2p-2} (e).
\end{align*}
A solar commensurability occurs when $(2 - 2p) \dot\omega + m \dot\Omega \approx 0$, or equivalently when $\dot\psi_{2 - 2 p, m, 0} \approx 0$. We give here the the explicit formula needed to calculate the widths of the 29 distinct curves of secular resonances, six of which are locations of lunisolar resonance (i.e., the inclination-dependent-only cases), appearing in Figure~\ref{fig:skeleton}.

The general Hansen coefficient $X_k^{l,m} (e)$, which permits the full disturbing function (prior to averaging) to be developed in terms of the mean anomalies, is a function of the orbit eccentricity and is given by the integral \citep{sH80}
\begin{align}
	X_k^{l, m} (e) = \frac{1}{2 \pi} \int_0^{2 \pi} \left( \frac{r}{a} \right)^l \cos (m f - k M)\, \mathrm{d} M.	
\end{align}
For $k = 0$, exact analytical expressions exist for the zero-order Hansen coefficients $X_0^{l,m}$ for all values of $l$ and $m$. The integral for $X_0^{l, m}$ can be written as
\begin{align}
	X_0^{l,m} & = \frac{1}{2 \pi} \int_0^{2 \pi} \left( \frac{r}{a} \right)^l \cos m f\, \mathrm{d} M.
\end{align}
Note that since cosine is an even function, it is only necessary to obtain expressions for $m > 0$ as $X_0^{l,m} = X_0^{l,-m}$. For $l \geq 1$ and $0 \leq m \leq l$, the integrals can be evaluated as
\begin{align}
	\label{eq:Hansen_pos}
	X_0^{l,m} & = \left( -\frac{e}{2} \right)^m 
	\setlength\arraycolsep{0.25pt} \renewcommand{\arraystretch}{0.75}	
	\left( \begin{array}{c} l + m + 1 \\ m \end{array} \right)
	F \left( \frac{m - l - 1}{2}, \frac{m - l}{2}, m + 1; e^2\right),
\end{align}
where $F (\alpha, \beta, \gamma; x)$ is a hypergeometric function in $e^2$. Several formulae of recurrence have been derived that greatly facilitate the calculation of these coefficients, however, for our purposes, $H_{2,p,2p-2} (e) = X_0^{2,2-2p} (e)$, for $p = 0, 1, 2$, can be easily evaluated as
\begin{align}
	\label{eq:Hansen}
	X_0^{2,\pm 2} = \frac{5}{2} e^2, \quad X_0^{2,0} = 1 + \frac{3}{2} e^2. 
\end{align}

The Kaula inclination functions $F_{2,m,p} (i)$ and $F_{2,s,1} (i_\M)$ are given by \citep{wK66}
\begin{align}
	F_{2,m,p} (i) \nonumber
	& = \sum_t \frac{(4 - 2t)!}{t! (2 -t)! (2 - m - 2t)! 2^{4 - 2t}} \sin^{2 - m - 2t} i \\
	& \hspace{12pt} 
		\sum_{s = 0}^m \left( \begin{array}{c} m \\ s \end{array} \right) \cos^s i 
		\sum_c \left( \begin{array}{c} 2 - m - 2t + s \\ c \end{array} \right)
		\left( \begin{array}{c} m - s \\ p - t - c \end{array} \right) (-1)^{c - k},
\end{align}
where $k$ is the integer part of $(2 - m)/2$, $t$ is summed from $0$ to the lesser of $p$ or $k$, and $c$ is summed over all values making the binomial coefficients nonzero. Expressions for $F_{2,m,p} (i)$ up to $m = 2$, $p = 2$, are given in Table~\ref{tab:Kaula}. 

\begin{table}[htp]
	\centering
	\caption{Kaula's inclination functions $F_{2,m,p} (i)$ from \citet{wK66}.}
	\label{tab:Kaula}
	\tabulinesep=0.3em	
	\begin{tabu}{ccc} \hline\hline
    \multicolumn{1}{c}{$m$} &
    \multicolumn{1}{c}{$p$} & 
    \multicolumn{1}{c}{$F_{2,m,p} (i)$} \\\hline    
    0 & 0 & $-3 \sin^2 i/8$ \\
    0 & 1 & $3 \sin^2 i/4 - 1/2$ \\
    0 & 2 & $-3 \sin^2 i/8$ \\
    1 & 0 & $3 \sin i (1 + \cos i)/4$ \\
    1 & 1 & $-3 \sin i \cos i/2$ \\
    1 & 2 & $-3 \sin i (1 - \cos i)/4$ \\
    2 & 0 & $3 (1 + \cos i)^2/4$ \\
    2 & 1 & $3 \sin^2 i/2$ \\
    2 & 2 & $3 (1 - \cos i)^2/4$ \\\hline
    \end{tabu}
\end{table} 

The expression for the Giacaglia functions, computed from Eq.~\ref{eq:Giacaglia}, are listed in Table~\ref{tab:Giacaglia}. 
\begin{table}[htp] 
	\captionsetup{justification=justified}
	\centering
	\caption{Giacaglia's tesseral harmonics rotation functions $U_2^{m,\mp s}$ for the Moon, where $C = \cos (\epsilon/2)$ and $S = \sin (\epsilon/2)$.}
	\label{tab:Giacaglia}
	\tabulinesep=0.3em	
	\begin{tabu}{rlcccccc}
    \multicolumn{2}{r}{$\mp s$} &
    \multicolumn{1}{c}{0} & 
    \multicolumn{1}{c}{-1} & 
    \multicolumn{1}{c}{1} & 
    \multicolumn{1}{c}{-2} & 
    \multicolumn{1}{c}{2} \\  
    $m$ \\[0.1em]
    0 	&
    		& $1 - 6 C^2 + 6 C^4$ 
		&  $-2 C S^{-1} (2 C^4 - 3 C^2 + 1)$ 
		& $-2 C S (1 - 2 C^2)$ 
    		& $C^2 S^{-2} (C^2 - 1)^2$ 
		& $C^2 S^2$ \\
    1 	&
    		& $-3 C S^{-1} (2 C^4 - 3 C^2 + 1)$ 
		& $S^{-2} (4 C^6 - 9 C^4 + 6 C^2 - 1)$ 
    		& $C^2 (4 C^2 - 3)$
		& $-C S^{-3} (C^2 - 1)^3$ 
		& $-C^3 S$ \\
	2 	& 
		& $6 C^2 S^{-2} (C^2 - 1)^2$ 
		& $-4 C S^{-3} (C^2 - 1)^3$
		& $-4 C^3 S^{-1} (C^2 - 1)$
		& $S^{-4} (C^2 - 1)^4$
		& $C^4$ 
    \end{tabu}
\end{table} 

The commensurability condition, harmonic angle, and associated harmonic coefficient for all resonance center curves are given in Tables~\ref{tab:lunar_harmonics} and \ref{tab:solar_harmonics}, color coded to match that of Fig.~\ref{fig:skeleton}. Note that when the angular arguments are the same (excepting a constant phase), as in each of the five resonance curves stemming from the critical inclination ($63.4^\circ$) at $e = 1$ or the lunisolar inclination-dependent-only resonances, the associated harmonic coefficients must be combined into one single term for the calculation of the resonant width, as detailed in Section~\ref{sec:widths}.

\begin{table}[htp]
	\captionsetup{justification=justified}
 	\centering
	\caption{Lunar secular resonance conditions, harmonic angles, and associated harmonic coefficients needed for the resonant width computations of Section~\ref{sec:widths}.}
	\label{tab:lunar_harmonics}	
	\tabulinesep=0.6em
    \begin{tabu}{llc}
    \multicolumn{3}{c}{\large\textsc{Lunar commensurabilities}} \\\hline
    \multicolumn{1}{l}{\cellcolor{Ivory2} $\dot\psi_{2-2p, m, \pm s}$} &        
    \multicolumn{1}{l}{\cellcolor{Ivory2} $\Phi_{2-2p, m, \pm s}^M$} &            
    \multicolumn{1}{c}{\cellcolor{Ivory2} $h_{2-2p, m, \pm s}^\M$}
    \\\hline
    \rowfont{\color{OrangeRed1}} \rowcolor{Ivory1}
    $\dot\psi_{2, 2, 0}$	& $2 \omega + 2 \Omega$
    		& $\ds -\frac{15 \mu_\M a^2 (3 \s^2 i_\M - 2)C^2 S^{-2} (C^2 - 1)^2}
		{32 a_\M^3 (1 - e_\M^2)^{3/2}} e^2 (1 + \text{c} i)^2$ \\
    \rowfont{\color{OrangeRed1}} \rowcolor{Ivory1}
    $\dot\psi_{2, 2, 1}$	& $2 \omega + 2 \Omega + \Omega_\M - \pi$	
    		& $\ds \frac{15 \mu_\M a^2 \s i_\M \text{c} i_\M C S^{-3} (C^2 - 1)^3}
		{16 a_\M^3 (1 - e_\M^2)^{3/2}} e^2 (1 + \text{c} i)^2$ \\  
    \rowfont{\color{OrangeRed1}} \rowcolor{Ivory1}
    $\dot\psi_{2, 2, \!^{\_}1}$	& $2 \omega + 2 \Omega - \Omega_\M$	
    		& $\ds \frac{15 \mu_\M a^2 \s i_\M \text{c} i_\M C^3 S^{-1} (C^2 - 1)}
		{16 a_\M^3 (1 - e_\M^2)^{3/2}} e^2 (1 + \text{c} i)^2$ \\
    \rowfont{\color{OrangeRed1}} \rowcolor{Ivory1}
    $\dot\psi_{2, 2, 2}$	& $2 \omega + 2 \Omega + 2 \Omega_\M - \pi$	
    		& $\ds -\frac{15 \mu_\M a^2 \s^2 i_\M S^{-4} (C^2 - 1)^4}
		{64 a_\M^3 (1 - e_\M^2)^{3/2}} e^2 (1 + \text{c} i)^2$ \\
    \rowfont{\color{OrangeRed1}} \rowcolor{Ivory1}
    $\dot\psi_{2, 2, \!^{\_}2}$ 	& $2 \omega + 2 \Omega - 2 \Omega_\M + \pi$	
    		& $\ds -\frac{15 \mu_\M a^2 \s^2 i_\M C^4}
		{64 a_\M^3 (1 - e_\M^2)^{3/2}} e^2 (1 + \text{c} i)^2$ \\
    \rowfont{\color{Green4}} \rowcolor{Ivory2}   
    $\dot\psi_{2, 1, 0}$ 	& $2 \omega + \Omega$	
    		& $\ds \frac{15 \mu_\M a^2 (3 \s^2 i_\M - 2) C S^{-1} (2 C^4 - 3 C^2 + 1)}
		{16 a_\M^3 (1 - e_\M^2)^{3/2}} e^2 \s i (1 + \text{c} i)$ \\
    \rowfont{\color{Green4}} \rowcolor{Ivory2}    
    $\dot\psi_{2, 1, 1}$	& $2 \omega + \Omega + \Omega_\M - \pi$ 		
    		& $\ds -\frac{15 \mu_\M a^2 \s i_\M \text{c} i_\M S^{-2} (4 C^6 - 9 C^4 + 6 C^2 - 1)}
		{16 a_\M^3 (1 - e_\M^2)^{3/2}} e^2 \s i (1 + \text{c} i)$ \\
    \rowfont{\color{Green4}} \rowcolor{Ivory2}    
    $\dot\psi_{2, 1, \!^{\_}1}$	& $2 \omega + \Omega - \Omega_\M$	 	
    		& $\ds -\frac{15 \mu_\M a^2 \s i_\M \text{c} i_\M C^2 (4 C^2 - 3)}
		{16 a_\M^3 (1 - e_\M^2)^{3/2}} e^2 \s i (1 + \text{c} i)$ \\
    \rowfont{\color{Green4}} \rowcolor{Ivory2}    
    $\dot\psi_{2, 1, 2}$	& $2 \omega + \Omega + 2 \Omega_\M - \pi$			
    		& $\ds \frac{15 \mu_\M a^2 \s^2 i_\M C S^{-3} (C^2 - 1)^3}
		{16 a_\M^3 (1 - e_\M^2)^{3/2}} e^2 \s i (1 + \text{c} i)$ \\
    \rowfont{\color{Green4}} \rowcolor{Ivory2}    
    $\dot\psi_{2, 1, \!^{\_}2}$	& $2 \omega + \Omega - 2 \Omega_\M + \pi$		 	
    		& $\ds \frac{15 \mu_\M a^2 \s^2 i_\M C^3 S}
		{16 a_\M^3 (1 - e_\M^2)^{3/2}} e^2 \s i (1 + \text{c} i)$ \\
    \rowfont{\color{Snow4}} \rowcolor{Ivory1}    
    $\dot\psi_{2, 0, 0}$	& $2 \omega$		
    		& $\ds -\frac{15 \mu_\M a^2 (3 \s^2 i_\M - 2) (1 - 6 C^2 + 6 C^4)}
		{64 a_\M^3 (1 - e_\M^2)^{3/2}} e^2 \s^2 i$ \\
    \rowfont{\color{Snow4}} \rowcolor{Ivory1}     
    $\dot\psi_{2, 0, 1}$	& $2 \omega + \Omega_\M - \pi$	
    		& $\ds \frac{45 \mu_\M a^2 \s i_\M \text{c} i_\M C S^{-1} (2 C^4 - 3 C^2 + 1)}
		{32 a_\M^3 (1 - e_\M^2)^{3/2}} e^2 \s^2 i$ \\
    \rowfont{\color{Snow4}} \rowcolor{Ivory1}     
    $\dot\psi_{2, 0, \!^{\_}1}$	& $2 \omega - \Omega_\M$	 	
    		& $\ds \frac{45 \mu_\M a^2 \s i_\M \text{c} i_\M C S (1 - 2 C^2)}
		{32 a_\M^3 (1 - e_\M^2)^{3/2}} e^2 \s^2 i$ \\
    \rowfont{\color{Snow4}} \rowcolor{Ivory1}     
    $\dot\psi_{2, 0, 2}$	& $2 \omega + 2 \Omega_\M - \pi$
    		& $\ds -\frac{45 \mu_\M a^2 \s^2 i_\M C^2 S^{-2} (C^2 - 1)^2}
		{64 a_\M^3 (1 - e_\M^2)^{3/2}} e^2 \s^2 i$ \\
    \rowfont{\color{Snow4}} \rowcolor{Ivory1}  
    $\dot\psi_{2, 0, \!^{\_}2}$	& $2 \omega - 2 \Omega_\M + \pi$
    		& $\ds  -\frac{45 \mu_\M a^2 \s^2 i_\M C^2 S^2}
		{64 a_\M^3 (1 - e_\M^2)^{3/2}} e^2 \s^2 i$ \\
    \rowfont{\color{Snow4}} \rowcolor{Ivory2}    
    $\dot\psi_{^{\_}2, 0, 0}$		& $-2 \omega$
    		& $\ds -\frac{15 \mu_\M a^2 (3 \s^2 i_\M - 2) (1 - 6 C^2 + 6 C^4)}
		{64 a_\M^3 (1 - e_\M^2)^{3/2}} e^2 \s^2 i$ \\
    \rowfont{\color{Snow4}} \rowcolor{Ivory2}     
    $\dot\psi_{^{\_}2, 0, 1}$		& $-2 \omega  + \Omega_\M - \pi$
    		& $\ds \frac{45 \mu_\M a^2 \s i_\M \text{c} i_\M C S^{-1} (2 C^4 - 3 C^2 + 1)}
		{32 a_\M^3 (1 - e_\M^2)^{3/2}} e^2 \s^2 i$ \\
    \rowfont{\color{Snow4}} \rowcolor{Ivory2}     
    $\dot\psi_{^{\_}2, 0, \!^{\_}1}$		& $-2 \omega - \Omega_\M$ 	
    		& $\ds \frac{45 \mu_\M a^2 \s i_\M \text{c} i_\M C S (1 - 2 C^2)}
		{32 a_\M^3 (1 - e_\M^2)^{3/2}} e^2 \s^2 i$ \\
    \rowfont{\color{Snow4}} \rowcolor{Ivory2}     
    $\dot\psi_{^{\_}2, 0, 2}$		& $-2 \omega + 2 \Omega_\M - \pi$
    		& $\ds -\frac{45 \mu_\M a^2 \s^2 i_\M C^2 S^{-2} (C^2 - 1)^2}
		{64 a_\M^3 (1 - e_\M^2)^{3/2}} e^2 \s^2 i$ \\
    \rowfont{\color{Snow4}} \rowcolor{Ivory2}  
    $\dot\psi_{^{\_}2, 0, \!^{\_}2}$		& $-2 \omega - 2 \Omega_\M + \pi$
    		& $\ds  -\frac{45 \mu_\M a^2 \s^2 i_\M C^2 S^2}
		{64 a_\M^3 (1 - e_\M^2)^{3/2}} e^2 \s^2 i$		
    \\[0.3em]\hline	
    \end{tabu}
\end{table}

\begin{table}[htp]
 	\centering
	\tabulinesep=0.6em    
	\begin{tabu}{llc}
	\multicolumn{3}{c}{\large\textsc{Lunar commensurabilities (cont.)}} \\\hline
    \multicolumn{1}{l}{\cellcolor{Ivory2} \color{black} $\dot\psi_{2-2p, m, \pm s}$} & 
    \multicolumn{1}{l}{\cellcolor{Ivory2} \color{black} $\Phi_{2-2p, m, \pm s}^\M$} &                       
    \multicolumn{1}{c}{\cellcolor{Ivory2} $h_{2-2p, m, \pm s}^\M$} \\\hline
    \rowfont{\color{Blue4}} \rowcolor{Ivory1}        
    $\dot\psi_{^{\_}2, 1, 0}$		& $-2 \omega + \Omega$		
    		& $\ds -\frac{15 \mu_\M a^2 (3 \s^2 i_\M - 2) C S^{-1} (2 C^4 - 3 C^2 + 1)}
		{16 a_\M^3 (1 - e_\M^2)^{3/2}} e^2 \s i (1 - \text{c} i)$ \\
    \rowfont{\color{Blue4}} \rowcolor{Ivory1}        
    $\dot\psi_{^{\_}2, 1, 1}$		& $-2 \omega + \Omega + \Omega_\M - \pi$			
    		& $\ds \frac{15 \mu_\M a^2 \s i_\M \text{c} i_\M S^{-2} (4 C^6 - 9 C^4 + 6 C^2 - 1)}
		{16 a_\M^3 (1 - e_\M^2)^{3/2}} e^2 \s i (1 - \text{c} i)$ \\
    \rowfont{\color{Blue4}} \rowcolor{Ivory1}        
    $\dot\psi_{^{\_}2, 1, \!^{\_}1}$ 	& $-2 \omega + \Omega - \Omega_\M$ 	
    		& $\ds \frac{15 \mu_\M a^2 \s i_\M \text{c} i_\M C^2 (4 C^2 - 3)}
		{16 a_\M^3 (1 - e_\M^2)^{3/2}} e^2 \s i (1 - \text{c} i)$ \\
    \rowfont{\color{Blue4}} \rowcolor{Ivory1}        
    $\dot\psi_{^{\_}2, 1, 2}$ 	& $-2 \omega + \Omega + 2 \Omega_\M - \pi$		
    		& $\ds -\frac{15 \mu_\M a^2 \s^2 i_\M C \S^{-3} (C^2 - 1)^3}
		{16 a_\M^3 (1 - e_\M^2)^{3/2}} e^2 \s i (1 - \text{c} i)$ \\
    \rowfont{\color{Blue4}} \rowcolor{Ivory1}     
    $\dot\psi_{^{\_}2, 1, \!^{\_}2}$ 	& $-2 \omega + \Omega - 2 \Omega_\M + \pi$	 	
    		& $\ds -\frac{15 \mu_\M a^2 \s^2 i_\M C^3 S}
		{16 a_\M^3 (1 - e_\M^2)^{3/2}} e^2 \s i (1 - \text{c} i)$ \\      
    \rowfont{\color{Red4}} \rowcolor{Ivory2}        
    $\dot\psi_{^{\_}2, 2, 0}$ 	& $-2 \omega + 2 \Omega$		
    		& $\ds -\frac{15 \mu_\M a^2 (3 \s^2 i_\M - 2) C^2 S^{-2} (C^2 - 1)^2}
		{32 a_\M^3 (1 - e_\M^2)^{3/2}} e^2 (1 - \text{c} i)^2$ \\
    \rowfont{\color{Red4}} \rowcolor{Ivory2}        
    $\dot\psi_{^{\_}2, 2, 1}$ 	& $-2 \omega + 2 \Omega + \Omega_\M - \pi$	
    		& $\ds \frac{15 \mu_\M a^2 \s i_\M \text{c} i_\M C S^{-3} (C^2 - 1)^3}
		{16 a_\M^3 (1 - e_\M^2)^{3/2}} e^2 (1 - \text{c} i)^2$ \\
    \rowfont{\color{Red4}} \rowcolor{Ivory2}        
    $\dot\psi_{^{\_}2, 2, \!^{\_}1}$ 	& $-2 \omega + 2 \Omega - \Omega_\M$
    		& $\ds \frac{15 \mu_\M a^2 \s i_\M \text{c} i_\M C^3 S^{-1} (C^2 - 1)}
		{16 a_\M^3 (1 - e_\M^2)^{3/2}} e^2 (1 - \text{c} i)^2$ \\
    \rowfont{\color{Red4}} \rowcolor{Ivory2}        
    $\dot\psi_{^{\_}2, 2, 2}$ 	& $-2 \omega + 2 \Omega + 2 \Omega_\M + \pi$
    		& $\ds -\frac{15 \mu_\M a^2 \s^2 i_\M S^{-4} (C^2 - 1)^4}
		{64 a_\M^3 (1 - e_\M^2)^{3/2}} e^2 (1 - \text{c} i)^2$ \\
    \rowfont{\color{Red4}} \rowcolor{Ivory2}    
    $\dot\psi_{^{\_}2, 2, \!^{\_}2}$ 	& $-2 \omega + 2 \Omega - 2 \Omega_\M + \pi$ 	
    		& $\ds -\frac{15 \mu_\M a^2 \s^2 i_\M C^4}
		{64 a_\M^3 (1 - e_\M^2)^{3/2}} e^2 (1 - \text{c} i)^2$ \\
    \rowfont{\color{Cyan4}} \rowcolor{Ivory1}        
    $\dot\psi_{0, 1, 0}$	& $\Omega$		
    		& $\ds -\frac{3 \mu_\M a^2 (3 \s^2 i_\M - 2) C S^{-1} (2 C^4 - 3 C^2 + 1)}
		{8 a_\M^3 (1 - e_\M^2)^{3/2}} (2 + 3 e^2) \s i \text{c} i$ \\
    \rowfont{\color{Cyan4}} \rowcolor{Ivory1}         
    $\dot\psi_{0, 1, \!^{\_}1}$ 	& $\Omega - \Omega_\M$
    		& $\ds \frac{3 \mu_\M a^2 \s i_\M \text{c} i_\M C^2 (4 C^2 - 3)}
		{8 a_\M^3 (1 - e_\M^2)^{3/2}} (2 + 3 e^2) \s i \text{c} i$ \\
    \rowfont{\color{Cyan4}} \rowcolor{Ivory1}         
    $\dot\psi_{0, 1, \!^{\_}2}$ 	& $\Omega - 2 \Omega_\M + \pi$ 	
    		& $\ds -\frac{3 \mu_\M a^2 \s^2 i_\M C^3 S}
		{8 a_\M^3 (1 - e_\M^2)^{3/2}} (2 + 3 e^2) \s i \text{c} i$ \\
    \rowfont{\color{Cyan4}} \rowcolor{Ivory1}       
    $\dot\psi_{0, 2, 0}$ 	& $2 \Omega$
    		& $\ds -\frac{3 \mu_\M a^2 (3 \s^2 i_\M - 2) C^2 S^{-2} (C^2 - 1)^2}
		{16 a_\M^3 (1 - e_\M^2)^{3/2}} (2 + 3 e^2) \s^2 i$ \\		
    \rowfont{\color{Cyan4}} \rowcolor{Ivory1}       
    $\dot\psi_{0, 2, \!^{\_}1}$ 	& $2 \Omega - \Omega_\M$
    		& $\ds \frac{3 \mu_\M a^2 \s i_\M \text{c} i_\M C^3 S^{-1} (C^2 - 1)}
		{8 a_\M^3 (1 - e_\M^2)^{3/2}} (2 + 3 e^2) \s^2 i$ \\
    \rowfont{\color{Cyan4}} \rowcolor{Ivory1}       
    $\dot\psi_{0, 2, \!^{\_}2}$ 	& $2 \Omega - 2 \Omega_\M + \pi$
    		& $\ds -\frac{3 \mu_\M a^2 \s^2 i_\M C^4}
		{32 a_\M^3 (1 - e_\M^2)^{3/2}} (2 + 3 e^2) \s^2 i$	   	           
    \\[0.3em]\hline	
    \end{tabu}
\end{table}

\begin{table}[htp]
	\captionsetup{justification=justified}
 	\centering
	\caption{Solar secular resonance conditions, harmonic angles, and associated harmonic coefficients needed for the resonant width computations of Section~\ref{sec:widths}.}
	\label{tab:solar_harmonics}		 
	\tabulinesep=0.6em    
	\begin{tabu}{lccc}
	\multicolumn{3}{c}{\large\textsc{Solar commensurabilities}} \\\hline
    \multicolumn{1}{l}{\cellcolor{Ivory2} \color{black} $\dot\psi_{2-2p, m}$} &        
    \multicolumn{1}{l}{\cellcolor{Ivory2} \color{black} $\Phi_{2-2p, m}^\S$} &                       
    \multicolumn{1}{c}{\cellcolor{Ivory2} $h_{2-2p, m}^\S$} \\\hline
    \rowfont{\color{OrangeRed1}} \rowcolor{Ivory1}        
    $\dot\psi_{2, 2}$		& $2 \omega + 2 (\Omega - \Omega_\S)$
    		& $\ds \frac{15 \mu_S a^2 \s^2 i_S}{64 a_S^3 (1 - e_S^2)^{3/2}} e^2 (1 + \text{c} i)^2$ \\
    \rowfont{\color{Green4}} \rowcolor{Ivory1}        
    $\dot\psi_{2, 1}$		& 	$2 \omega + \Omega - \Omega_\S$
    		& $\ds -\frac{15 \mu_S a^2 \s i_S \text{c} i_S}{16 a_S^3 (1 - e_S^2)^{3/2}} e^2 \s i (1 + \text{c} i)$ \\
    \rowfont{\color{Snow4}} \rowcolor{Ivory1}        
    $\dot\psi_{2, 0}$ 	& $2 \omega$
    		& $\ds -\frac{15 \mu_S a^2 (3 \s^2 i_S - 2)}{64 a_S^3 (1 - e_S^2)^{3/2}} e^2 \s^2 i$ \\
    \rowfont{\color{Snow4}} \rowcolor{Ivory1}        
    $\dot\psi_{^{\_}2, 0}$ 	& $-2 \omega$
    		& $\ds -\frac{15 \mu_S a^2 (3 \s^2 i_S - 2)}{64 a_S^3 (1 - e_S^2)^{3/2}} e^2 \s^2 i$ \\		
    \rowfont{\color{Blue4}} \rowcolor{Ivory1}        
    $\dot\psi_{^{\_}2, 1}$		& $-2 \omega + \Omega - \Omega_\S$
    		& $\ds \frac{15 \mu_S a^2 \s i_S \text{c} i_S}{16 a_S^3 (1 - e_S^2)^{3/2}} e^2 \s i (1 - \text{c} i)$ \\
    \rowfont{\color{Red4}} \rowcolor{Ivory1}    
    $\dot\psi_{^{\_}2, 2}$ 	& $-2 \omega + 2 (\Omega - \Omega_\S)$
    		& $\ds \frac{15 \mu_S a^2 \s^2 i_S}{64 a_S^3 (1 - e_S^2)^{3/2}} e^2 (1 - \text{c} i)^2$ \\
    \rowfont{\color{Cyan4}} \rowcolor{Ivory1}    
    $\dot\psi_{0, 1}$ 	& $\Omega - \Omega_\S$
    		& $\ds \frac{3 \mu_S a^2 \s i_S \text{c} i_S}{8 a_S^3 (1 - e_S^2)^{3/2}} (2 + 3 e^2) \s i \text{c} i$ \\ 
    \rowfont{\color{Cyan4}} \rowcolor{Ivory1}    
    $\dot\psi_{0, 2}$ 	& $2 (\Omega - \Omega_\S)$
    		& $\ds \frac{3 \mu_S a^2 \s^2 i_S}{32 a_S^3 (1 - e_S^2)^{3/2}} (2 + 3 e^2) \s^2 i$           
    \\[0.3em]\hline	
    \end{tabu}
\end{table}

\section{Inclination-dependent-only lunisolar resonances} \label{sec:lunisolar_res}

\begin{figure}[htp]
	\captionsetup{justification=justified}
	\centering
	\includegraphics[scale=1.0]{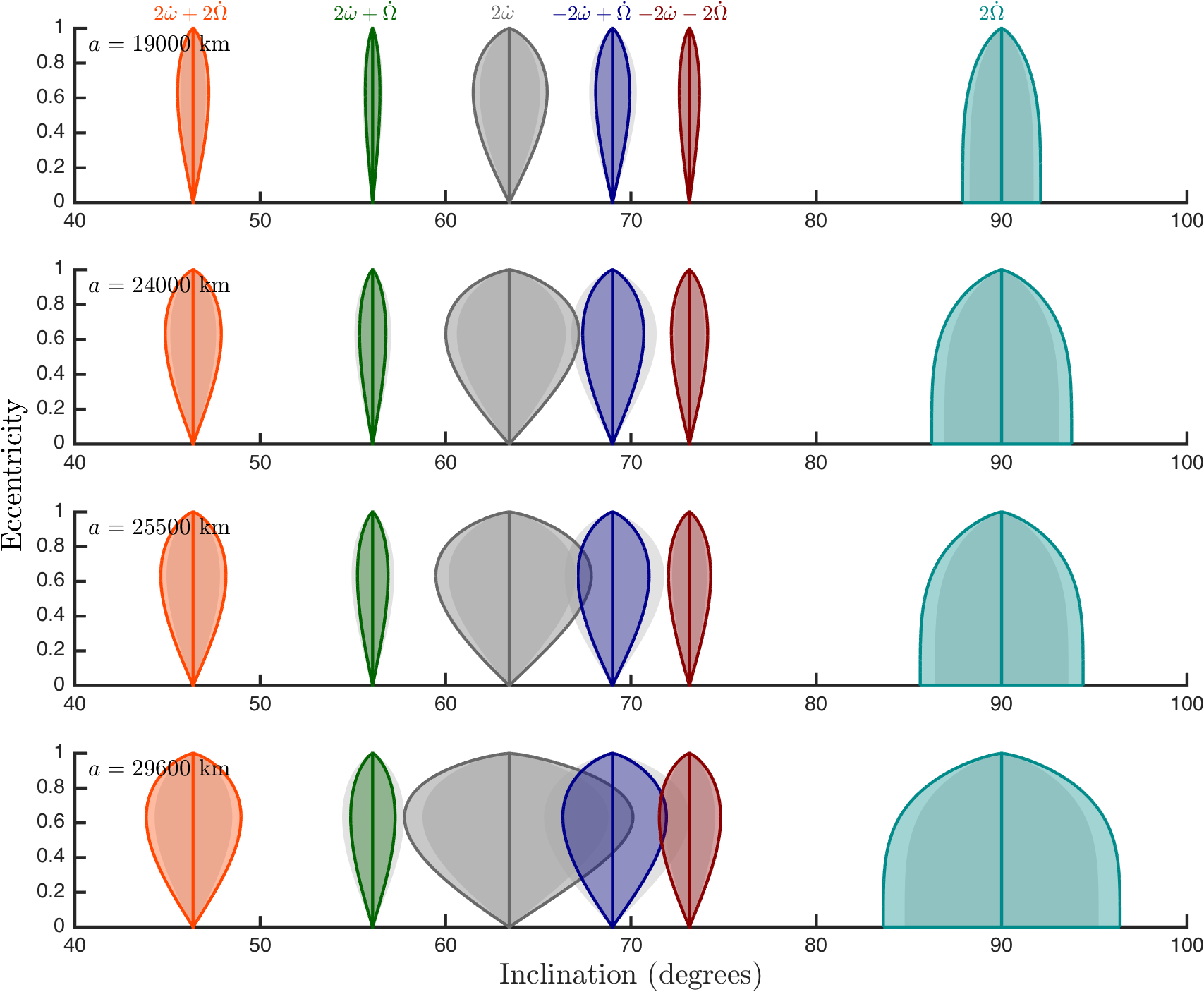} \\\vspace{12pt}
	\caption{Resonance centers and widths for the special class of inclination-dependent-only lunisolar resonances \citep[q.v.][]{sH80}, plotted both with and without the solar contribution.}
	\label{fig:chirikov_lunisolar}
\end{figure}

Figure~\ref{fig:chirikov_lunisolar} shows the lunisolar inclination-dependent-only resonance centers (solid lines) and widths (transparent lobes) for increasing values of the satellites's semi-major axis, computed with the solar contribution and without (shaded gray and without an edge color). The solar harmonics, neglected in previous works \citep{tEkH97}, play an increasingly important role as the semi-major axis is increased, either by widening or shrinking the resonance domain. The lobe shape of the lunisolar apsidal resonances \citep[q.v.][]{sB01} and the rocket shape of the nodal resonance are easily explained by the dependence of the widths on the orbit eccentricity. The widths depend on $e$ through their associated harmonic coefficients (Appendix~\ref{sec:harmonic_coefficients}), which comes in through the zero-order Hansen coefficients, and through Eq.~\ref{eq:second_partial_h}. From Eqs.~
\ref{eq:coeff}, \ref{eq:harm_coeff_lunar}, \ref{eq:harm_coeff_solar}, and \ref{eq:Hansen}, $\nu \rightarrow 0$ as $e \rightarrow 0$ for $p = 0$ or $p=2$, but not for $p = 1$. 
To extend the result to the case where $e \rightarrow 1$, let us note that 
the equation 
\begin{align}
	n_{1}\dot{\omega} + n_{2} \dot{\Omega} 
	& = \pm 2 \sqrt{\nu . \partial^2 \H^\text{sec} / \partial X^2 \big\vert_{X = X_\star}}
	\intertext{is equivalent to}
	\label{eq:lobe-e-1}
	\kappa P_{\bold{n}}(i) 
	& = \pm 2 u(e) \sqrt{\nu . \partial^2 \H^\text{sec} / \partial X^2 \big\vert_{X = X_\star}}
\end{align}
where $\kappa$ a constant (independent of $e$ and $i$), $u(e)=(1-e^{2})^{2}$ and $P_{\bold{n}}(i)=n_1 (5 \cos^{2} i -1)/2 - n_2 \cos i$. When $e \rightarrow 1$, the right-hand side  of Eq. (\ref{eq:lobe-e-1}) goes to zero, and, as a result, $i$ must be a root of $P_{\bold{n}}$ in the limit $e \rightarrow 1$.  

\section{Numerical setups} \label{sec:numerical_setup}

The dynamical model we used in the propagation of the orbits accounts for the perturbations stemming from the Earth, the Moon, and the Sun. To simplify the mathematical problem, all of these gravitational perturbations have been analytically averaged over the mean anomaly $M$ of the satellite (the fast variable), and propagated using numerical integrations, which is much faster thanks to the absence of short-periodic variations. This is a well known and efficient semi-analytical approach to study the qualitative evolution of orbits over very long timespans. The averaging procedure has been done for the Earth's geopotential up to $J_5$. For tesseral resonances (located for certain semi-major axes), where there exists a commensurability between the frequency of the satellite's mean motion and the sidereal time, a partial averaging method was applied to retain only the long-periodic perturbations. This is equivalent to retaining in the Earth's geopotential only the slowly varying quantities associated with the critical resonant angle, as it is technically detailed in \citet{vM13}. Tesseral resonances are analytically well modeled by a $1.5$-DOF Hamiltonian and they primary affect the semi-major axis of the satellite, leading to a thin chaotic response of the system (the $1.5$-DOF bounded chaos is physically equivalent to intermittency phenomenon on the semi-major axis). Despite the fact that chaotic response of the system is very confined in phase-space (on the order of kilometers), we have included these perturbations in the full system to check if such effects may couple with the lunisolar effects on longer timescales. The coefficients of the Earth's gravity field come from the GRIM5-S1 model. Our semi-analytical propagator is configurable and the third-body perturbations from the Moon and the Sun have been developed up to degree $4$ and $3$, respectively. All of the equations of motion have been formulated through the equinoctial elements, related to the Keplerian elements by
\begin{align}
	\label{eq:Equinoctial}
	\begin{array}{l}
	a, \\
	e_{y} =  e \sin(\omega + \Omega), \\
	e_{x} = e \cos(\omega + \Omega) , \\
	i_{y} = \sin(i/2) \sin(\Omega), \\	
 	i_{x} = \sin(i/2) \cos(\Omega), \\
	\xi = M + \omega + \Omega, 
	\end{array}
\end{align}
which are suitable for all considered dynamical configurations. The variational system (i.e., the equations of motion for the state and tangent vector) are then propagated using a fixed step size (set to $1$ day) Runge-Kutta 6-th order integration algorithm.

\begin{figure}[!htb]
	\captionsetup{justification=justified}
	\minipage{0.33\textwidth}
	\includegraphics[width=\linewidth]{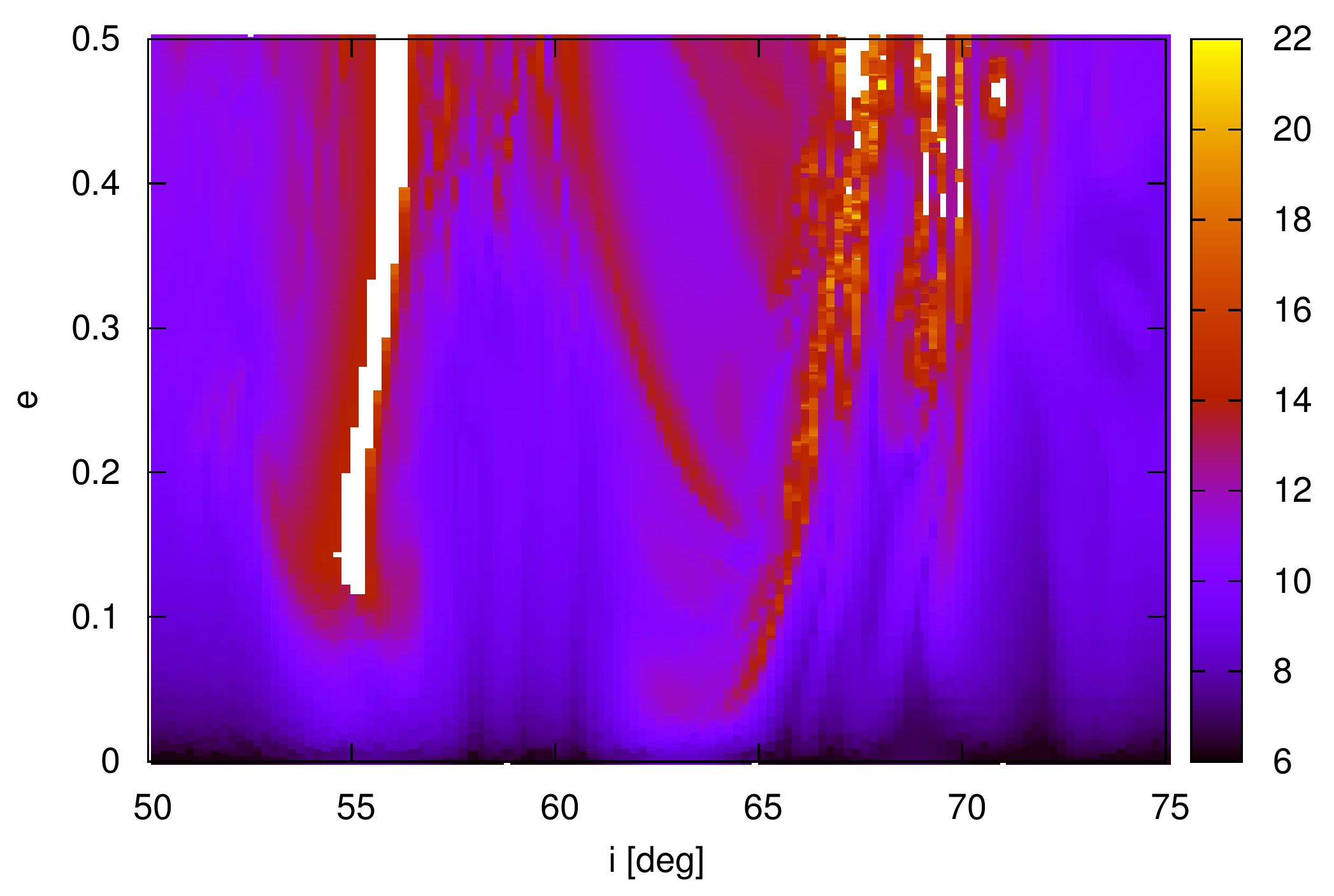}
	\endminipage\hfill
	\minipage{0.33\textwidth}
	\includegraphics[width=\linewidth]{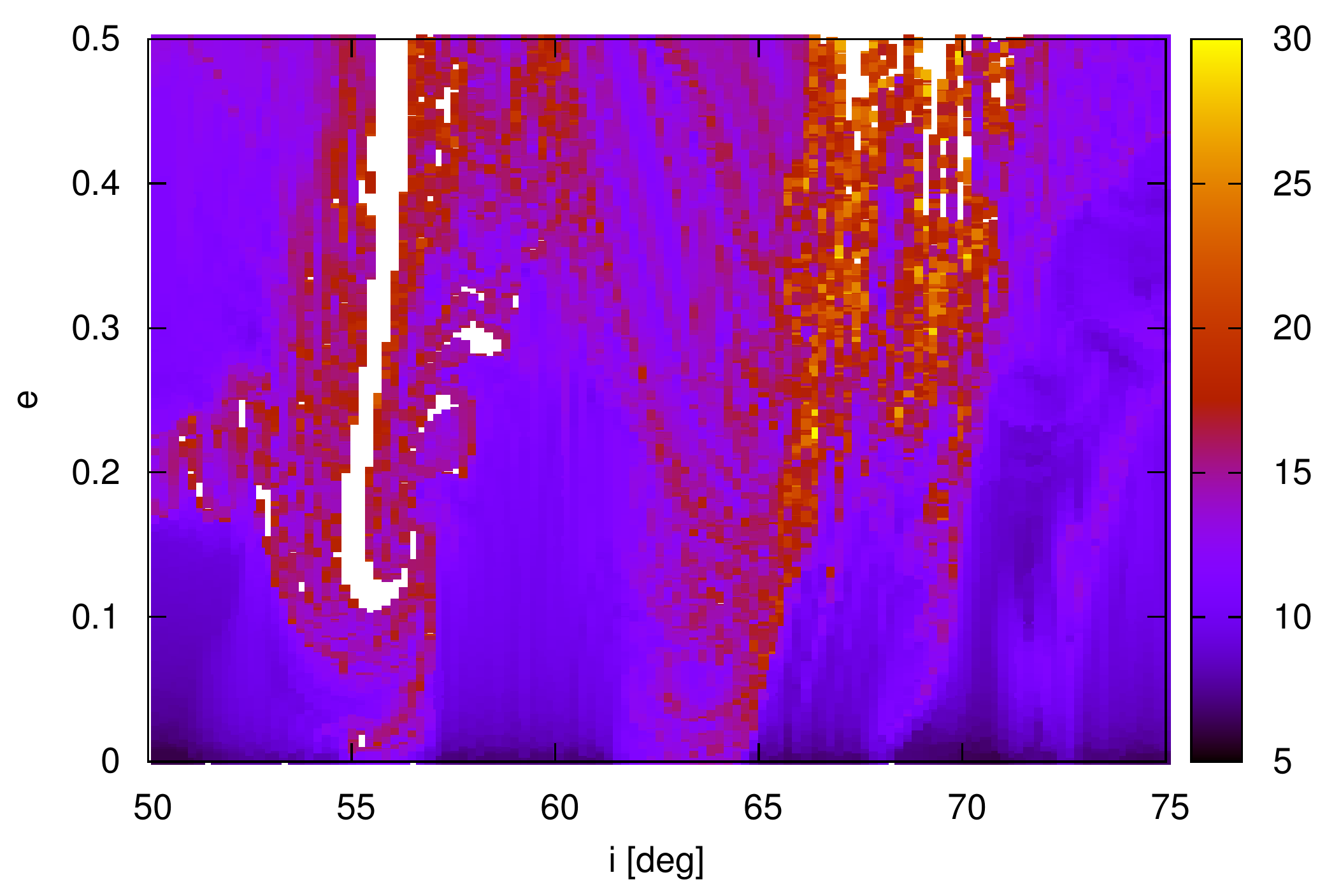}
	\endminipage\hfill
	\minipage{0.33\textwidth}%
	\includegraphics[width=\linewidth]{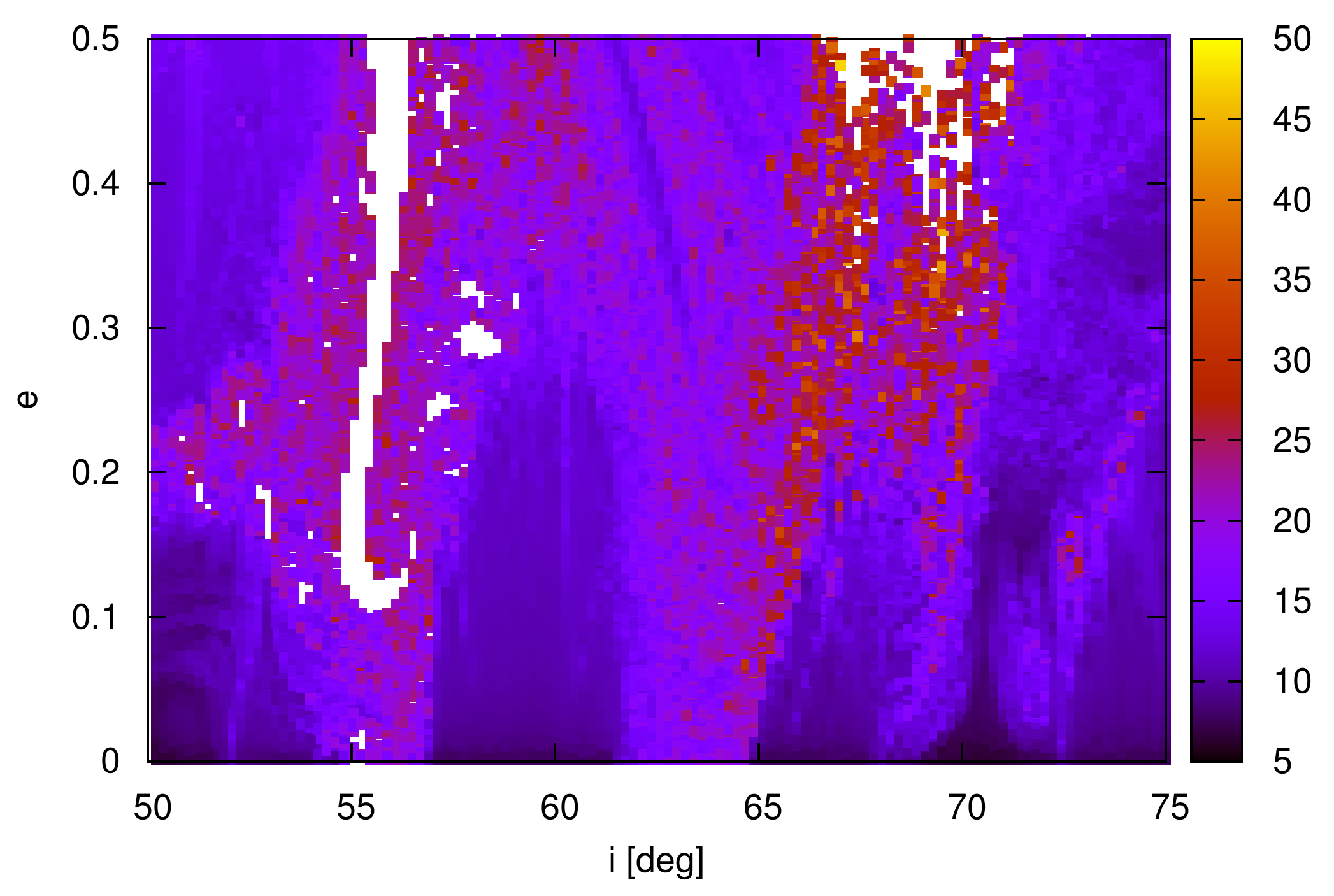}
	\endminipage
	\caption{ \label{fig:FLI-map-time} Time-calibration for the FLI maps. Here we show the typical results obtained after $100, 200$ and $530$ years respectively for $a_{0}=25,500$ km, under force model $2$.}
\end{figure}

To produce the different maps of the atlas, only the initial eccentricity and inclination were varried, distributed in a regular grid of $320 \times 115$ initial conditions for the four semi-major axes of Fig.~\ref{fig:apertures_zoom}. The computation of the FLIs on a grid of initial conditions allows a clear distinction, in short CPU time, of invariant tori and resonances \citep{eL10_numerical,mG02}. The number of initial conditions chosen was a good balance between CPU time and the final pictures offered by the resolution. Concerning the simulations, after a calibration procedure, we decided to present the results of the FLI analysis after $530$ years of propagation. By increasing the iteration time, the basic features of the FLI maps are not changed, but small higher order resonances can be detected in a proper resolution. Our chosen timescale may seem prohibitive and might also not be the best trade-off between clear distinction of the separation of nearby orbits (if ever) and CPU time. However, as showed by Fig.~\ref{fig:FLI-map-time}, a $250$ years propagation may leads to partially erroneous conclusions concerning the stability in MEO. Moreover, for weakly hyperbolic orbits (usually for moderately eccentric orbits), the time necessary to detect the divergence is longer. Consequently, there is no risk about the conclusions by presenting the results after this propagation time, at the cost of somewhat more CPU time \citep{AstroFica}. For the zoomed-in portion, the resolution and propagation duration have been increased to get finer details, structures, and more contrast in the maps. Unless explicitly stated, all others parameters (initial phases, initial epoch, initial tangent vector) have been fixed for each map. Concerning the initial tangent vector, we used the same for the whole map, a fixed normed vector orthogonal to the flow, $f$, in Eq.~\ref{eq-varia}. The robustness of our results with respect to the choice of $w_{0}$ was tested by computing the maximal Lyapunov exponent. The Lyapunov exponent, which is first of all a time average, is a property of the orbit, independent of the initial point of that orbit. By drawing the mLCE maps, we observe a nice agreement with the FLI maps in terms of the detected structures (chaotic or stable). The only (obvious) disagreement was the contrast in the maps, easily explained: Lyapunov exponents are slow to converge. This argument is strong enough to attest the relaxation that we can have with respect of $w_{0}$. Finally, the final values of the FLIs are coded, and projected into the plane of $i$--$e$ initial conditions, using a color palette ranging from black to yellow. Darker points correspond to stable orbits, while lighter colors correspond to chaotic orbits. The FLIs of orbits whose propagation stop before reaching the final time, $t_{f}$, have not been considered in the maps: they appear as a white to avoid the introduction of spurious structures.\footnote{Just before the integration stops, the FLI value of a re-entry orbit is still small but the slope of the FLI curve increases exponentially all of the sudden. Thus, there is a risk to consider a chaotic orbit as stable.} This is the reason why some maps seems to be perforated. Re-entry orbits were in practice  declared if there exist a time $t < t_{f}$ such that $a\big(1-e\big) < 120$ km. 

\end{appendices}



\begin{thebibliography}{}

\bibitem[Arnold et al.(2006)]{vA06}
Arnold, V.I., Kozlov, V.V., Neishtadt, A.I.:
Mathematical Aspects of Classical and Celestial Mechanics, 3rd edn.
Springer-Verlag, Berlin (2006)

\bibitem[Barrio et al.(2009)]{rB09_spurious} 
Barrio, R., Borczyk, W., Breiter, S.:
Spurious structures in chaos indicators maps.
Chaos Solit. Fract. \textbf{40}, 1697--1714 (2009)

\bibitem[Batygin et al.(2015)]{kB15}
Batygin, K., Morbidelli, A., Holman, M.J.:
Chaotic disintegration of the inner Solar System. 
Astrophys. J. \textbf{799}, 120--135 (2015)

\bibitem[Breiter(1999)]{sB99}
Breiter, S.:
Lunisolar apsidal resonances at low satellite orbits.
Celest. Mech. Dyn. Astr. \textbf{74}, 253--274 (1999)

\bibitem[Breiter(2001)]{sB01}
Breiter, S.:
Lunisolar resonances revisited.
Celest. Mech. Dyn. Astr. \textbf{81}, 81--91 (2001)

\bibitem[Breiter(2003)]{sB03_fundamental}
Breiter, S.:
Fundamental models of resonance.
Monograf{\'\i}as de la Real Academia de Ciencias de Zaragoza \textbf{22}, 83--92 (2003)

\bibitem[Celletti and Gale{\c s}(2014)]{aCcG14}
Celletti, A., Gale\c{s}, C.:
On the dynamics of space debris: 1:1 and 2:1 resonances.
J. Nonlinear Sci. \textbf{24}, 1231--1262 (2014)

\bibitem[Chirikov(1979)]{bC79_universal}
Chirikov, B.V.:
A universal instability of many-dimensional oscillator systems.
Phys. Rep. \textbf{52}, 263--379 (1979)

\bibitem[Cook(1962)]{gC62}
Cook, G.E.:
Luni-solar perturbations of the orbit of an Earth satellite. 
Geophys. J. \textbf{6}, 271--291 (1962)

\bibitem[Daquin(2015)]{jD15}
Daquin, J., Deleflie, F., P\'{e}rez, J.:
Comparison of mean and osculating stability in the vicinity of the (2: 1) tesseral resonant surface
Acta Astronautica. \textbf{111}, 170--177 (2015)

\bibitem[Deleflie et al.(2011)]{fD11}
Deleflie, F., Rossi, A., Portmann, C., M\'{e}tris, G., Barlier, F.:
Semi-analytical investigations of the long term evolution of the eccentricity of Galileo and GPS-like orbits.
Adv. Space Res. \textbf{47}, 811--821 (2011)

\bibitem[Delhaise and Morbidelli(1993)]{fDaM93}
Delhaise, F. Morbidelli, A.:
Luni-solar effects of geosynchronous orbits at the critical inclination.
Celest. Mech. Dyn. Astr. \textbf{57}, 155--173 (1993)

\bibitem[Ely(2002)]{tE02}
Ely, T.A.:
Eccentricity impact on east-west stationkeeping for Global Position System class orbits.
J. Guid. Cont. Dyn. \textbf{25}, 352--357 (2002)

\bibitem[Ely and Howell(1997)]{tEkH97}
Ely, T.A., Howell, K.C.:
Dynamics of artificial satellite orbits with tesseral resonances including the effects of luni-solar perturbations.
Int. J. Dyn. Stab. Syst. \textbf{12}, 243--269 (1997)

\bibitem[Froeschl\'{e} et al.(1997)]{cF97}
Froeschl\'{e}, C., Gonczi, R., Lega, E.:
The fast Lyapunov indicator: a simple tool to detect weak chaos. Application to the structure of the main asteroidal belt. 
Planet. Space Sci. \textbf{45}, 881--886 (1997)

\bibitem[Froeschl\'{e} and Lega(2000)]{cF00_structure}
Froeschl\'{e}, C., Lega, E.:
On the structure of symplectic mappings. The fast Lyapunov indicator: a very sensitive tool. 
Celest. Mech. Dyn. Astr. \textbf{78}, 167--195 (2000)

\bibitem[Froeschl\'{e} et al.(2000)]{cF00_graphical}
Froeschl\'{e}, C., Guzzo, M., Lega, E.:
Graphical evolution of the Arnold web: from order to chaos.
Science \textbf{289}, 2108--2110 (2000)

\bibitem[Garfinkel(1966)]{bG66}
Garfinkel, B.:
Formal solution in the problem of small divisors. 
Astron. J. \textbf{71}, 657--669 (1966)

\bibitem[Giacaglia(1974)]{gG74}
Giacaglia, G.E.O.:
Lunar perturbations of artificial satellites of the Earth. 
Celest. Mech. \textbf{9}, 239--267 (1974)

\bibitem[Guzzo(2002)]{mG02}
Guzzo, M., Lega, E., Froeschl\'{e}, C.:
On the numerical detection of the effective stability of chaotic motions in quasi-integrable systems
Physica D: Nonlinear Phenomena \textbf{163}, 1--25 (2002)

\bibitem[Haller(1999)]{gH99}
Haller, G.:
Chaos Near Resonance. 
Springer-Verlag, New York (1999)  

\bibitem[Hughes(1980)]{sH80}
Hughes, S.: 
Earth satellite orbits with resonant lunisolar perturbations. I. Resonances dependent only on inclination.
Proc. R. Soc. Lond. A \textbf{372}, 243--264 (1980)


\bibitem[Jupp(1988)]{aJ88}
Jupp, A.H.:
The critical inclination problem -- 30 years of progress.
Celest. Mech. \textbf{43}, 127--138 (1988)

\bibitem[Kaula(1966)]{wK66}
Kaula, W.M.:
Theory of Satellite Geodesy, 
Blaisdell, Waltham (1966)

\bibitem[Lane(1988)]{mL88}
Lane, M.T.:
An analytical treatment of resonance effects on satellite orbits.
Celest. Mech. \textbf{42}, 3--38 (1988)

\bibitem[Lane(1989)]{mL89}
Lane, M.T.:
An analytical modeling of lunar perturbations of artificial satellites of the Earth.
Celest. Mech. Dyn. Astr. \textbf{46}, 287--305 (1989)

\bibitem[Lega et al.(2010)]{eL10_numerical}
Lega, E., Guzzo, M., Froeschl\'{e}, C.:
A numerical study of the hyperbolic manifolds in a priori unstable systems. A comparison with Melnikov approximations. 
Celest. Mech. Dyn. Astr. \textbf{107}, 115--127 (2010)


\bibitem[Lithwick and Wu(2011)]{yLyW11}
Lithwick, Y., Wu, Y.:
Theory of secular chaos and Mercury's orbit.
Astrophys. J. \textbf{739}, 31--47 (2011)

\bibitem[Mardling(2008)]{rM08}
Mardling, R.A.:
Resonances, chaos and stability: The three-body problem in astrophysics. 
Lect. Notes. Phys. \textbf{760}, 59--96 (2008)

\bibitem[Morand(2013)]{vM13}
Morand, V.:
Semi analytical implementation of tesseral harmonics perturbations for high eccentricity orbits.
In Proceedings of the AAS/AIAA Astrodynamics Specialist Conference, Hilton Head, South Carolina, Paper AAS 13--749 (2013)

\bibitem[Milani(1992)]{aM92}
Milani, A., Nobili, A.M:
An example of stable chaos in the Solar System.
Nature \textbf{6379}, 569--571 (1992)

\bibitem[Morbidelli(2002)]{aM02}
Morbidelli, A.:
Modern Celestial Mechanics: Aspects of Solar System Dynamics.
Taylor \& Francis, London (2002)  

\bibitem[Morbidelli and Froeschl\'{e}(1996)]{aMcF96}
Morbidelli, A., Froeschl\'{e}, C.:
On the relationship between Lyapunov times and macroscopic instability times. 
Celest. Mech. Dyn. Astr. \textbf{63}, 227--239 (1996)

\bibitem[Morbidelli and Giorgilli(1995)]{aM95}
Morbidelli, A., Giorgilli, A.:
On a connection between KAM and Nekhoroshev's theorems. 
Physica D \textbf{86}, 514--516 (1995)

\bibitem[Morbidelli and Guzzo(1996)]{aMmG96}
Morbidelli, A., Guzzo, M.:
The Nekhoroshev theorem and the asteroid belt dynamical system. 
Celest. Mech. Dyn. Astr. \textbf{65}, 107--136 (1996)

\bibitem[Murray et al.(1997)]{nM97}
Murray, N., Holman, M.: 
Diffusive chaos in the outer asteroid belt.
Astron. J. \textbf{114}, 1246--1259 (1997)

\bibitem[Richter et al.(2014)]{mR14}
Richter, M., Lange, S., B\"{a}cker, A., Ketzmerick, R.:
Visualization and comparison of classical structures and quantum states of four-dimensional maps.
Phys. Rev. E \textbf{89}, 022902 (2014)

\bibitem[Rosengren et al.(2015)]{aR15}
Rosengren, A.J., Alessi, E.M., Rossi, A., Valsecchi, G.B.:
Chaos in navigation satellite orbits caused by the perturbed motion of the Moon.
Mon. Not. R. Astron. Soc. \textbf{449}, 3522--3526 (2015)

\bibitem[Skokos(2010)]{chS10}
Skokos, Ch.:
The Lyapunov characteristic exponents and their computation. 
Lect. Notes Phys. \textbf{790}, 63--135 (2010)

\bibitem[Todorovi\'{c} et al.(2008)]{AstroFica}
Todorovi\'{c}, N., Lega, E., Froeschl\'{e}, C.:
Local and global diffusion in the Arnold web of a priori unstable systems.
Celest. Mech. Dyn. Astr. \textbf{102}, 13--27 (2008)

\bibitem[Upton et al.(1959)]{pM59}
Upton, E., Bailie, A., Musen, P.:
Lunar and solar perturbations on satellite orbits.
Science \textbf{130}, 1710--1711 (1959)

\bibitem[Varvolglis(2004)]{hV04}
Varvoglis, H.:
Diffusion in the asteroid belt.
Proceedings of the International Astronomical Union. 
\textbf{IAUC197}, 157--170 (2004)

\end{thebibliography}
\end{document}